\documentclass[twocolumn]{aastex6}

\usepackage{amsmath}

\usepackage{array}
\newcolumntype{P}[1]{>{\centering\arraybackslash}p{#1}}
\usepackage[normalem]{ulem}

\begin{document}


\title{Probing the Intergalactic Medium with L\lowercase{y}$\mathrm{\alpha}$ and 21 \lowercase{cm} Fluctuations}

%% Use \author, \affil, plus the \and command to format author and affiliation 
%% information.  If done correctly the peer review system will be able to
%% automatically put the author and affiliation information from the manuscript
%% and save the corresponding author the trouble of entering it by hand.
%%
%% The \affil should be used to document primary affiliations and the
%% \altaffil should be used for secondary affiliations, titles, or email.

%% Authors with the same affiliation can be grouped in a single
\author{Caroline Heneka\altaffilmark{1,2,3}, Asantha Cooray\altaffilmark{2}, Chang Feng\altaffilmark{2}}
\affil{$^{1}$ Dark Cosmology Center, Niels Bohr Institute, University of Copenhagen, Juliane Maries Vej 30, DK-2100 Copenhagen, Denmark \\
$^{2}$ Department of Physics $\&$ Astronomy, University of California, Irvine, CA 92697, USA \\
$^{3}$ Institute of Theoretical Physics, University of Heidelberg, Philosophenweg 16, D-69120 Heidelberg, Germany}

%\and
%\author{Chang Feng\altaffilmark{2}}
%\affil{TeXnology Inc}
%
%\author{Julie Steffen\altaffilmark{4}}
%\affil{American Astronomical Society \\
%2000 Florida Ave., NW, Suite 300 \\
%Washington, DC 20009-1231, USA}
%
%%% Use the \and command so offset the last author.
%\and
%
%\author{Jeff Lewandowski\altaffilmark{5}}
%\affil{IOP Publishing, Washington, DC 20005}


%\altaffiltext{1}{AAS Journals Data Scientist}
%\altaffiltext{2}{greg.schwarz@aas.org}
%\altaffiltext{3}{AAS Journals Associate Editor-in-Chief}
%\altaffiltext{4}{AAS Director of Publishing}
%\altaffiltext{5}{IOP Senior Publisher for the AAS Journals}

\begin{abstract}

We study 21cm and Ly$\mathrm{\alpha}$ fluctuations, as well as H$\mathrm{\alpha}$, while distinguishing between Ly$\mathrm{\alpha}$ emission of galactic, diffuse, and scattered intergalactic medium (IGM) origin.
Cross-correlation information about the state of the IGM is obtained, testing neutral versus ionized medium cases with different tracers in a seminumerical simulation setup. In order to pave the way toward constraints on reionization history and modeling beyond power spectrum information, we explore parameter dependencies of the cross-power signal between 21$\,$cm and Ly$\mathrm{\alpha}$, which displays a characteristic morphology and a turnover from negative to positive correlation at scales of a couple Mpc$^{-1}$. In a proof of concept for the extraction of further information on the state of the IGM using different tracers, we demonstrate the use of the 21$\,$cm and H$\mathrm{\alpha}$ cross-correlation signal to determine the relative strength of  galactic and IGM emission in Ly$\mathrm{\alpha}$. We conclude by showing the detectability of the 21$\,$cm and Ly$\mathrm{\alpha}$ cross-correlation signal over more than one decade in scale at high signal-to-noise ratio for upcoming probes like SKA and the proposed all-sky intensity mapping satellites SPHEREx and CDIM, while also including the Ly$\mathrm{\alpha}$ damping tail and 21cm foreground avoidance in the modeling.

\end{abstract}


\keywords{cosmology: theory --- dark ages, reionization, first stars --- diffuse radiation --- intergalactic medium --- large-scale structure of universe}



\section{Introduction} \label{sec:intro}
At the epoch of reionization (EoR) the first galaxies emerged some 100 million years after the Big Bang, and their radiation reionized the then cold, neutral hydrogen that makes up for most of the intergalactic medium (IGM).
Regions of ionized hydrogen increased more and more in size, until they completely overlapped at the end of reionization. 
Constraints from observations of the Ly$\mathrm{\alpha}$ forest toward quasars put the end of this epoch at about one billion years after the Big Bang, or at a redshift of $z\approx 6$~\citep{2006AJ....132..117F, 2015MNRAS.447..499M}. 
 The exact reionization model itself is currently very uncertain regarding, for example, ionizing sources that drive it, spatial structure, and the onset of reionization.  Intensity mapping of emission-line fluctuations provides a powerful future avenue to test reionization models and sources, star and galaxy formation, and the structure and composition of the IGM at high redshifts. It enables us to test a wide range of scales, with the measurement of line fluctuation power spectra being feasible with future probes.

 One prominent example is the emission of the forbidden spin-flip transition of neutral hydrogen, the so-called 21cm line. Interferometers such as the Low Frequency Array~\citep{2013A&A...556A...2V} and the Murchison Widefield Array~\citep[MWA;][]{2013PASA...30...31B} aim to detect the global 21cm signal; the MWA is predicted to measure the 21cm power spectrum over more than a decade in scale~\citep{2008ApJ...680..962L,2013MNRAS.429L...5B}. Future probes such as the Hydrogen Epoch of Reionization Array (HERA)
 and the Square Kilometre Array (SKA)\footnote{https://skatelescope.org/} will be able to detect power spectra of 21cm fluctuations at high redshifts over up to two decades in scale, mapping most of the sky, as well as constrain the timing and morphology
of reionization, the properties of early galaxies, and the early sources of heating~\citep{2015aska.confE...1K,2015PritchardSKA,2017PASP..129d5001D}.
 A lot of work has gone into modeling and preparing these detections, using seminumerical simulations, such as 21cmFAST~\citep{Mesinger10} or SimFast21~\citep{SimFast21}, and hydrodynamical simulations to explore the parameter space for reionization models; see, for example,~\citet{2016MNRAS.463.1462O}.

In addition to the 21cm line,  intensity mapping of emission lines  like CO, C {\scriptsize II}, O {\scriptsize II}, N {\scriptsize II} or H$\mathrm{\alpha}$ is a promising tool at high redshifts, testing the nature of the IGM and of star and of galaxy formation~\citep{2011ApJ...741...70L, 2012ApJ...745...49G,Serra:2016jzs}.  Intensity mapping of the Ly$\mathrm{\alpha}$ line, a tracer for the ionized medium, has been explored and modeled for high redshifts in~\citet{Silva12} and \citet{Pullen:2013dir}. Not only will intensity mapping at higher redshifts prove to be important, but so too will the mapping of lines like CO and C {\scriptsize II} at low redshifts, providing a wealth of information about the galactic and IGM. Low-redshift intensity mapping will be able to disentangle foregrounds for high-redshift measurements via cross-correlation of different tracers~\citep{Comaschi:2016soe}.

When constraining reionization, the cross-correlation of different tracers, that is, emission lines tracing the neutral versus ionized medium, provides important additional information. For example, as shown in~\citet{Hutter:2016}, when coupling {\it N}-body/SPH simulations ~\citep{Springel:2000yr,Springel:2005mi} with radiative transfer code~\citep{pCRASH}, a negative cross-correlation shows up when cross-correlating 21cm and Ly$\mathrm{\alpha}$ fluctuations that breaks the parameter degeneracies present in reionization models for power spectra alone. Also, the cross-correlation of 21cm emission and Ly$\mathrm{\alpha}$ emitters improves constraints on the mean ionized fraction~\citep{Sobacchi:2016mhx}. Encouragingly, the measurement of line fluctuations beyond 21cm will be feasible with future missions, as for example the all-sky infrared intensity mapping satellites SPHEREx and the Cosmic Dawn Intensity Mapper (CDIM) proposed in~\citet{2014spherex} and~\citet{CDIM}, respectively. 

 In this paper, we want to show how robust information on reionization is obtained with tools other than the power spectrum, when cross-correlating intensity maps of line emission for tracers of galactic emission and of neutral and ionized media. The cross-correlation signal of intensity maps is less prone to suffer from systematics or incomplete foreground removal and is quite independent of the exact modeling of line-emitting galaxies. We therefore explore in detail, including a wealth of physical effects in the simulations, the cross-correlation signal for 21cm (tracer of neutral IGM) versus Ly$\mathrm{\alpha}$ (tracer of ionized medium), as well as Ly$\mathrm{\alpha}$ versus H$\mathrm{\alpha}$ (tracer of galactic emission). We demonstrate the measurability of the cross-correlation signal, which is highly sensitive to the structure of the ionized versus neutral medium and therefore crucial in constraining reionization history and models.

Our paper is organized as follows. We start in Section~\ref{sec:sim} with a detailed discussion of our simulation of intensity maps for 21cm fluctuations, for different Ly$\mathrm{\alpha}$ emission components, and for H$\mathrm{\alpha}$ emission, and we show the respective power spectra. In Section~\ref{sec:cross} we present the cross-correlation signals of 21cm and Ly$\mathrm{\alpha}$, as well as Ly$\mathrm{\alpha}$ and H$\mathrm{\alpha}$, and vary some of the model parameters. We conclude with signal-to-noise ratio calculations for both 21cm and Ly$\mathrm{\alpha}$ auto spectra as well as their cross-power spectra for a combined measurement with SKA stage one and SPHEREx as well as CDIM in Section~\ref{sec:StoN}.

\section{Simulation of Line Fluctuations}\label{sec:sim}

\subsection{21 cm Fluctuations}\label{sec:21cm}
In this section, we briefly discuss the simulated 21cm line emission, which traces the neutral IGM and will be used for cross-correlation studies in later sections. By 21cm temperature, we mean the brightness temperature for the forbidden spin-flip transition of neutral hydrogen in its ground state.

Seminumerical codes efficiently simulate ionization and 21cm temperature maps, while showing good agreement with both {\it N}-body/radiative transfer codes and analytical modeling at redshifts relevant for the EoR~\citep{Santos:2007dn,Trac:2008yz}. We aim to achieve a relatively time-efficient exploration of the model parameter space, especially when coupling the simulation of 21cm and Ly$\mathrm{\alpha}$ fluctuations for cross-correlations studies, while modeling relevant effects as physically accurately as possible and improving the modeling with parameterizations from observations. For the simulation of galactic Ly$\mathrm{\alpha}$ and H$\mathrm{\alpha}$ emission contributions in later sections, we also want to create halo catalogs beyond density fields created in Lagrangian perturbation theory (as used for the 21cm maps). We therefore use the parent code to 21cmFAST, DexM~\citep{DexM07},\footnote{http://homepage.sns.it/mesinger/Download.html} to create linear density, linear velocity, and evolved velocity fields at first order in Lagrangian perturbation theory~\citep[Zel'dovich approximation,][]{Zel70} and ionization fields in the framework of an excursion set approach, while having a halo finder option to create a corresponding halo catalog. 

With density, velocity, and ionization fields, the 21cm brightness temperature offset $\delta T_\mathrm{b}$ of the spin gas temperature $T_\mathrm{S}$ from the cosmic microwave background (CMB) temperature $T_{\gamma}$ at redshift $z$ is obtained via
\begin{align}
\delta T_\mathrm{b} \left( z \right) &= \frac{T_\mathrm{S}-T_{\gamma}}{1+z}\left(1- e^{-\tau_{\nu_0}} \right) \nonumber \\
& \approx  27x_\mathrm{HI}\left( 1+\delta_\mathrm{nl}\right)\left( \frac{H}{\mathrm{d}v_\mathrm{r}/\mathrm{d}r + H}\right)\left( 1- \frac{T_{\gamma}}{T_\mathrm{S}}\right) \nonumber \\ & \vspace{0.2cm}\times \left( \frac{1+z}{10} \frac{0.15}{\Omega_\mathrm{m} h^2}\right) \left( \frac{\Omega_\mathrm{b} h^2}{0.023}\right) \mathrm{mK} , \label{eq:Tb}
\end{align}
where redshift $z$ is related to observed frequency $\nu$ as $z = \nu_0/\nu -1$, with optical depth $\tau_{\nu_0}$ at rest-frame frequency $\nu_0$, ionization fraction $x_\mathrm{HI}$, nonlinear density contrast $\delta_\mathrm{nl}=\rho / \bar{\rho}_0 -1$, Hubble parameter $H\left( z\right)$, comoving gradient of line-of-sight velocity $\mathrm{d}v_\mathrm{r}/\mathrm{d}r$, as well as present-day matter density $\Omega_\mathrm{m}$, present-day baryonic density $\Omega_\mathrm{b}$, and Hubble factor $h$. The approximation in Equation~(\ref{eq:Tb}) assumes a postheating regime with the CMB background temperature being much smaller than the spin gas temperature $T_{\gamma} \ll T_\mathrm{S}$, so that the full spin gas temperature evolution with redshift can be neglected when calculating the brightness temperature offset $\delta T_\mathrm{b}$. For the simulation results shown in this study, we nevertheless ran the full spin temperature evolution from redshift $z=35$ down to $z=6$, which is more computationally costly, for consistency with the calculations of Ly$\mathrm{\alpha}$ intensity fluctuations in the IGM in Section~\ref{sec:IGMdiff}, where the full gas temperature evolution is required.

Throughout this paper, our fiducial cosmology assumes $\Lambda$CDM with parameters
\begin{align}
w &=-1, \ \Omega_\mathrm{m}=0.32, \ \Omega_\mathrm{K}=0, \ \Omega_\mathrm{b}=0.049, \nonumber \\
h &=0.67, \ \sigma_8=0.83, \ n_\mathrm{s}=0.96, \ \Omega_\mathrm{r}=8.6\times10^{-5} \,, \nonumber
\end{align}
as well as $N_\mathrm{eff}=3.046$ and $Y_\mathrm{He}=0.24$. Reionization model parameters are the ionizing photon mean free path $R_\mathrm{mfp}^\mathrm{UV}$, the minimal virial temperature of halos contributing ionizing photons $T_\mathrm{vir}$, the efficiency parameter for the number of X-ray photons per solar mass of stars $\zeta_\mathrm{x}$, the fraction of baryons converted to stars $f_{*}$, and the efficiency factor for ionized bubbles $\zeta$. A bubble of radius $R$ is said to be ionized when the collapse fraction smoothed on scale $R$ fulfills the criterium $f_\mathrm{coll} \geq \zeta^{-1}$.
The fiducial reionization model parameters used throughout this paper, unless stated otherwise, are
\begin{align}
R_\mathrm{mfp}^\mathrm{UV} &=40 \, \mathrm{Mpc}, \  T_\mathrm{vir}=10^4\, \mathrm{K}, \nonumber \\
\zeta_\mathrm{x} = &10^{56}, \  f_{*} =0.1, \ \zeta=10 . \nonumber 
\end{align}
All distances and scales are expressed as comoving in units of Mpc and Mpc$^{-1}$, respectively. 

\begin{figure*}
%\figurenum{2}
%\hfill \hspace{0.02cm}
\includegraphics[width=0.9\columnwidth]{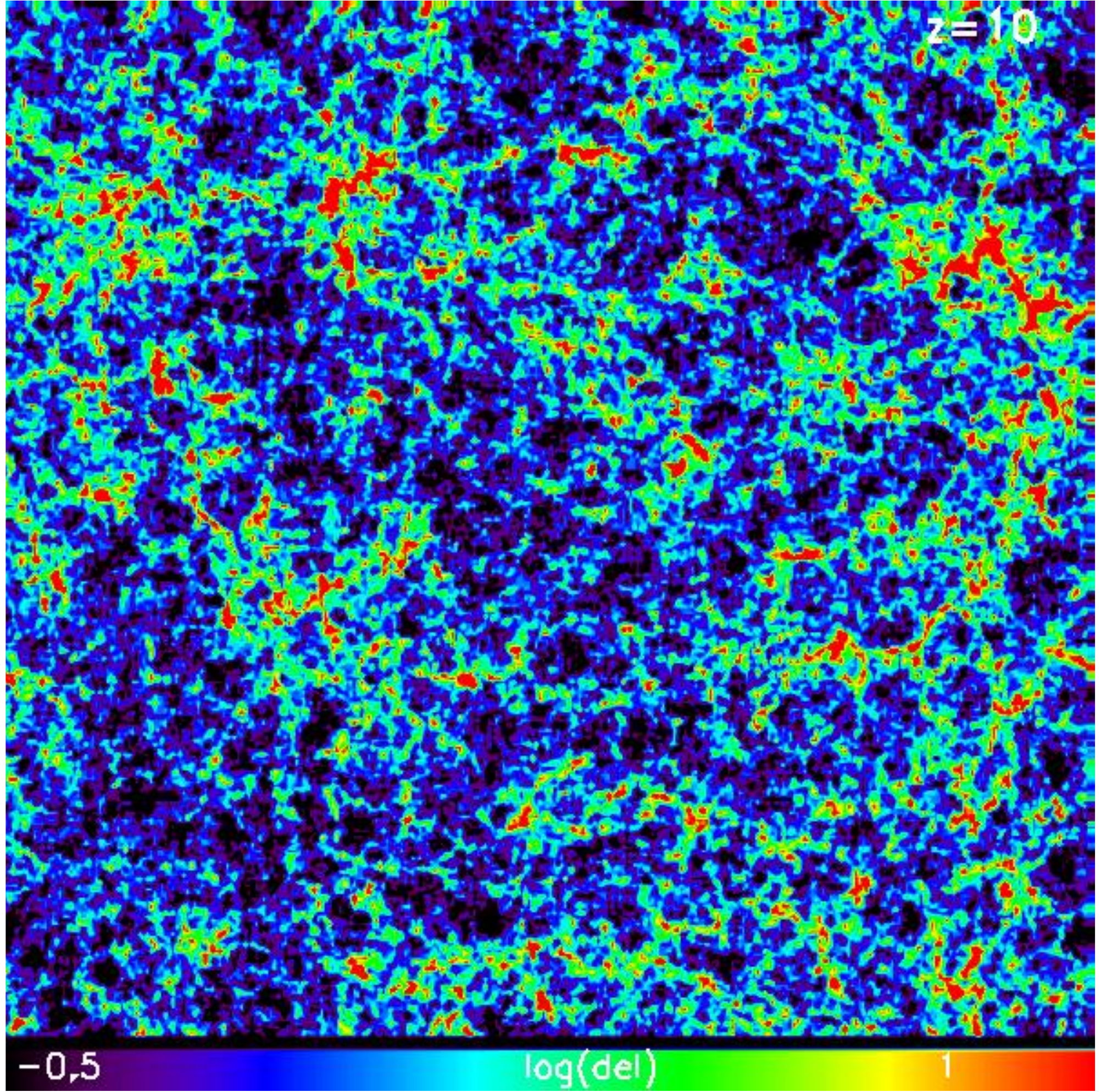}
%\hfill
\includegraphics[width=0.9\columnwidth]{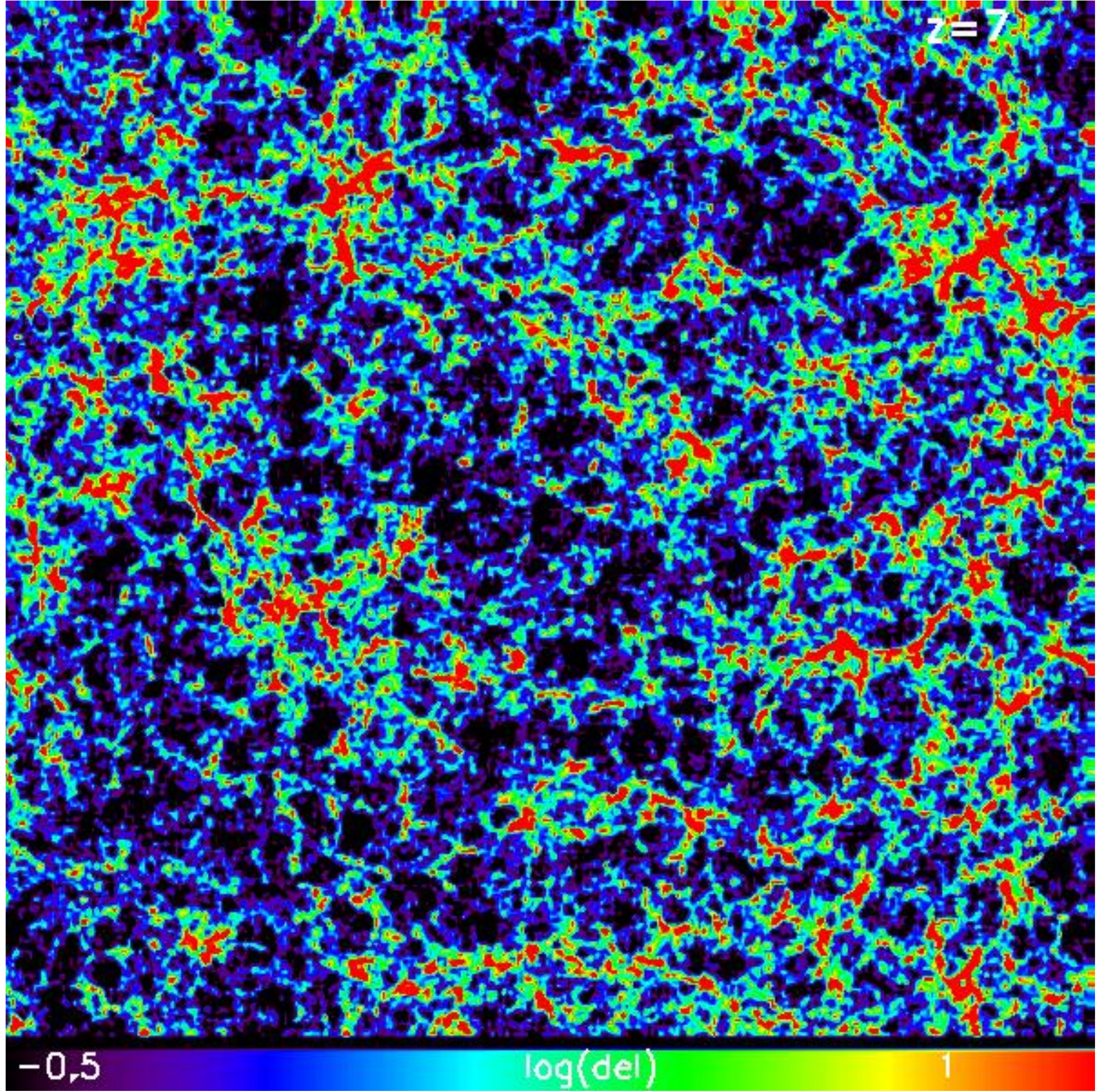}
%\hfill
%\vspace{0.5cm}
\includegraphics[width=0.9\columnwidth]{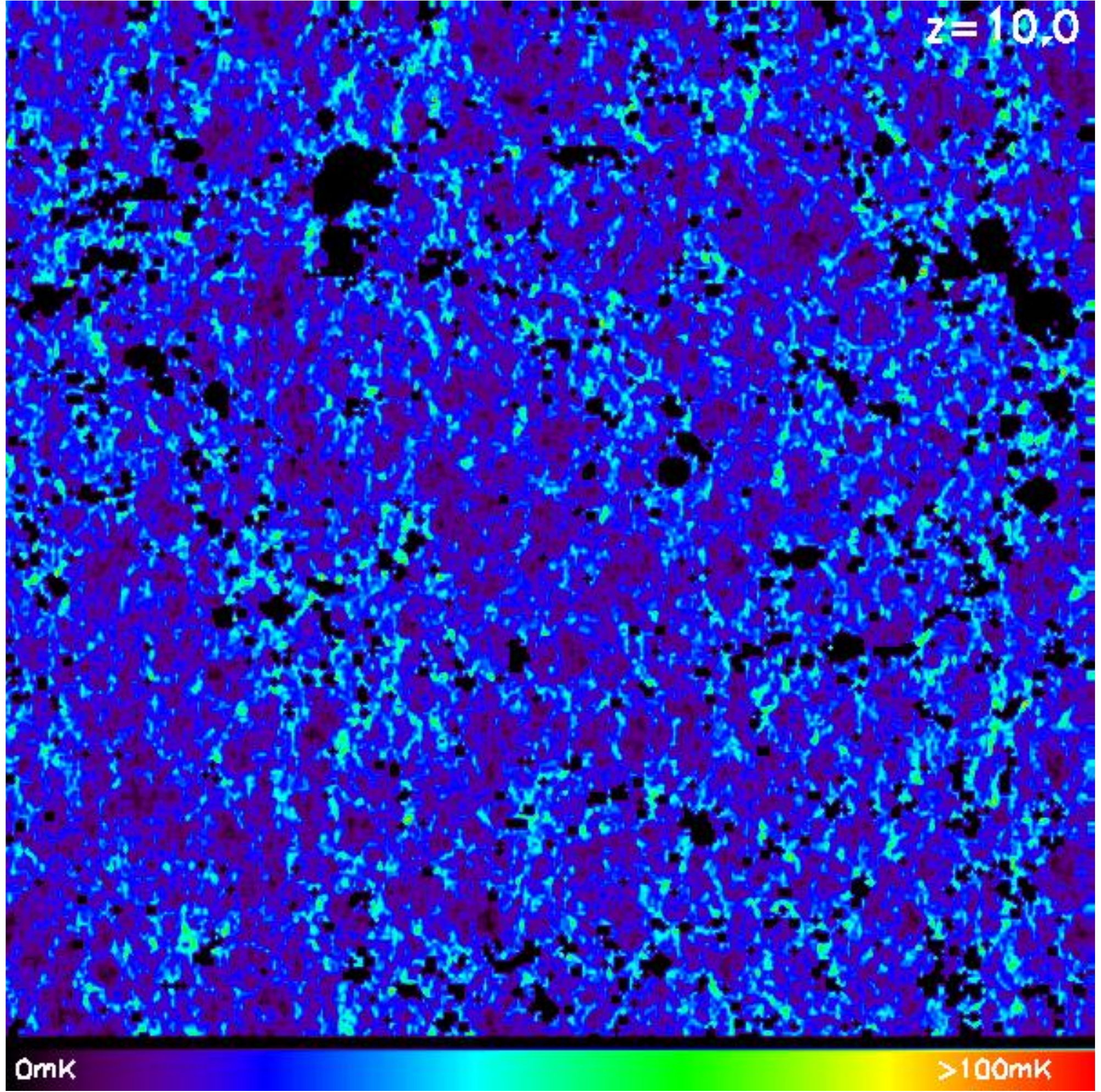}
\includegraphics[width=0.9\columnwidth]{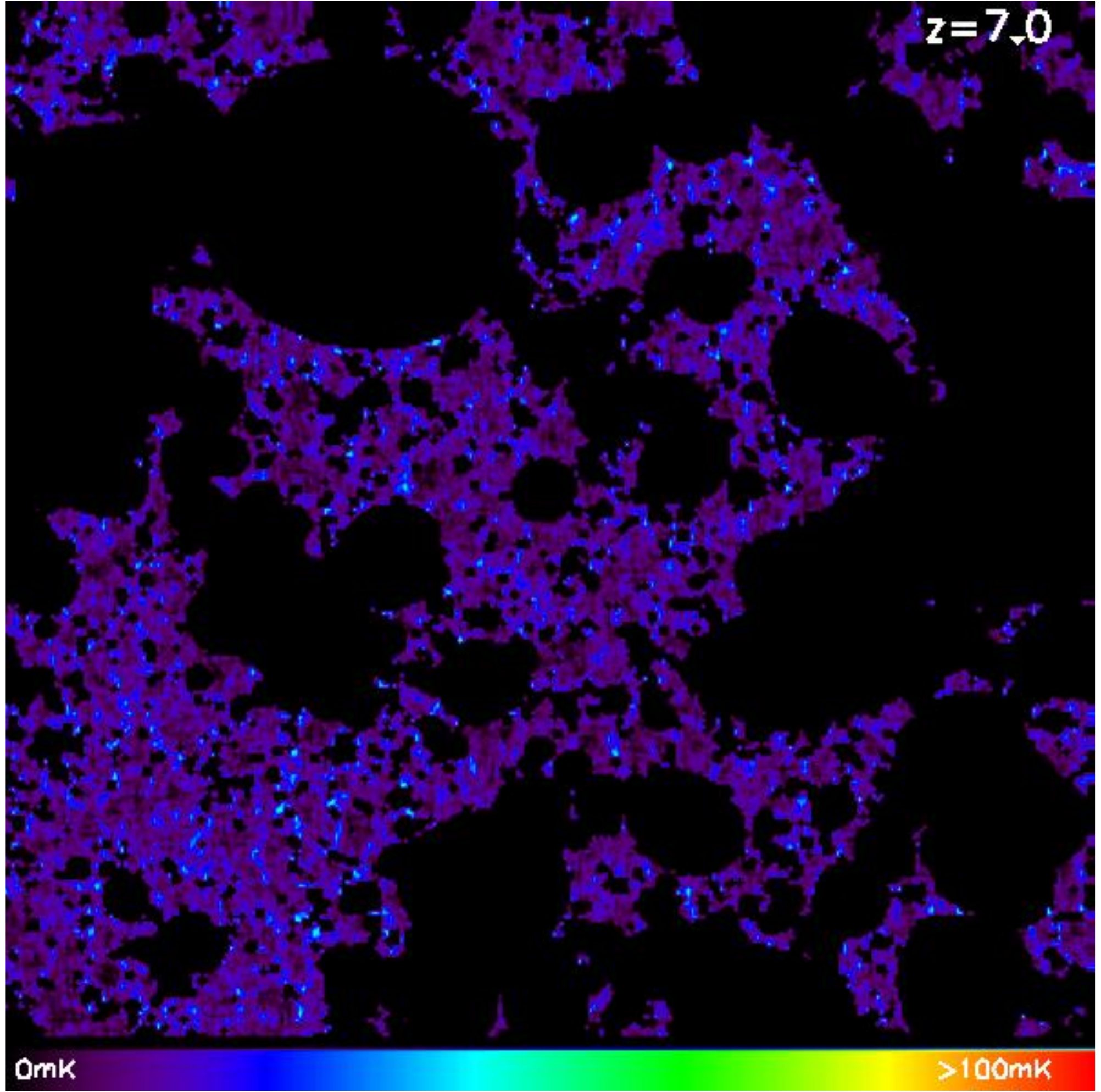}
%\hfill
\includegraphics[width=0.9\columnwidth]{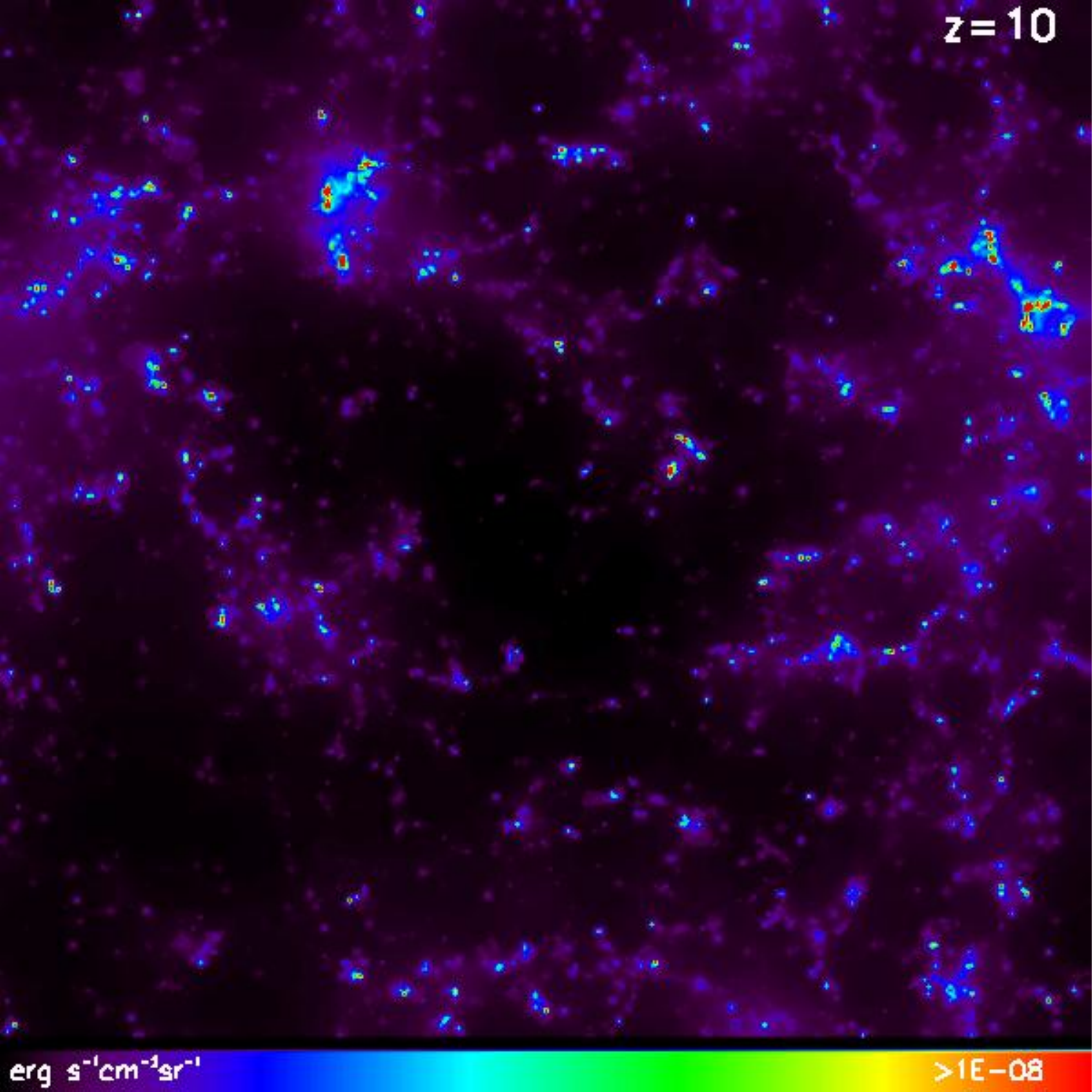}
\hfill
\includegraphics[width=0.9\columnwidth]{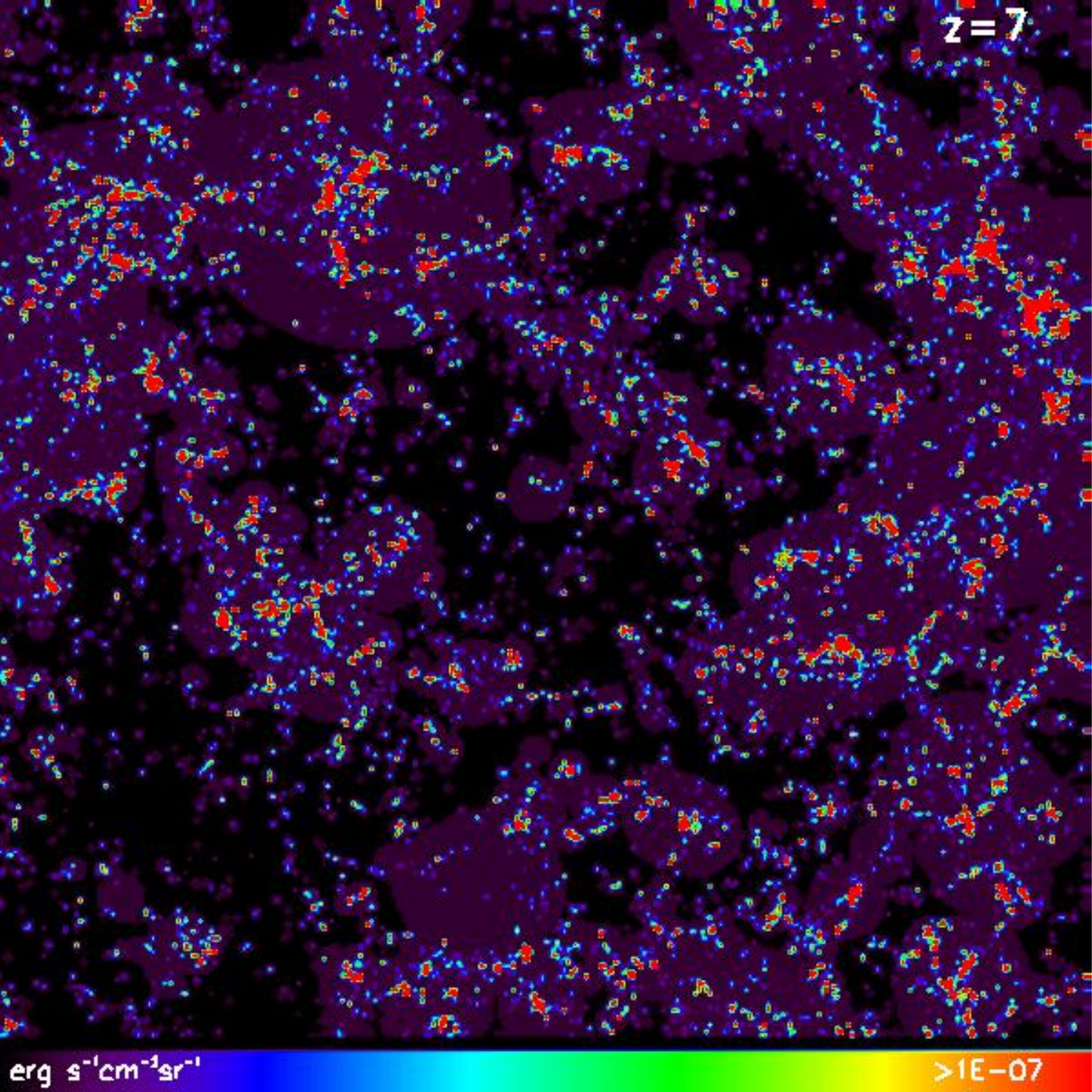}
%\vspace{0.2cm}
\caption{Slices of simulated density (top) and corresponding 21cm brightness temperature offset $\delta T_\mathrm{b}$ (middle) in a 200 Mpc box. Left: redshift $z=10$ and mean neutral fraction of $\bar{x}_\mathrm{HI}=0.87$; Right: redshift $z=7$ and $\bar{x}_\mathrm{HI}=0.27$; parameter settings as in Section~\ref{sec:21cm}. The two bottom panels show for comparison the total simulated Ly$\mathrm{\alpha}$ surface brightness in erg$\,$s$^{-1}$cm$^{-2}$sr$^{-1}$; for a detailed description of these simulations and a description of different contributions to Ly$\mathrm{\alpha}$ emission taken into account, see Section~\ref{sec:lya}. \label{fig:21cm}}
\end{figure*}

 Figure~\ref{fig:21cm} shows the simulated density field (top panels) and 21cm brightness temperature offset (middle panels) in a simulation box slice of (200 x 200) Mpc  at redshift $z=10$ for mean neutral fraction $\bar{x}_\mathrm{HI}=0.87$ (left panels) and at $z=7$ for $\bar{x}_\mathrm{HI}=0.27$ (right panels). Going from $z=10$ to $z=7$, i.e., from high to low redshift, a more peaked density field is obvious, as well as the growth of ionized patches with negligible 21cm emission, as 21cm emission is tracing neutral hydrogen. The two bottom panels show for comparison the corresponding simulation box of total Ly$\mathrm{\alpha}$ surface brightness for the same density field; the simulation of  Ly$\mathrm{\alpha}$ emission is discussed in detail in Section~\ref{sec:lya}.

We calculate temperature fluctuations on the grid $\delta_{21} \left(\bf{x},z\right) $ as
\begin{equation}
\delta_{21} \left(\bf{x},z\right) = \frac{\delta T_\mathrm{b}\left( \bf{x},z\right)}{\bar{T}_{21}\left( z\right)}-1
\end{equation}
with average temperature $\bar{T}_{21}\left( z\right)$; analogous for fluctuations in surface brightness. In the following, we define the dimensionless 21cm power spectrum as 
$\tilde{\Delta}_{21}\left( k\right) = k^3/\left( 2\pi^2 V\right) \left< |\delta_{21}|^2\right>_k$ and the dimensional power spectrum as $\Delta_{21}\left( k\right) = \bar{T}_{21 }^2 \tilde{\Delta}_{21}\left( k\right) $.

\vspace{1cm}
%\clearpage
\subsection{Ly$\mathrm{\alpha}$ Fluctuations}\label{sec:lya}
 The simulation of Ly$\mathrm{\alpha}$ fluctuations during reionization for both the galactic contribution and the emission stemming from the IGM  is described in this section. By galactic component we mean the contribution coming from within the virial radius of Ly$\mathrm{\alpha}$-emitting galaxies (LAE) themselves; the IGM component comprises both the Ly$\mathrm{\alpha}$ background caused by X-ray/UV heating and scattering of Lyman-{\it n} photons, as well as the diffuse ionized IGM around galaxies where hydrogen recombines. Ly$\mathrm{\alpha}$ emission itself is the transition of the electron in neutral hydrogen to the lowest energy state $n=1$ from $n=2$.

\subsubsection{Parametrized Ly$\mathrm{\alpha}$ Luminosities}\label{sec:Llya}
 We start by describing our procedure for modeling the Ly$\mathrm{\alpha}$ emission from galaxies.
The different  contributions to the Ly$\mathrm{\alpha}$ emission from galaxies are closely related to star formation and therefore can be connected to the star formation rate (SFR) of galaxies as a function of redshift and halo mass. The dominant source of Ly$\mathrm{\alpha}$ galactic emission is mainly hydrogen recombination, as well as collisional excitation. Two more subdominant contributors to galactic Ly$\mathrm{\alpha}$ emission are continuum emission via stellar, free-free, free-bound, and two-photon emission, as well as gas cooling via collisions and excitations in gas of temperatures smaller than $T_\mathrm{K} \approx 10^4\,$K~\citep{Fardal:2001tr, Dopita:2002cj, Fernandez:2005gx,Guo:2011}.

We start with recombination as a source of galactic Ly$\mathrm{\alpha}$ emission. Ionizing equilibrium in the interstellar gas is assumed, so that a fraction $f_\mathrm{rec}\approx 66\%$ of hydrogen recombinations result in the emission of one Ly$\mathrm{\alpha}$ photon, for spherical clouds of about $10^4\,$K~\citep{Gould:1995sj}. The fraction of Ly$\mathrm{\alpha}$ photons not absorbed by dust is parameterized as in~\citet{Hayes:2011}:
\begin{equation}
f_\mathrm{Ly\alpha}\left( z\right) = C_\mathrm{dust} 10^{-3}\left( 1+z\right)^{\zeta}\,,
\end{equation}
with $C_\mathrm{dust}=3.34$ and $\zeta=2.57$. From simulations, the escape fraction of ionizing photons can be fitted by
\begin{equation}
f_\mathrm{esc}\left( z\right) = \exp\left[ -\alpha\left( z\right) M^{\beta\left( z\right)}\right],
\end{equation}
with halo mass $M$. Parameters $\mathrm{\alpha}$ and $\beta$ are functions of redshift as in~\citet{Razoumov:2010}.
The number of Ly$\mathrm{\alpha}$ photons emitted in a galaxy per second can then be expressed as
\begin{equation}
\dot{N}_\mathrm{Ly\alpha} = A_\mathrm{He} f_\mathrm{rec} f_\mathrm{Ly\alpha} \left( 1 - f_\mathrm{esc}\right) \dot{N}_\mathrm{ion}\,,
\label{eq:NLya}
\end{equation}
with the photon fraction that goes into helium ionization $A_\mathrm{He} = \left(4-Y_\mathrm{He}\right)/\left(4-3Y_\mathrm{He}\right)$, with helium mass fraction $Y_\mathrm{He}$, and the rate of ionizing photons emitted by stars $\dot{N}_\mathrm{ion} = Q_\mathrm{ion}\times \mathrm{SFR}$. The average number of ionizing photons emitted per solar mass of star formation is taken to be $Q_\mathrm{ion}\approx 6 \times 10^{60}\,M_{\odot}^{-1}$. This value is obtained by modeling the stellar lifetime and number of ionizing photons emitted per unit time as in~\citet{Schaerer:2001jc} for a population II stellar spectral energy distribution (SED) of solar metallicity and integrating over a Salpeter initial mass function. The galactic component of Ly$\mathrm{\alpha}$ luminosity due to recombination is then simply given by
\begin{equation}
L_\mathrm{rec}^\mathrm{gal} = E_\mathrm{Ly\alpha}\dot{N}_\mathrm{Ly\alpha}\,, \label{eq:Lrec}
\end{equation}
where we assume emission at the Ly$\mathrm{\alpha}$ rest frequency $\nu_0 = 2.47\times10^{15}\,$Hz at energy $E_\mathrm{Ly\alpha} = 1.637\times10^{-11}\,$erg.

The Ly$\mathrm{\alpha}$ emission from excitation during hydrogen ionization is estimated in~\citet{Silva12} for thermal equilibrium, taking SED results from~\citet{Maraston:2004em} to get an average ionizing photon energy of $E_{\nu} = 21.4\,$eV. This energy relates to the energy emitted as Ly$\mathrm{\alpha}$ radiation due to collisional excitation as $E_{exc}/E_{\nu} \approx 0.1$~\citep{Gould:1995sj}.
The Ly$\mathrm{\alpha}$ luminosity from excitations of the interstellar medium then reads as
\begin{equation}
L_\mathrm{exc}^\mathrm{gal} = f_\mathrm{Ly\alpha} \left( 1- f_\mathrm{esc}\right) A_\mathrm{He} E_{exc} \dot{N}_\mathrm{ion} \,,
\end{equation}
again, as in the recombination case, depending on the parameterization of the SFR as a function of mass and redshift via the rate of ionizing photons $\dot{N}_\mathrm{ion}$.

\begin{figure*}
\includegraphics[width=0.95\columnwidth]{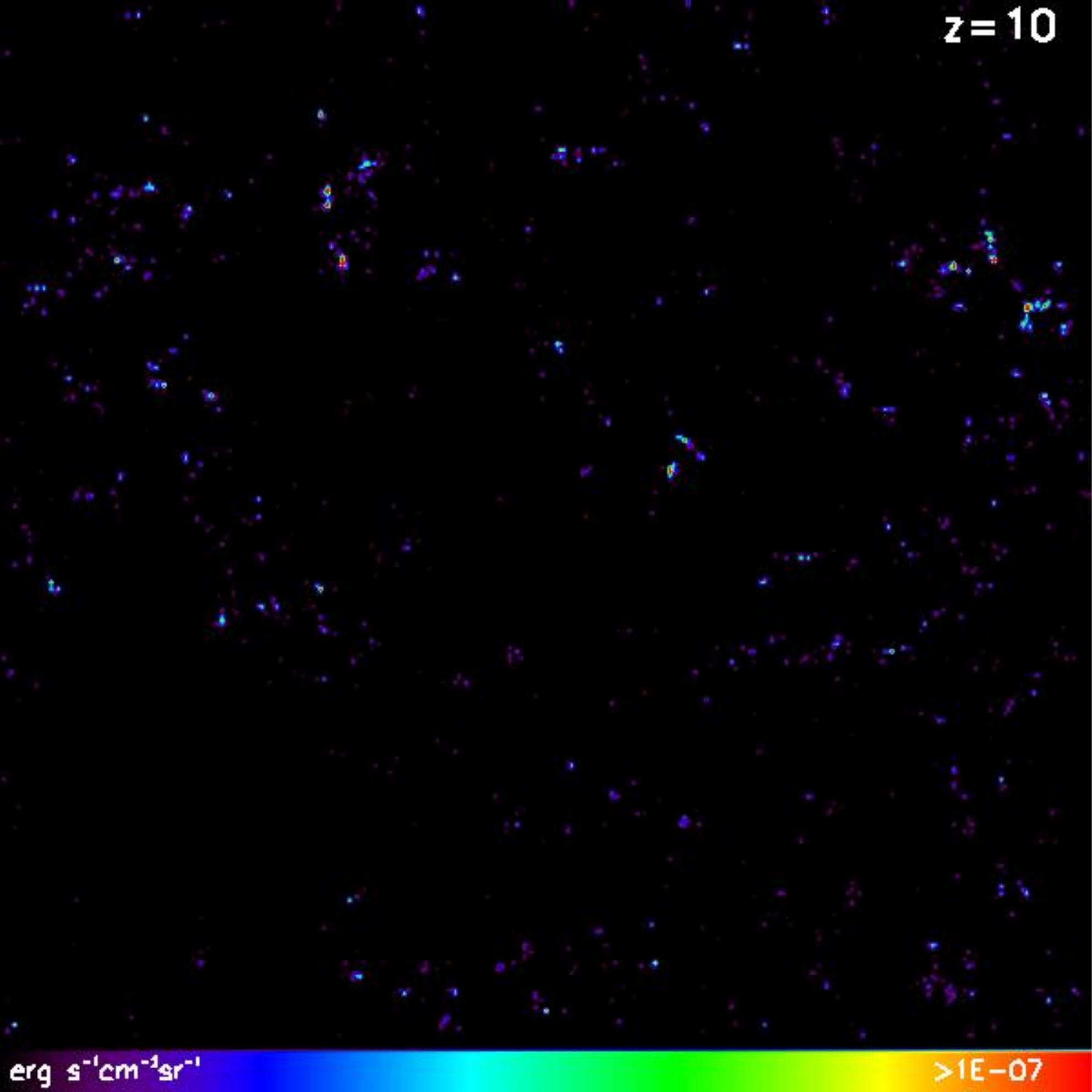}
\hfill \hspace{1.5cm}
\includegraphics[width=0.95\columnwidth]{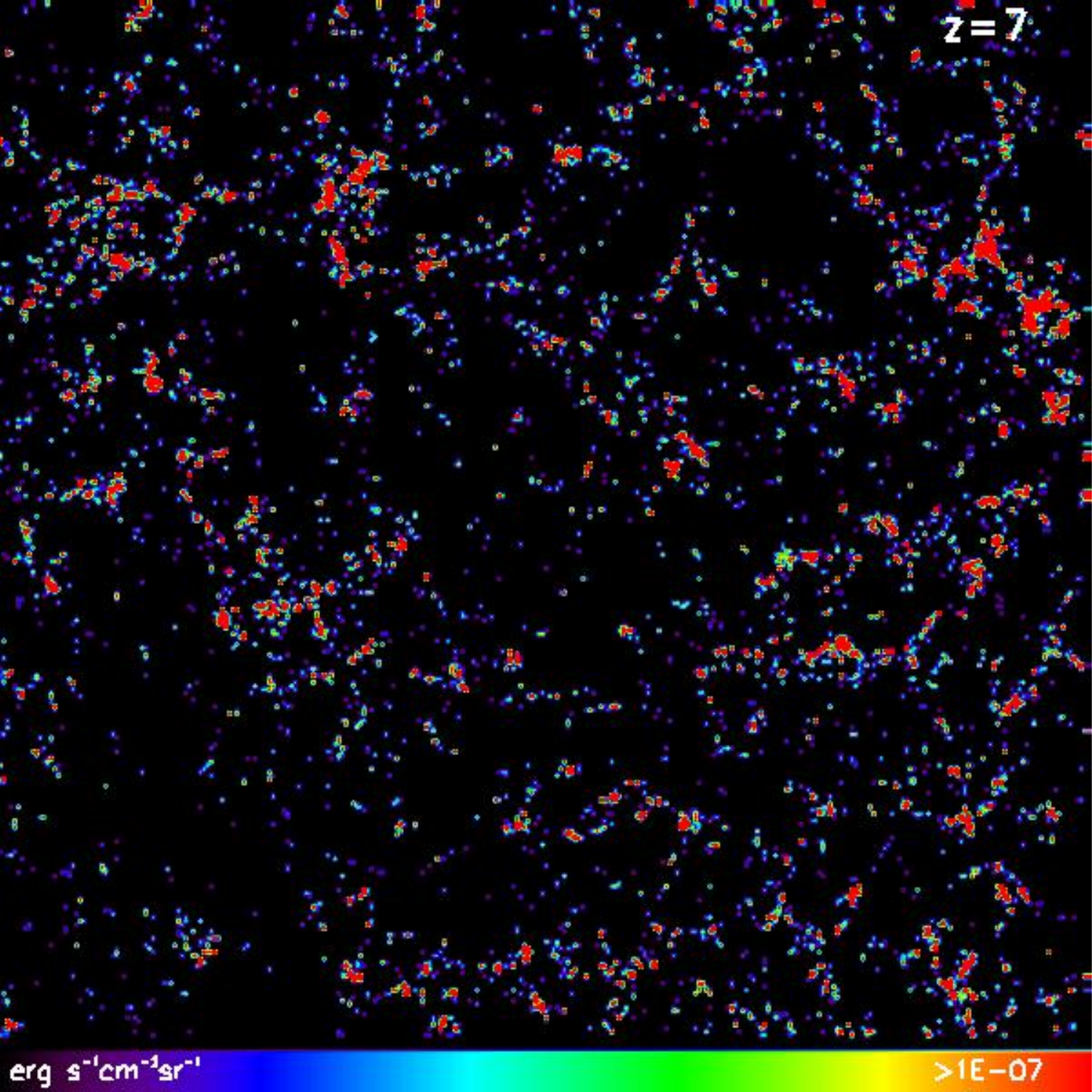}
\hfill \vspace{0.5cm}
\includegraphics[width=0.95\columnwidth]{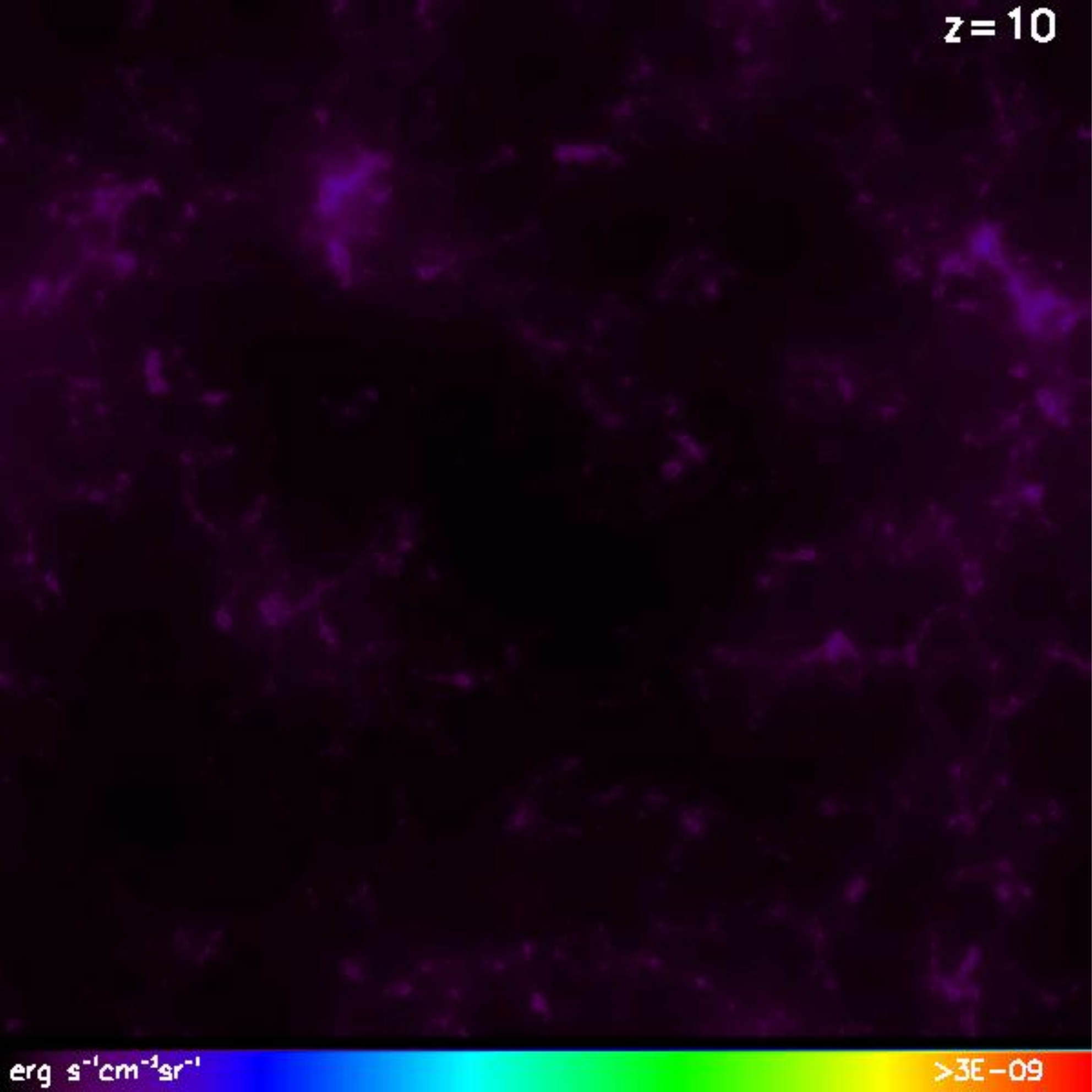}
\hfill
\includegraphics[width=0.95\columnwidth]{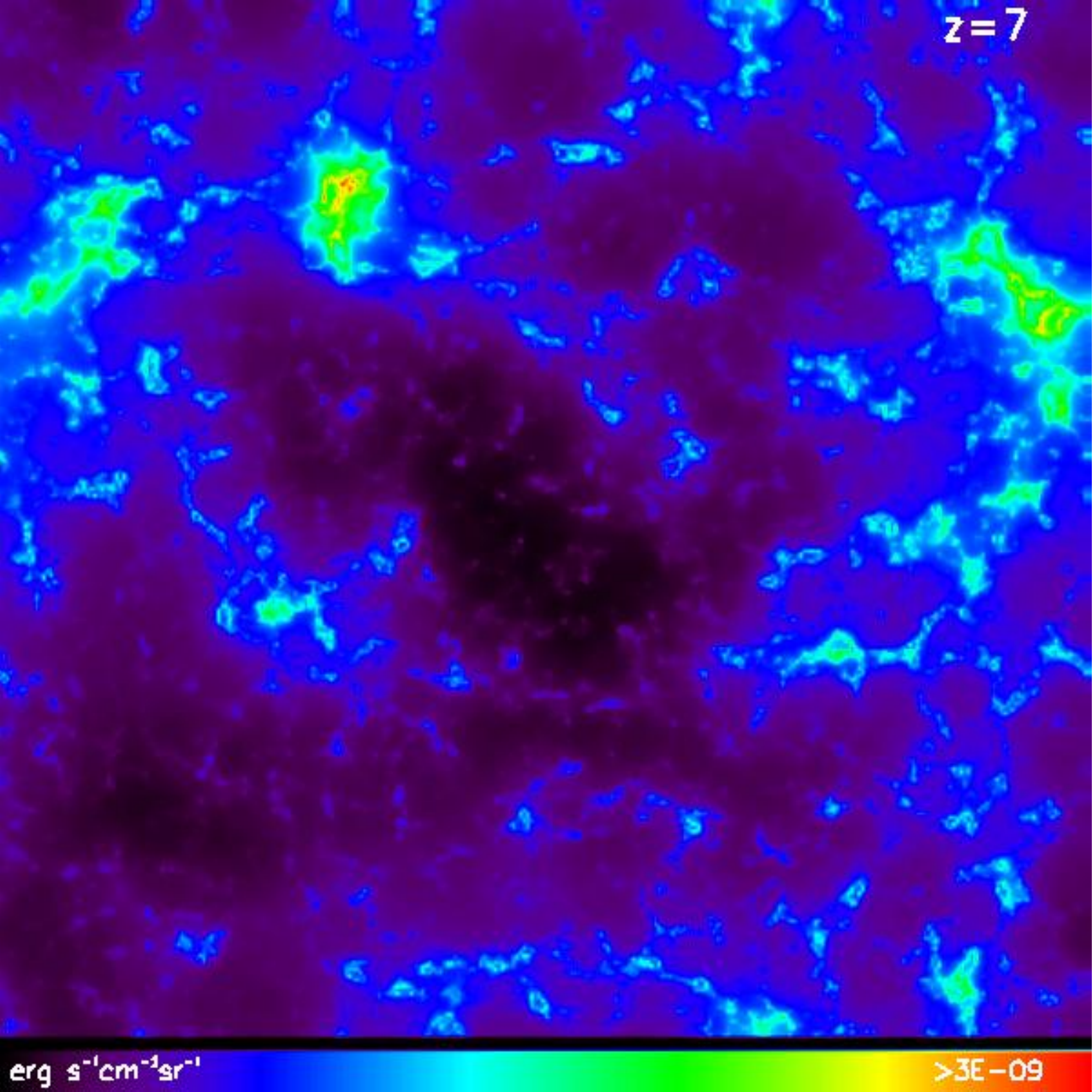}
\caption{Slices of simulations of Ly$\mathrm{\alpha}$ surface brightness in erg$\,$s$^{-1}$cm$^{-2}$sr$^{-1}$ at $z=10$ and $\bar{x}_\mathrm{HI}=0.87$ (left) and $z=7$ and $\bar{x}_\mathrm{HI}=0.27$ (right), with 200$\,$Mpc box length; Top: galactic Ly$\mathrm{\alpha}$ emission $\nu I_{\nu}^\mathrm{gal}\left( {\bf x} , z \right)$ as described in Section~\ref{sec:Llya}; bottom: scattered IGM component $\nu I_{\nu}^\mathrm{sIGM}\left( {\bf x} , z \right)$ as described in Section~\ref{sec:IGMdiff}.} \label{fig:lyabgsim}
\end{figure*}

The crucial relation between SFR and halo mass for the calculation of Ly$\mathrm{\alpha}$ luminosities is parameterized to match the observed trend of an increasing SFR for smaller mass halos, becoming almost constant for larger halo masses with $M>10^{11}M_{\odot}$~\citep{Conroy:2008dx, Popesso:2012}. The parameterization we use throughout this paper is taken from~\citet{Silva12} and was obtained by fitting to a reasonable reionization history, together with a Ly$\mathrm{\alpha}$ luminosity function compatible with observations.
This SFR reads as
\begin{equation}
\frac{\mathrm{SFR}}{M_{\odot}/\mathrm{yr}} = \left( 2.8 \times 10^{-28}\right) \left(\frac{M} {M_{\odot}}\right)^a \left( 1+ \frac{M}{c_1}\right)^b \left( 1+ \frac{M}{c_2}\right)^d  , 
\label{eq:SFR}
 \end{equation}
with fitting parameters $a=-0.94$, $d=-1.7$, $c_1 = 10^9 M_{\odot}$, and $c_2 = 7\times10^{10} M_{\odot}$. Plugging this SFR into the rate of ionizing photons in Equation~(\ref{eq:NLya}) gives the dependence of Ly$\mathrm{\alpha}$ luminosity Equation~(\ref{eq:Lrec}) on halo mass at a fixed redshift. The redshift evolution of Ly$\mathrm{\alpha}$ galactic emission depends on the escape fraction $f_\mathrm{esc}\left( z\right)$, the fraction of Ly$\alpha$ photons not absorbed by dust $f_\mathrm{Ly\alpha}\left( z\right)$, as well as halo number, mass, and distribution (also creating a spatial distribution of galactic luminosities). The total galactic Ly$\mathrm{\alpha}$ luminosity due to recombination and excitation is given by 
\begin{equation}
L^\mathrm{gal} \left(M , z \right)= L^\mathrm{gal}_\mathrm{rec} \left( M , z \right) + L^\mathrm{gal}_\mathrm{exc} \left( M , z \right)  \,, \label{eq:Lintr}
\end{equation}
for each halo of mass $M$ at redshift $z$. For simulation boxes with each voxel defined by position ${\bf x}$ and redshift $z$, one can sum the luminosities per voxel and divide by the comoving voxel volume, in order to get a smoothed luminosity density (per comoving volume) on the grid $L^\mathrm{gal} \left( {\bf x} , z\right)$. For the luminosities per voxel, we smoothed the Ly$\mathrm{\alpha}$ emission over virial radii. The comoving luminosity density then can easily be converted to surface brightness $I_{\nu}^\mathrm{gal} \left( {\bf x} , z \right)$ via
\begin{equation}
I_{\nu}^\mathrm{gal} \left( {\bf x} , z \right) = y\left( z\right) d_\mathrm{A}^2 \left( z\right) \frac{L^\mathrm{gal} \left( {\bf x} , z\right)}{4 \pi d_\mathrm{L}^2} \,, 
\label{eq:LtoI}
\end{equation}
with comoving angular diameter distance $d_\mathrm{A}$, proper luminosity distance $d_\mathrm{L}$, and $y\left( z\right)=\mathrm{d}\chi/\mathrm{d}\nu = \lambda_{0}\left( 1+z\right)^2 / H\left( z\right)$ (for comoving distance $\chi$, observed frequency $\nu$ and rest-frame wavelength $\lambda_0 = 2.46 \times 10^{-15}$m of Ly$\mathrm{\alpha}$ radiation). By assigning Ly$\mathrm{\alpha}$ luminosities to host halos depending on halo masses, we have created a spatial distribution of galactic luminosities in our simulation that follows the halo distribution and therefore is naturally position-dependent, as can clearly be seen in Figure~\ref{fig:lyabgsim} (top panels). Here we show the Ly$\mathrm{\alpha}$ surface brightness for the direct galactic emission component $I_{\nu}^\mathrm{gal}\left( {\bf x} , z \right)$ in slices through our simulation, box length 200$\,$Mpc, at redshift $z=10$ (left) and  $z=7$ (right), with more halos emitting in the Ly$\mathrm{\alpha}$ regime as reionization progresses.

\subsubsection{Ly$\mathrm{\alpha}$ Emission from the Diffuse IGM}\label{sec:IGMs}
 In addition to direct galactic emission, the Ly$\mathrm{\alpha}$ emission region is also composed of the ionized diffuse IGM around halos~\citep{Pullen:2013dir}.
Here ionizing radiation escapes the halos of Ly$\mathrm{\alpha}$-emitting galaxies and can ionize neutral hydrogen in the diffuse IGM. Similar to the emission from within halos, Ly$\mathrm{\alpha}$ radiation is then reemitted through recombinations. The comoving number density of  recombinations in the diffuse IGM reads as
\begin{equation}
\dot{n}_\mathrm{rec}\left({\bf x} ,z \right) = \alpha_\mathrm{A} n_\mathrm{e}\left( z \right) n_\mathrm{HII} \left( z \right),
\end{equation}
with the case A recombination coefficient $\alpha_\mathrm{A}$ for moderately high redshifts, free electron density $n_\mathrm{e} = x_\mathrm{i} n_\mathrm{b}$ (depending on ionization fraction $x_\mathrm{i}$ and baryonic comoving number density $n_\mathrm{b}$), and with $n_\mathrm{HII} = x_\mathrm{i} n_\mathrm{b} \left( 4-4Y_\mathrm{He}\right)/\left(4-3 Y_\mathrm{He}\right)$, the comoving number density of ionized hydrogen ($Y_\mathrm{He} $ is the helium mass fraction).
The comoving recombination coefficient  $\alpha_\mathrm{A}$ depends on the IGM gas temperature $T_\mathrm{K}$ via~\citep{Abel:1996kh,Furlanetto:2006jb}

\begin{equation}
\alpha_\mathrm{A} \approx 4.2 \times 10^{-13} \left( T_\mathrm{K}/10^4 \mathrm{K}\right)^{-0.7} \left( 1+z\right)^3 \mathrm{cm^3 s^{-1}}.
\end{equation}
The Ly$\mathrm{\alpha}$ luminosity density due to recombinations in the IGM is given by
\begin{equation}
l_\mathrm{rec}^\mathrm{IGM} \left({\bf x} , z \right) = f_\mathrm{rec} \dot{n}_\mathrm{rec}\left({\bf x} , z \right) E_\mathrm{Ly\alpha} , 
\label{eq:lrecIGM}
\end{equation}
where we insert $f_\mathrm{rec} \approx 0.66$ for the  fraction of Ly$\mathrm{\alpha}$ photons emitted per hydrogen recombination as in Section~\ref{sec:Llya} for the galactic contribution and a Ly$\mathrm{\alpha}$ rest-frame energy of $E_\mathrm{Ly\alpha} = 1.637\times 10^{-11}$erg.

We simulate the number density of recombinations per pixel by evolving gas temperature $T_\mathrm{K}$, baryonic comoving number density $n_\mathrm{b}$, and ionization fraction $x_\mathrm{i}$ in the IGM and by calculating the Ly$\mathrm{\alpha}$ luminosity density for each pixel in our simulation box. The baryonic comoving number density $n_\mathrm{b} \left({\bf x} , z \right)$ is calculated making use of the nonlinear density contrast generated by the DexM code~\citep{DexM07}, see also Section~\ref{sec:21cm}, via $n_\mathrm{b} \left({\bf x} , z \right) = \bar{n}_\mathrm{b,0} \left(1+z \right)^3\left[ 1+\delta_\mathrm{nl}\left({\bf x} , z \right)\right]$, where we take the present-day mean baryonic number density to be $\bar{n}_\mathrm{b,0} =1.905\times 10^{-7}$cm$^{-3}$. When evolving gas temperature fluctuations, we extract the gas temperature $T_\mathrm{K} \left({\bf x} , z \right)$ from the evolution equations for the full spin temperature evolution in the DexM code, which keeps track of the inhomogeneous heating history of the gas. Alternatively, we can make a conservative estimate for Ly$\mathrm{\alpha}$ brightness fluctuations by neglecting fluctuations in gas temperature $T_\mathrm{K}$ and in baryonic density $n_\mathrm{b}$. When ignoring density perturbations, we can set the comoving baryonic number density to  $\bar{n}_\mathrm{b}\left( z\right) = 1.905\times 10^{-7} \left( 1+z\right)^3$cm$^{-3}$. For ionized regions we set $T_\mathrm{K} = 10^4\,$K, corresponding to typical halo virial temperatures. This is similar to the assumption of ionized pixels being completely ionized. We do so as our code neglects photoionization heating from reionization itself when determining the temperature. For Ly$\alpha$ this is a good-enough approximation given the weak temperature dependence entering via the recombination coefficient.

The luminosity density $l_\mathrm{rec}^\mathrm{IGM} \left( {\bf x} , z\right)$ can easily be converted into surface brightness $I_\mathrm{\nu, rec}^\mathrm{IGM}\left( {\bf x} , z \right)$ of the diffuse IGM via
\begin{equation}
I_\mathrm{\nu, rec}^\mathrm{IGM}\left( {\bf x} , z \right) = y\left( z\right) d_\mathrm{A}^2 \left( z\right) \frac{l_\mathrm{rec}^\mathrm{IGM} \left( {\bf x} , z\right)}{4 \pi d_\mathrm{L}^2} , \label{eq:IGMs}
\end{equation}
as was done in Equation~(\ref{eq:LtoI}) for the galactic contribution to the total Ly$\mathrm{\alpha}$ surface brightness.

\begin{figure*}
\includegraphics[width=0.9\columnwidth]{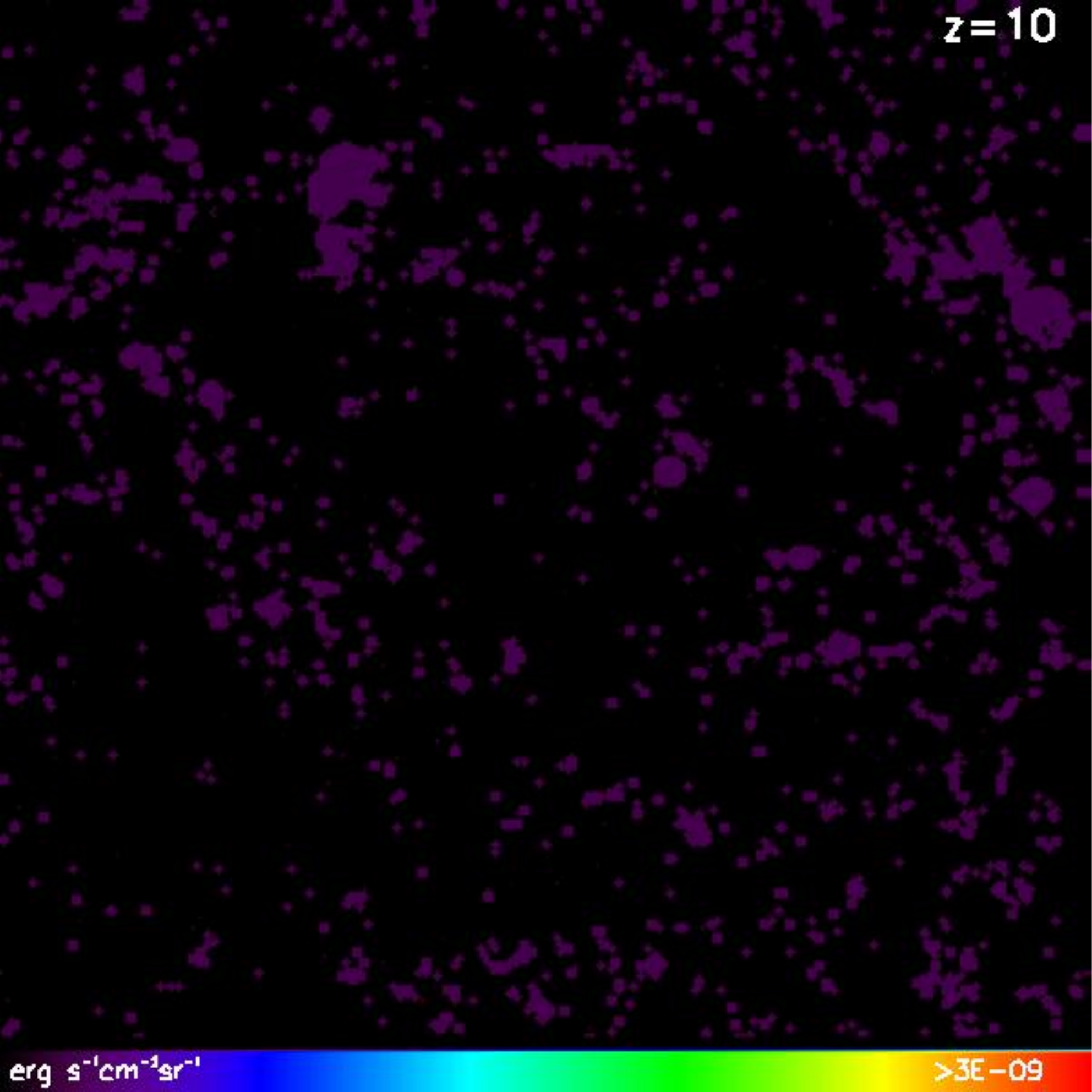}
\includegraphics[width=0.9\columnwidth]{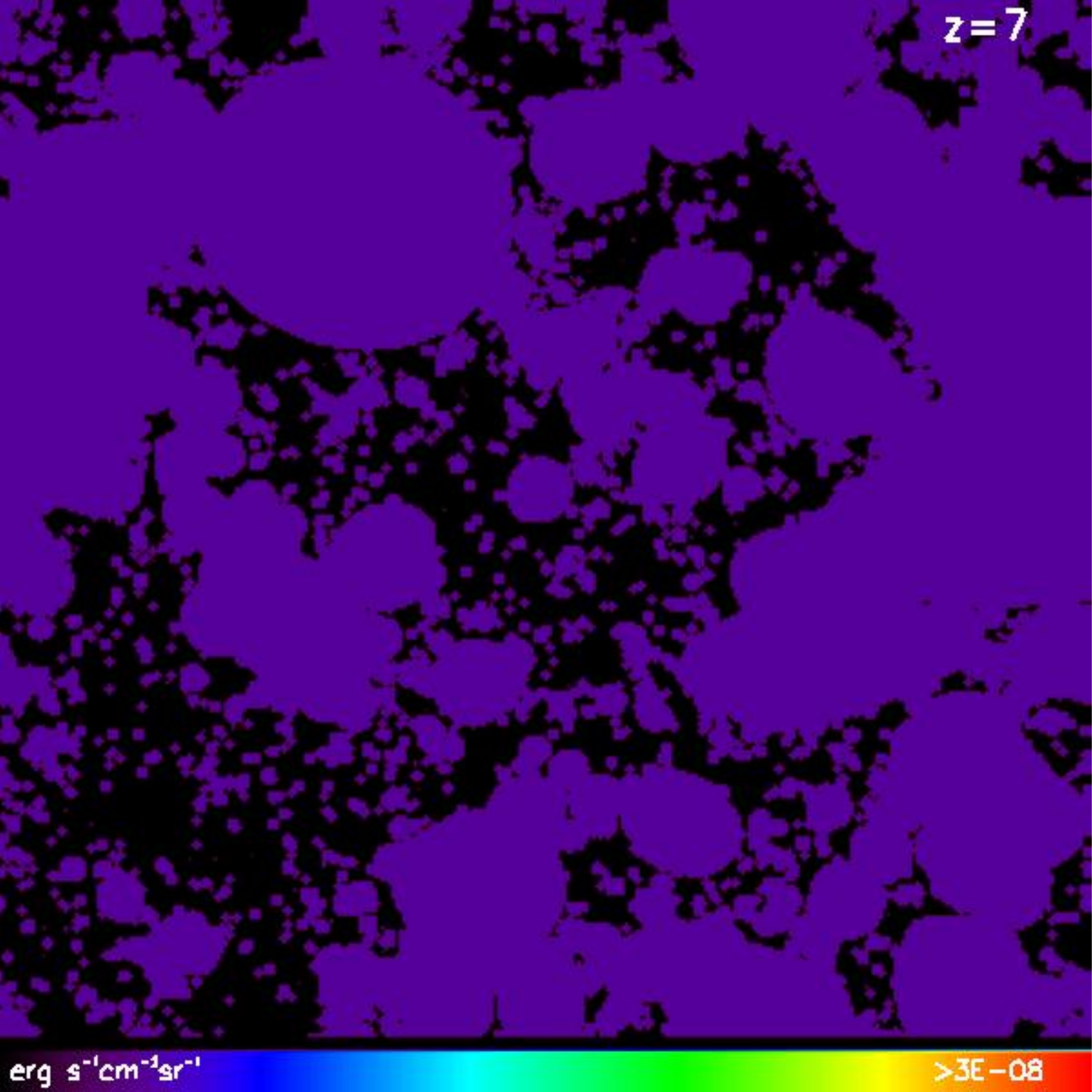}
\includegraphics[width=0.9\columnwidth]{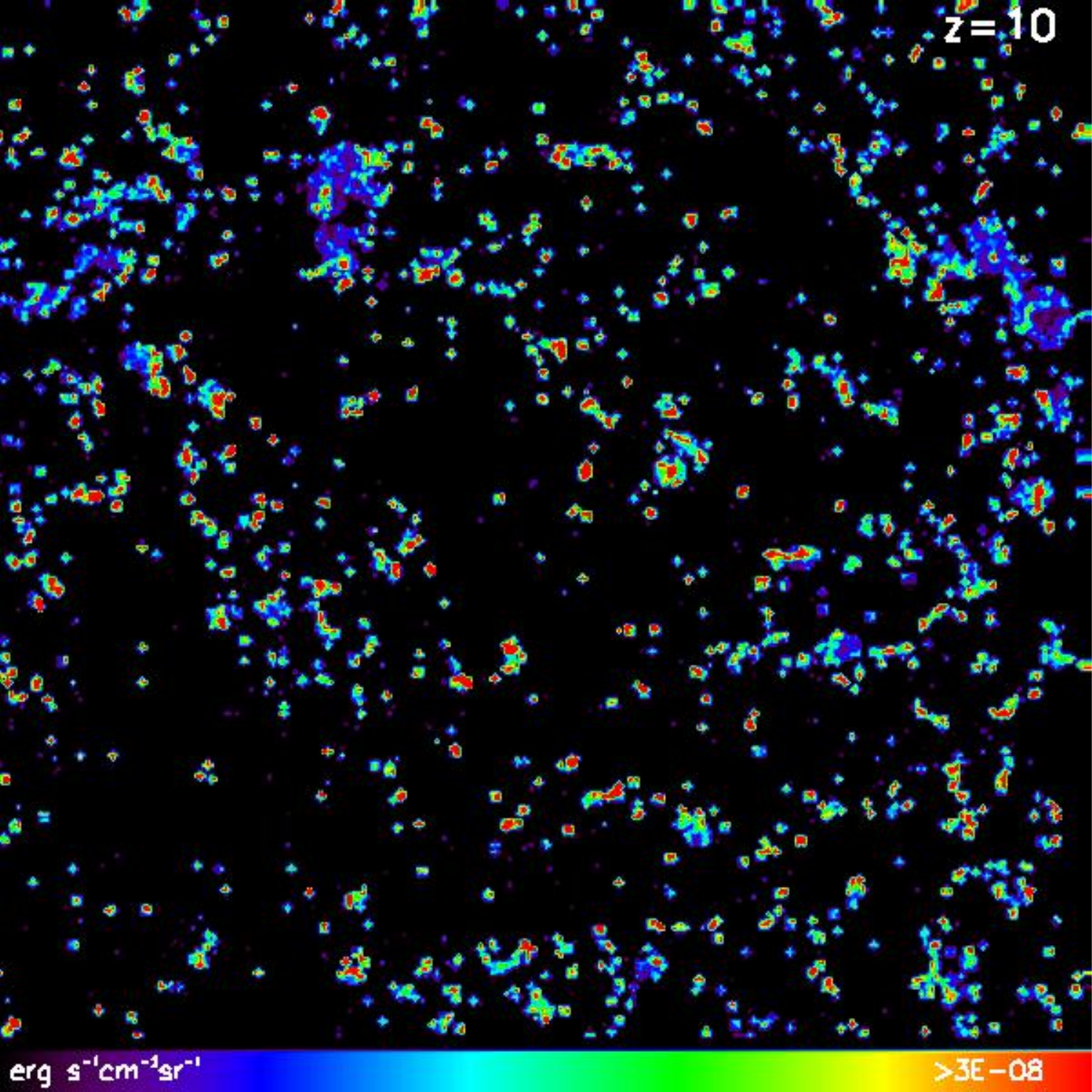}
\hfill
\includegraphics[width=0.9\columnwidth]{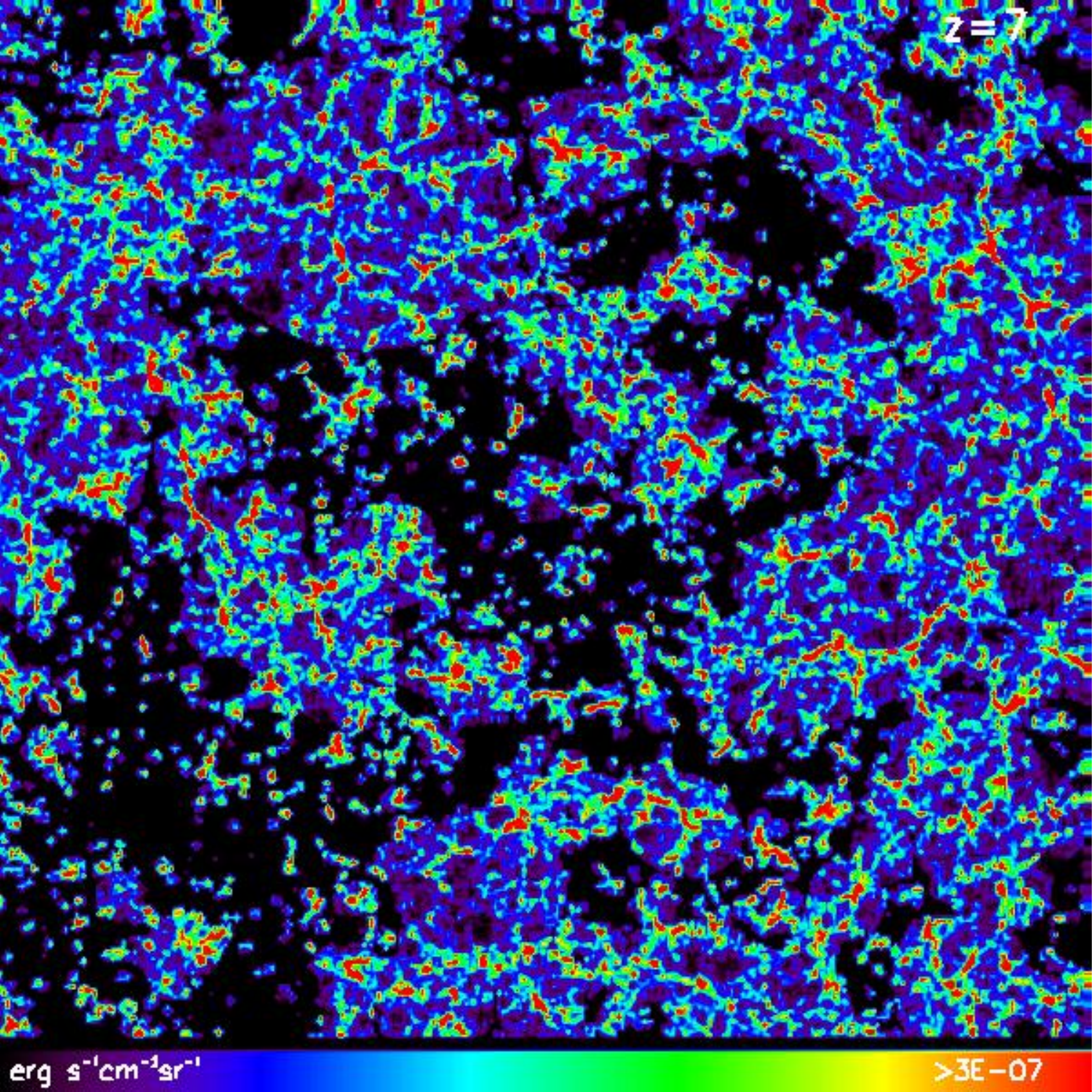}
\caption{Slices of simulations of 200$\,$Mpc box length at $z=10$ and $\bar{x}_\mathrm{HI}=0.87$ (left) and $z=7$ and $\bar{x}_\mathrm{HI}=0.27$ (right) of Ly$\mathrm{\alpha}$ surface brightness in erg$\,$s$^{-1}$cm$^{-2}$sr$^{-1}$ for the diffuse IGM  $I_\mathrm{\nu , rec}^\mathrm{IGM}\left( {\bf x} , z \right)$. Top panels depict the brightness fluctuations for constant gas temperature in ionized regions and constant comoving baryonic density, and bottom panels for a varying comoving baryonic density.} \label{fig:lyaIGMs} 
\end{figure*}

In Figure~\ref{fig:lyaIGMs} we compare simulations of the Ly$\mathrm{\alpha}$ surface brightness for the diffuse IGM component when making a conservative estimate of the brightness fluctuations, by neglecting fluctuations in gas temperature $T_\mathrm{K}$ and in comoving baryonic density $n_\mathrm{b}$ (top panels), and when taking into account fluctuations in the comoving baryonic density $n_\mathrm{b}$ (bottom panels), for the cases of redshift $z=10$ (left panels) and $z=7$ (right panels). As expected, fluctuations in surface brightness become more pronounced when taking into account fluctuations in the baryonic density.

\subsubsection{Ly$\mathrm{\alpha}$ Emission from the Scattered IGM}\label{sec:IGMdiff}
 In this section, we briefly describe the scattered IGM Ly$\mathrm{\alpha}$ background during reionization. The main contributors are X-ray and UV heating, as well as direct stellar emission via scattering in the IGM of Lyman-{\it n} photons emitted from galaxies. Unlike the galactic contribution in Section~\ref{sec:Llya}, where the parameterization boils down to a dependence on halo mass via the SFR, for the scattered IGM Ly$\mathrm{\alpha}$ emission we need to follow the evolution of gas temperature and ionization state at each point $\left({\bf x},z \right)$ in the simulation box, as done for the diffuse IGM in the previous section. We make use of the Ly$\mathrm{\alpha}$ background that has been evolved as described in~\citet{Mesinger10} for 21cmFAST/DexM. It takes into account X-ray excitation of neutral hydrogen, with X-ray heating balanced by photons redshifting out of Ly$\mathrm{\alpha}$ resonance~\citep{Pritchard:2006sq}, as well as direct stellar emission of UV photons emitted between the Ly$\mathrm{\alpha}$ frequency and the Lyman limit, which redshift into Lyman-{\it n} resonance and are absorbed by the IGM. The emission due to stellar emissivity is estimated as a sum over Lyman resonances, as, for example, in~\citet{Barkana:2004vb}.
Snapshots of the spherically averaged Ly$\mathrm{\alpha}$ photon counts per unit area, unit time, unit frequency, and unit steradian $J_{\alpha}$, due to X-ray heating and direct stellar emission in the UV, are extracted and converted to Ly$\mathrm{\alpha}$ surface brightness of the scattered IGM $I_{\nu}^\mathrm{sIGM}\left( {\bf x} , z \right)$ via~\citep{Silva12}
\begin{equation}
I_{\nu}^\mathrm{sIGM}\left( {\bf x} , z \right) = \frac{6 E_\mathrm{Ly\alpha} d_\mathrm{A}^2}{\left( 1+z\right)^2 d_\mathrm{L}^2} J_{\alpha} . 
\end{equation}
We note, that, in the setup used here, the Ly$\alpha$ background does not include soft-UV sources such as quasars. It is also important to mention that the same density fields, and therefore ionization and halo fields derived, are used for both the diffuse and scattered IGM components shown, along with the galactic emission in Ly$\alpha$.
Figure~\ref{fig:lyabgsim} (bottom panels) shows the extracted IGM component in Ly$\mathrm{\alpha}$ surface brightness at $z=10$ and $z=7$. Between $z=10$ and $z=7$, the scattered IGM is clearly lit up by Ly$\mathrm{\alpha}$, with filamentary structures more pronounced at lower redshift.

\subsubsection{Power Spectra and Summary Ly$\mathrm{\alpha}$ Simulation}  \label{sec:lyasim}
The steps taken to simulate the Ly$\mathrm{\alpha}$ surface brightness fluctuations are summed up in the following. 

After parameterizing the Ly$\mathrm{\alpha}$ luminosities as a function of redshift and halo mass in Section~\ref{sec:Llya}, we need to assign luminosities to host halos. We run a halo finder on the density field at a given redshift, evolved from one set of initial density fluctuations. Then luminosities are assigned to galaxy host halos with halo masses above a minimum mass $M_\mathrm{min}$ (corresponding for example to $M_\mathrm{min}=1.3\times10^8\,M_{\odot}$ at $z=7$), equivalent to a minimum virial temperature $T_\mathrm{vir}=10^4\,$K needed for sufficient efficiency of baryonic cooling when forming galaxies. Maximum halo masses found correspond to $\approx 3\times 10^{11}\,M_{\odot}$ at $z=10$ and $\approx 2\times 10^{12}\,M_{\odot}$ at $z=7$. As mentioned in Section~\ref{sec:Llya}, Equation~(\ref{eq:SFR}) is a parameterization of the SFR that captures a reionization history and luminosity function compatible with observations, fitting the abundance of Ly$\mathrm{\alpha}$ emitters. A possible further tuning of the simulated luminosities to an observed luminosity function can be obtained in this step by varying the duty cycle $f_\mathrm{duty}$, which randomly assigns $f_\mathrm{duty}$-percent of halos as hosting a galaxy. 
A duty cycle $f_\mathrm{duty}=1$ means that all halos above $M_\mathrm{min}$ are assumed to host a galaxy that emits in Ly$\mathrm{\alpha}$; a duty cycle smaller than one takes into account that not all halos might host a galaxy bright in Ly$\mathrm{\alpha}$. We set $f_\mathrm{duty}$ to one here, as our $\mathrm{SFR}$ was tuned to fit luminosity functions from observations, but will briefly show the impact of introducing a duty cycle smaller than one in Section~\ref{sec:modelp}. Also, one could account for the distribution of satellite galaxies to further refine the distribution of Ly$\mathrm{\alpha}$ emitters in future analyses. After assigning Ly$\mathrm{\alpha}$ luminosities to host halos, we build the smoothed field of the galactic contribution $I_{\nu}^\mathrm{gal}\left( {\bf x} , z \right)$ to Ly$\mathrm{\alpha}$ surface brightness as in Equation~(\ref{eq:LtoI}), shown in Figure~\ref{fig:lyabgsim} (top panels) for redshift $z=10$ (left) and $z=7$ (right). 

In addition to the surface brightness due to direct galactic emission, the emitting region is also composed of ionized, diffuse IGM around halos, as discussed in Section~\ref{sec:IGMs}. The resulting Ly$\mathrm{\alpha}$ surface brightness $I_\mathrm{\nu , rec}^\mathrm{IGM}\left( {\bf x} , z \right)$ is given by Equation~(\ref{eq:IGMs}) and presented in Figure~\ref{fig:lyaIGMs} for redshift $z=10$ (left panels) and $z=7$ (right panels), when neglecting fluctuations in gas temperature and comoving baryonic density  (top panels), and when taking into account fluctuations in the comoving baryonic density (bottom panels). 

Alongside with the modeling of galactic emission from the halo and emission from the surrounding diffuse IGM, we run the evolution of the scattered Ly$\mathrm{\alpha}$ background for the same density, ionization, and halo fields, taking into account UV/X-ray heating and scattering of Lyman-{\it n} photons. We therefore only treat one realization of density, luminosity, and brightness fields. The UV/X-ray heating and scattering of Lyman-{\it n} photons gives the scattered IGM contribution to the Ly$\mathrm{\alpha}$ surface brightness $I_\mathrm{\nu, diff}^\mathrm{IGM}\left( {\bf x} , z \right)$, as described in Section~\ref{sec:IGMdiff} and shown in Figure~\ref{fig:lyabgsim} (bottom panels) for redshift $z=10$ (left) and $z=7$ (right). For the simulation of emission from both the scattered and the diffuse IGM, we run the full evolution of gas temperature and gas density, as well as ionization fraction of the IGM.

Having simulated the different contributions to Ly$\mathrm{\alpha}$ surface brightness, the fluctuations in the smoothed surface brightness field read as
\begin{equation}
\delta_{I_{\nu}} \left( {\bf x} ,z \right)= \sum_{i} \frac{\nu I_{\nu , i}\left( {\bf x},z\right)}{\nu \bar{I}_{\nu , i}\left( z\right)}-1\,,
\label{eq:deltaInu}
\end{equation}
summing, when wanted, pixelwise at observed frequency $\nu$, over Ly$\mathrm{\alpha}$ contributions to the surface brightness, that is, galactic, diffuse, and scattered IGM, with mean Ly$\mathrm{\alpha}$ surface brightness $\bar{I}_{\nu}\left( z \right)$. We express the dimensionless power spectrum as $\tilde{\Delta}_\mathrm{Ly\alpha}\left( k\right) = k^3/\left( 2\pi^2 V\right) \left< |\delta_{I_{\nu}}|^2\right>_k$ and, when a comparison of absolute emission strength is desirable, we use the dimensional power spectrum $\Delta_\mathrm{Ly\alpha}\left( k\right) = \left(\nu \bar{I}_{\nu }\right)^2 \tilde{\Delta}_\mathrm{Ly\alpha}\left( k\right)$.

\begin{figure}[!h]
\includegraphics[width=0.95\columnwidth]{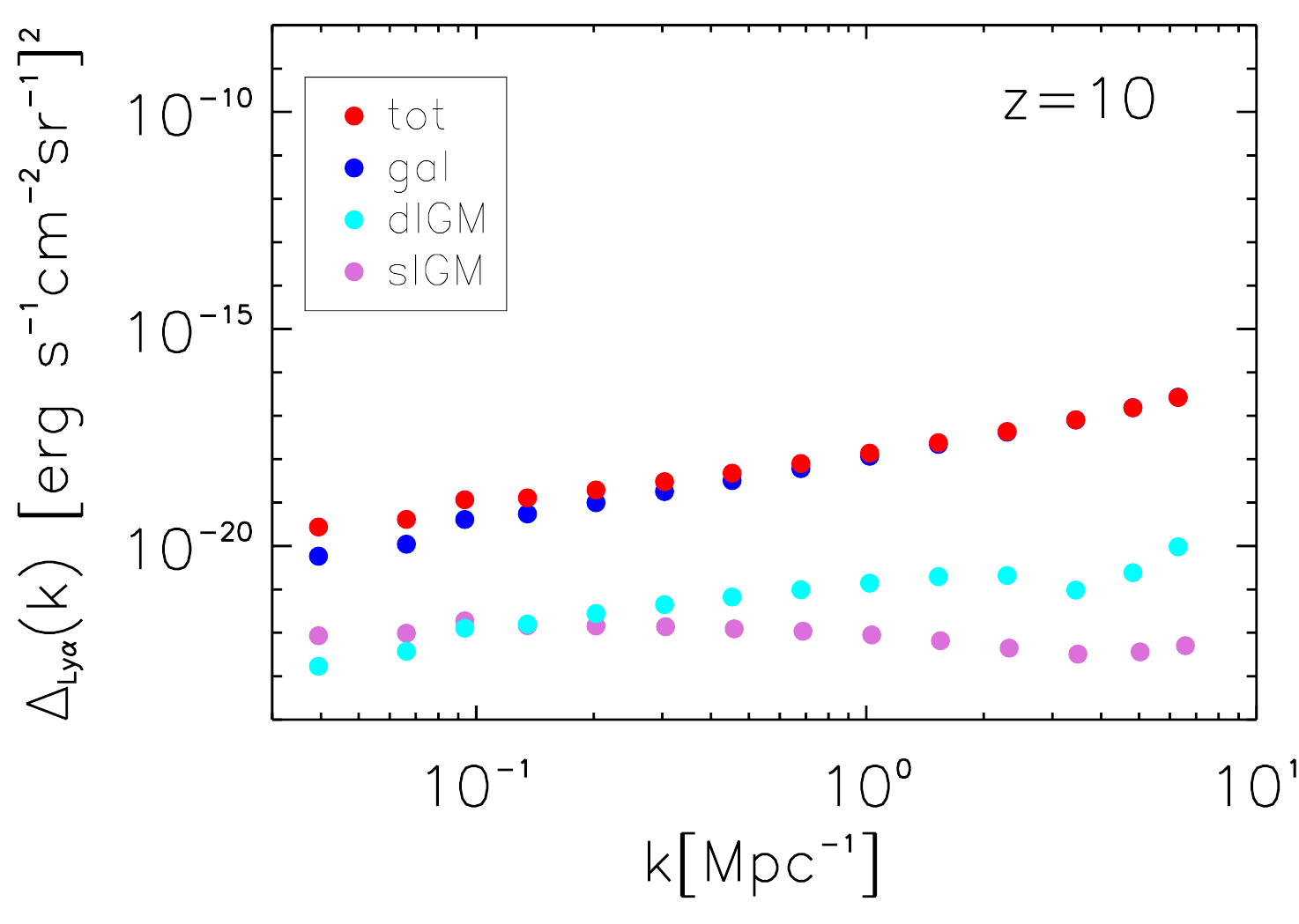}
\vfill \vspace{0.1cm}
\includegraphics[width=0.95\columnwidth]{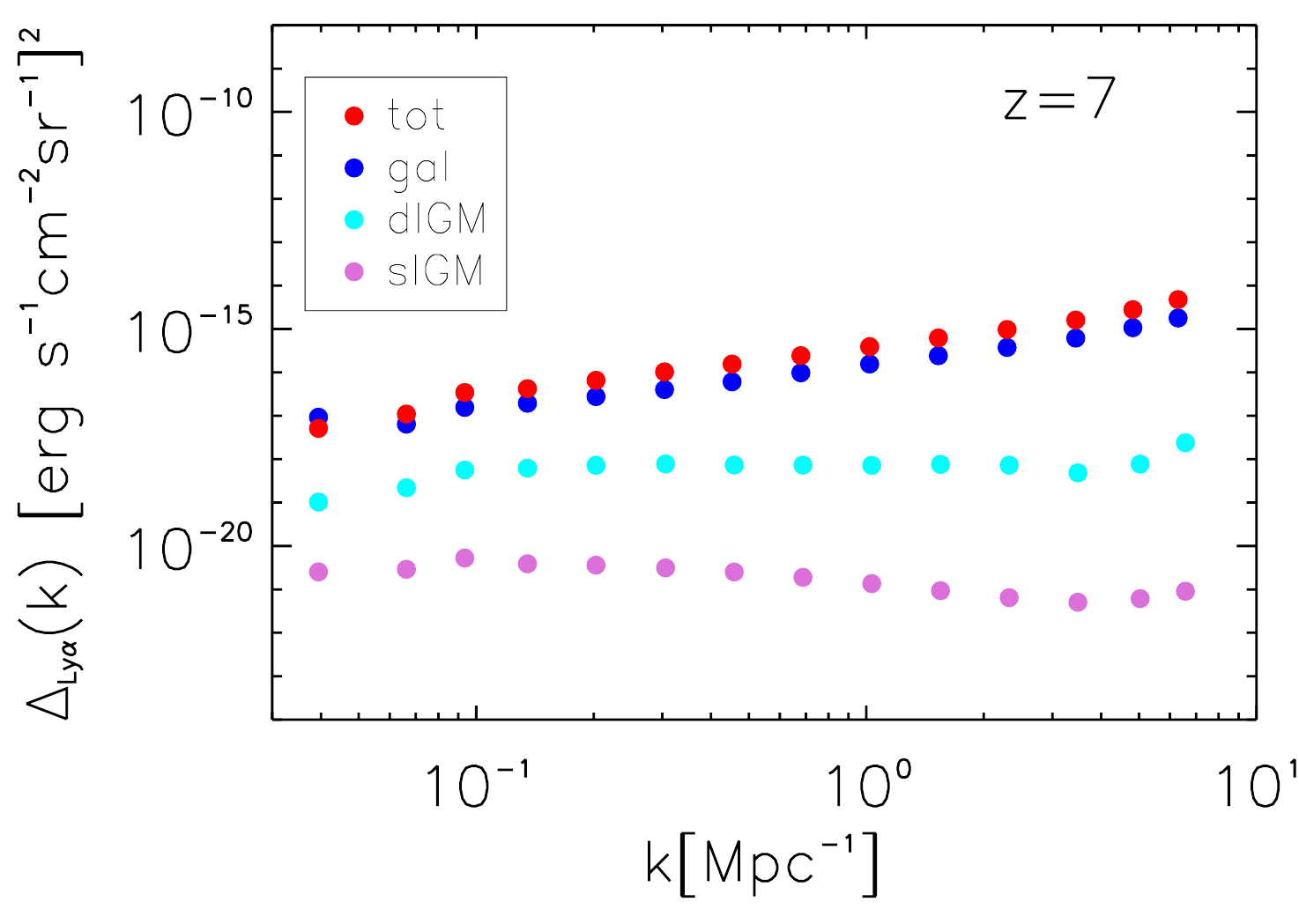}
\caption{Ly$\mathrm{\alpha}$ power spectra in surface brightness $\left( \nu I_{\nu}\right)$: total emission (tot, red), galaxy (gal, blue), diffuse IGM  (dIGM, cyan), and scattered IGM (sIGM, orchid) contributions for redshift $z=10$ (top panel) and $z=7$ (bottom panel).} \label{fig:IGMgalz}
\end{figure}

Figure~\ref{fig:IGMgalz} shows the power spectra at redshift $z=10$ (top panel) and $z=7$ (bottom panel) for the three dominant contributions to Ly$\mathrm{\alpha}$ surface brightness fluctuations, that is, for direct galactic emission (gal), for diffuse IGM emission (dIGM), when neglecting fluctuations in gas temperature and  comoving baryonic density, and for scattered IGM emission (sIGM) and total emission (tot). The Ly$\mathrm{\alpha}$ surface brightness of the IGM components proves to be subdominant and less {\it k}-dependent in comparison to the galactic emission component, and the power increases at lower redshift toward a fully ionized universe. Table~\ref{tab:Inu} sums up the corresponding mean intensities for each emission component. To check consistency, we compare with Ly$\alpha$ power spectrum results from other work in Appendix~\ref{app:Lya}.

\begin{table}[th!]
\centering
\caption{Mean surface brightness of Ly$\alpha$ emission for different sources at redshift $z=10$ and $z=7$}
\centering
\begin{tabular}{P{3.0cm} P{2.0cm} P{2.0cm} }
\hline \hline
\multicolumn{1}{l}{Source of Emission}  & $\nu I_{\nu} \left( z=10\right)$  &  $\nu I_{\nu} \left( z=7\right)$ \\
 (erg s$^{-1}$ cm$^{-2}$ sr$^{-1}$)  &   &  \\
\hline
 \multicolumn{1}{l}{Total} &  $3.1\times 10^{-9}$  & $1.8\times 10^{-8}$  \\ 
 \multicolumn{1}{l}{Galactic} &   $3.3\times 10^{-10}$  &  $1.0\times 10^{-8}$ \\ 
  \multicolumn{1}{l}{Diffuse IGM} &   $2.7\times 10^{-9}$  &  $5.1\times 10^{-9}$  \\ 
  \multicolumn{1}{l}{ Scattered IGM} &   $2.5\times 10^{-11}$  &  $2.9\times 10^{-9}$  \\ 
  \hline
\multicolumn{3}{c}{NOTE. - See Figure~\ref{fig:IGMgalz} for corresponding power spectra.}
\end{tabular}
\label{tab:Inu}
\end{table}

\begin{figure}
\includegraphics[width=\columnwidth]{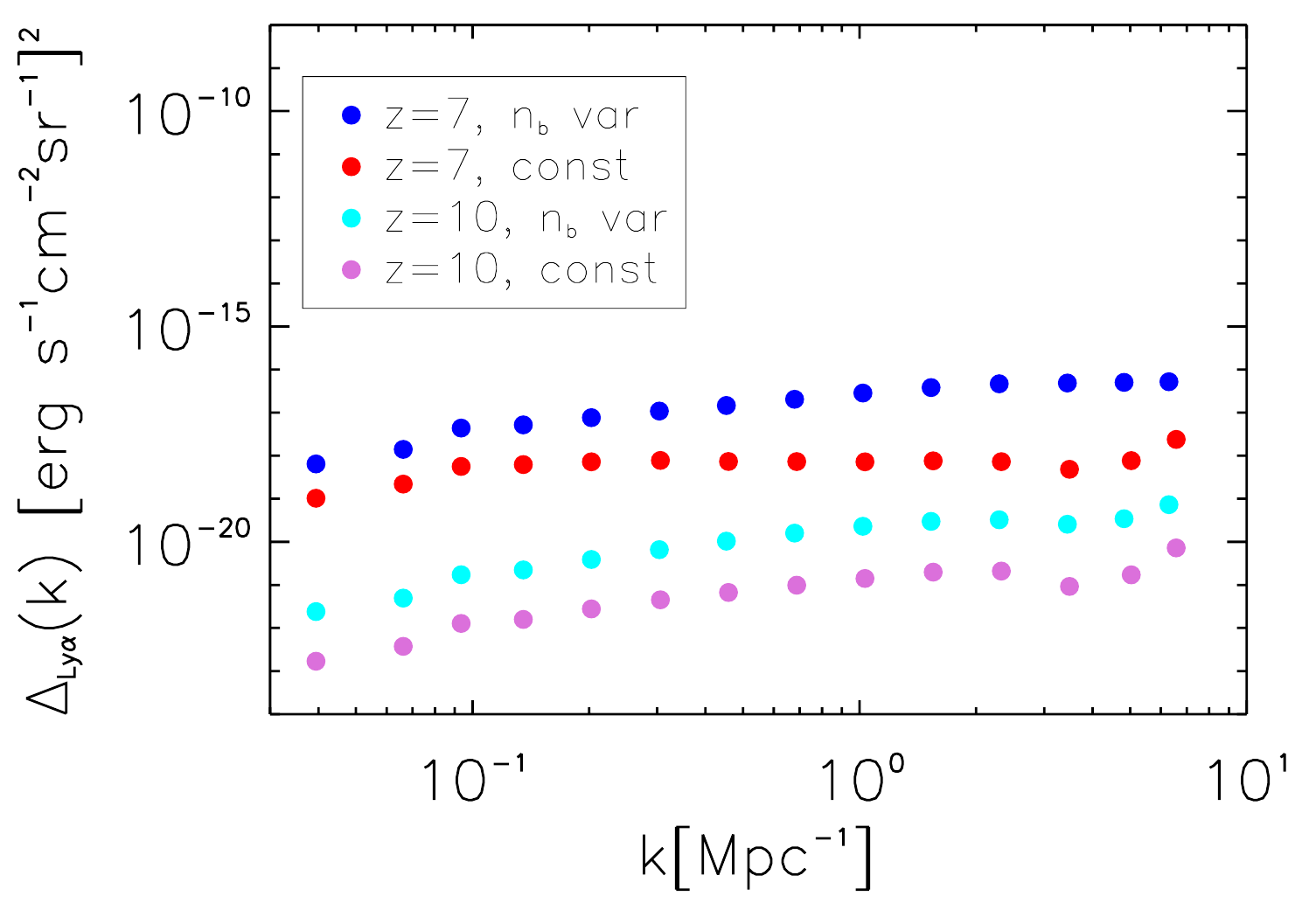}
\caption{Ly$\mathrm{\alpha}$ power spectra in surface brightness $\left( \nu I_{\nu}\right)$ for the diffuse IGM contribution: taking into account fluctuations in the comoving baryonic density $n_\mathrm{b}$ (``$n_\mathrm{b}$ var") and for constant $n_\mathrm{b}$ ``const" at redshift $z=10$ and $z=7$.} \label{fig:dIGM}
\end{figure}

Figure~\ref{fig:dIGM} depicts the power spectra of Ly$\mathrm{\alpha}$ surface brightness for the diffuse IGM both when neglecting and when taking into account fluctuations in comoving baryonic density for redshift $z=10$ and $z=7$. As expected, taking into account fluctuations increases the power. We will take the simulation of the Ly$\mathrm{\alpha}$ emission in the diffuse IGM for constant gas temperature and constant baryonic density as a conservative lower bound for our cross-correlation studies in the following sections, as also for the ionization fields in the simulation of 21cm emission each pixel is assigned to be either fully ionized or neutral.

\subsection{H$\mathrm{\alpha}$ Fluctuations and Power Spectra}\label{Ha}
Unlike Ly$\mathrm{\alpha}$, which also has a significant IGM component, both diffuse and scattered, H$\mathrm{\alpha}$ emission can be assumed to be of mostly galactic origin. It traces the ionized hydrogen component in galaxies. Thus H$\mathrm{\alpha}$ is an interesting tracer of the galaxy-only component in emission, as compared to Ly$\mathrm{\alpha}$, and can be used to single out the amount of the galactic contribution versus IGM contribution in Ly$\mathrm{\alpha}$ brightness via cross-correlation of the two tracers. We checked that the small diffuse IGM component nevertheless present for H$\alpha$ does not spoil this idea, by including this component in our simulation, analogous to Equation~(\ref{eq:lrecIGM}) for Ly$\alpha$, with $f_\mathrm{rec}\approx 1/2$ and $E_\mathrm{H\alpha}\approx 6.626\times10^{-14}$erg.

Similar to the assignment of Ly$\mathrm{\alpha}$ luminosities depending on halo mass and redshift in Section~\ref{sec:Llya}, we also parameterize the H$\mathrm{\alpha}$ luminosities to ultimately depend on halo mass and redshift.
We use the relation between total SFR and H$\mathrm{\alpha}$ luminosity from~\citet{Kennicutt:1997ng}, which reads as
\begin{equation}
L_\mathrm{H \alpha} = 1.26\times10^{41}\left(\mathrm{erg}\, \mathrm{s^{-1} } \right) \times \mathrm{SFR}\left( M_{\odot} \mathrm{yr^{-1}}\right) , 
\label{eq:KS}
\end{equation}
and assign intrinsic H$\mathrm{\alpha}$ luminosities to host halos according to their mass. Again, as for the modeling of Ly$\mathrm{\alpha}$ emission, we assume a minimum host halo virial temperature of $T_\mathrm{vir} = 10^4\,$K for baryonic cooling to be efficient and halos to be able to host a galaxy. 

\begin{figure}
\hspace{0.45cm}\includegraphics[width=0.9\columnwidth]{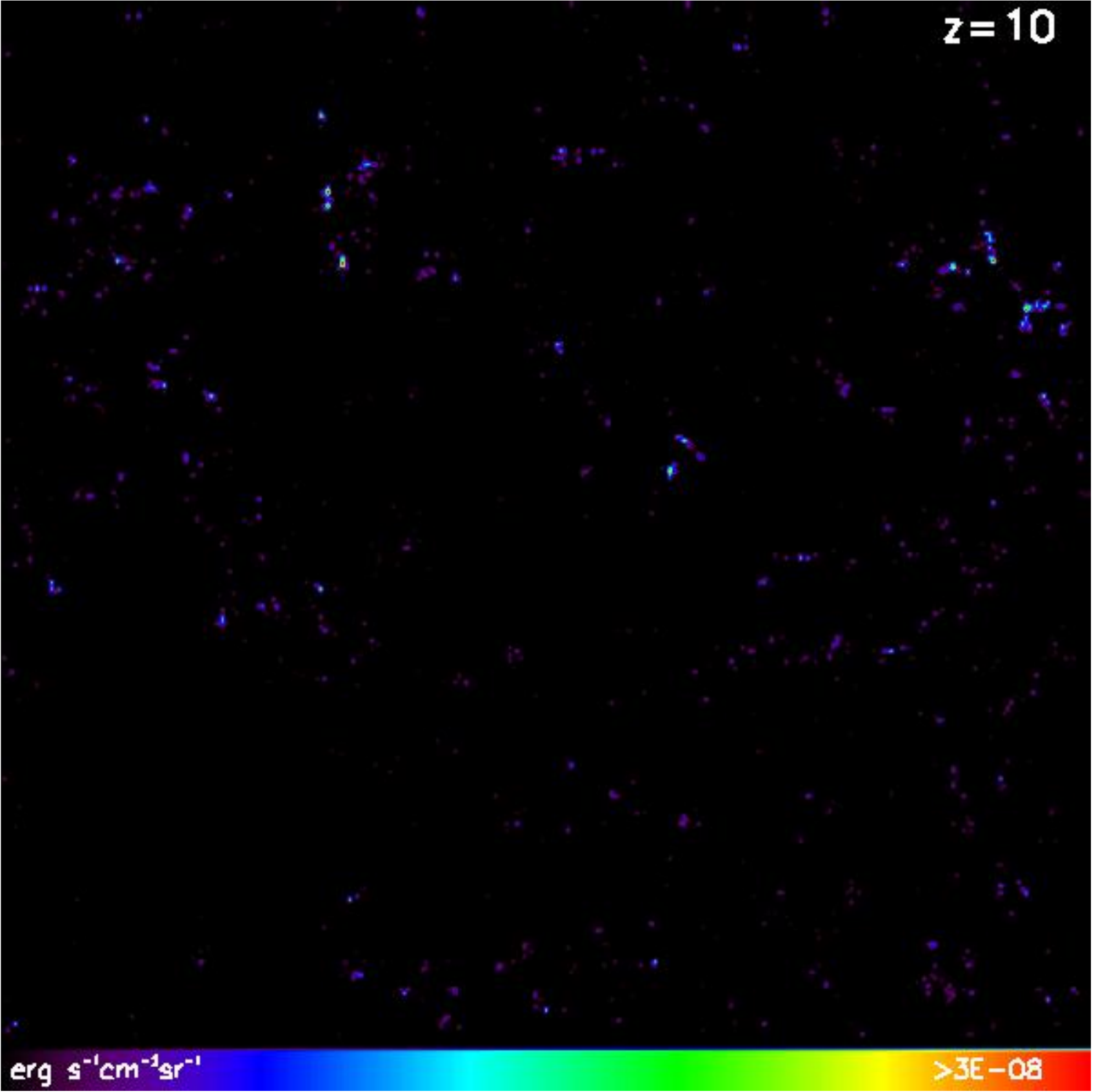}
\vspace{0.2cm}
\hspace{0.45cm}\includegraphics[width=0.9\columnwidth]{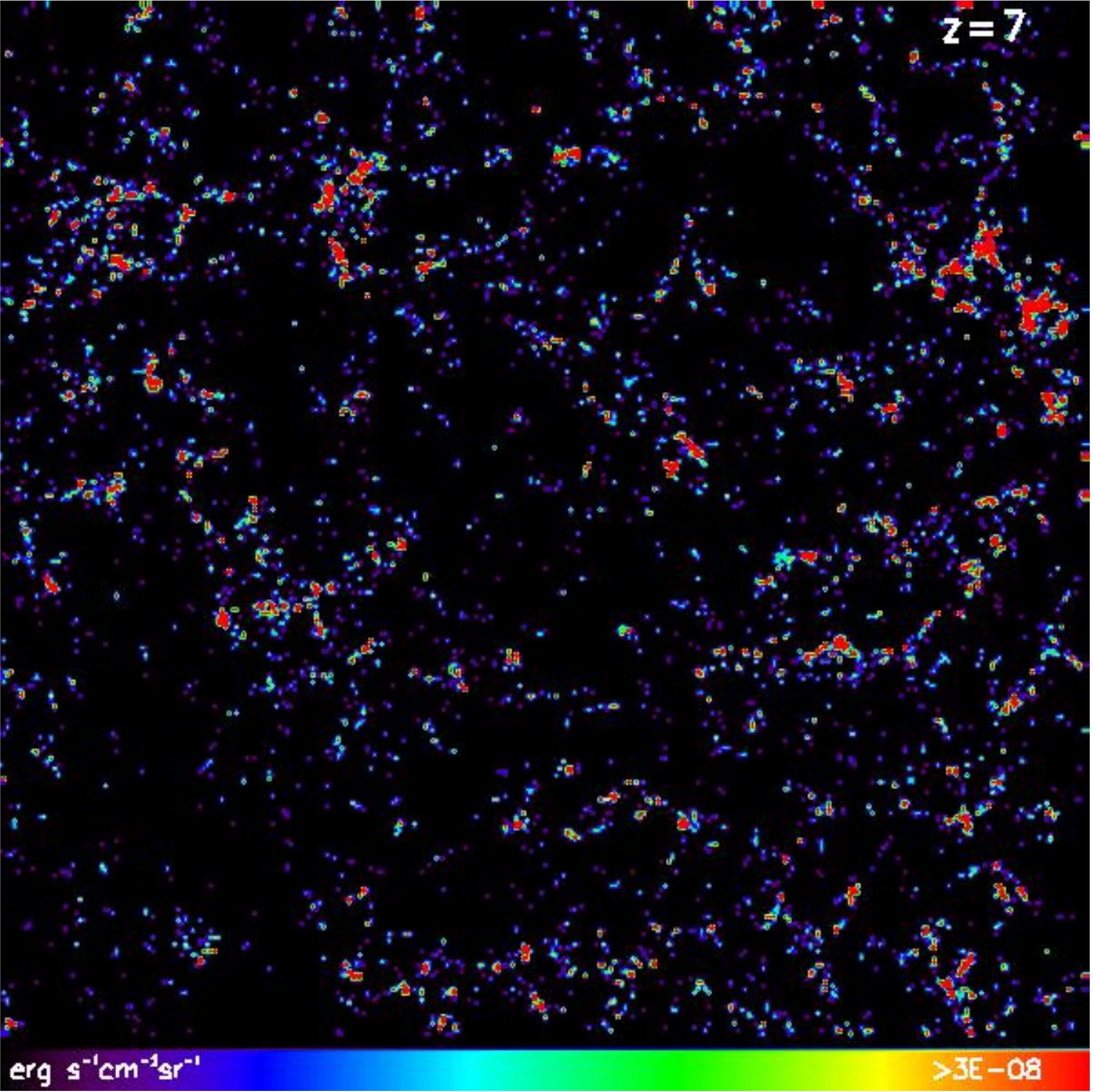}
\vspace{0.2cm}
\includegraphics[width=\columnwidth]{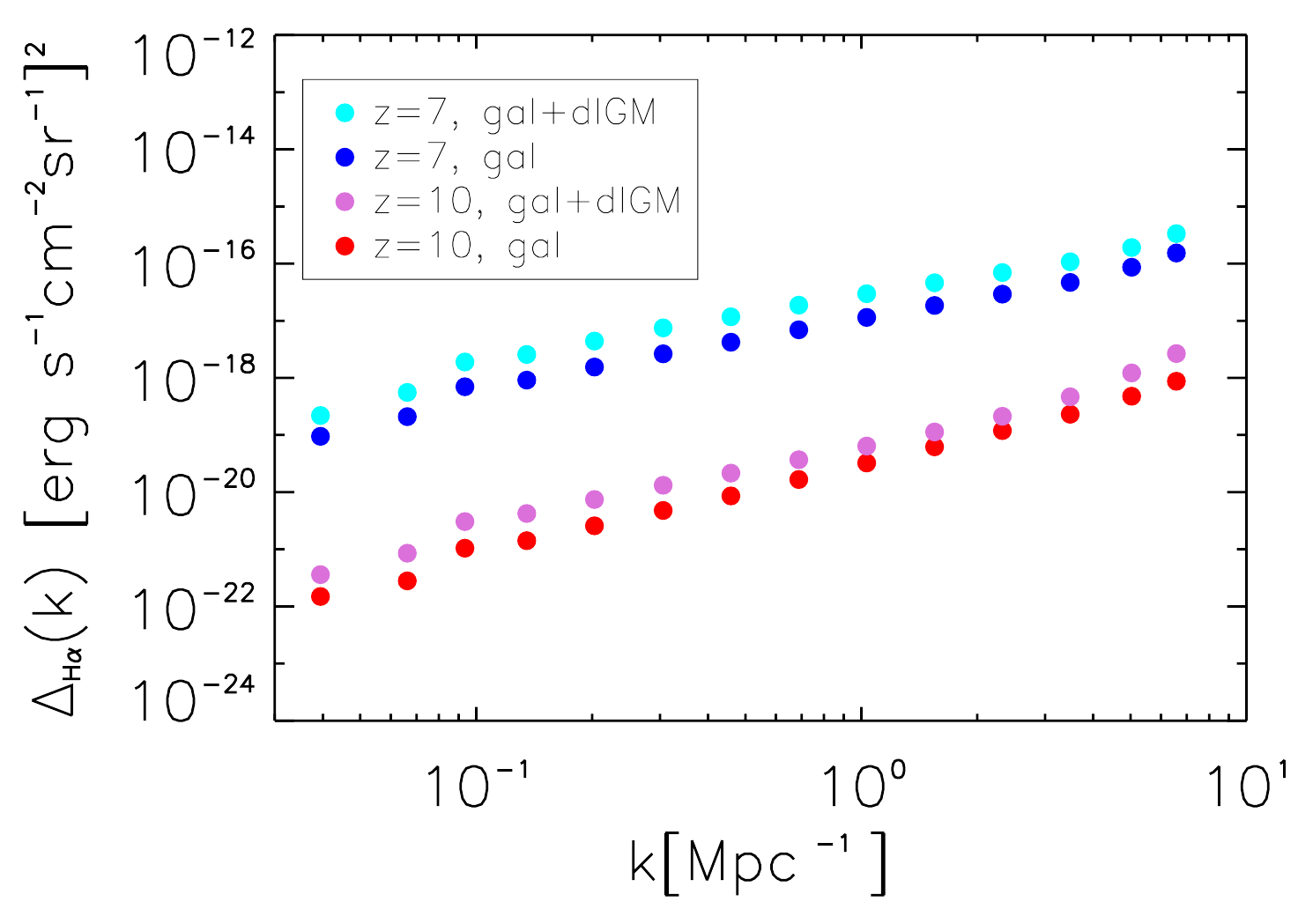}
\caption{Top and middle: simulated box slices of (200 x 200)$\,$Mpc at $z=10$ (top) and $z=7$ (middle) of H$\mathrm{\alpha}$ intrinsic surface brightness (not corrected for dust absorption) in erg$\,$s$^{-1}$cm$^{-2}$sr$^{-1}$, with both galactic and diffuse IGM emission in H$\alpha$. Bottom: corresponding power spectra at $z=7$ and $z=10$ for galactic only ``gal'' and galactic plus diffuse IGM contribution ``gal + dIGM''.} \label{fig:Ha}
\end{figure}

For the power spectrum, we calculate surface brightness fluctuations per pixel smoothed over virial radii for the galactic component, analogous to Equation~(\ref{eq:LtoI}) for Ly$\mathrm{\alpha}$ galactic emission, and add the diffuse IGM component pixelwise.
The power spectrum (for fluctuations in brightness intensity) is shown together with the distribution of luminous halos at redshift $z=10$ and $z=7$ in Figure~\ref{fig:Ha}. Note that the intrinsic power in H$\mathrm{\alpha}$ is about two orders of magnitude lower than for Ly$\mathrm{\alpha}$, which approximately reflects the intrinsic line ratio of about $8.7$~\citep{1971Brockhurst,1987MNRAS.224..801H} between the two emission lines.
We neglect for now dust obscuration of H$\mathrm{\alpha}$ sources, as we aim in Section~\ref{sec:crossHa} at a proof of concept for singling out the IGM part of Ly$\mathrm{\alpha}$ emission via cross-correlation with H$\mathrm{\alpha}$ emission.

\section{Cross-correlation Studies}\label{sec:cross}
In this section, we present results for the cross-correlation signal of brightness fluctuations in 21cm, Ly$\mathrm{\alpha}$ and H$\mathrm{\alpha}$ emission; their simulation has been described in the previous sections. The goal is to explore robust methods beyond the power spectrum, which will enable us to probe the state of the IGM during reionization.
We start with the cross-correlation signal for 21cm and different components of Ly$\mathrm{\alpha}$ brightness fluctuations in Section~\ref{sec:IGMcomp}. We proceed to show the impact on the cross-correlation signal when varying some of the model parameters in Section~\ref{sec:modelp}, and when including Ly$\alpha$ damping in Section~\ref{sec:damp}. We finish by presenting a method to single out the IGM component in Ly$\mathrm{\alpha}$ brightness  fluctuations by cross-correlating with H$\mathrm{\alpha}$ fluctuations in Section~\ref{sec:crossHa}.

 We define the dimensionless cross-power spectrum as $\tilde{\Delta}_{I,J} = k^3/\left( 2\pi^2 V\right) \Re \left< \delta_{I}\delta_{J}^*\right>_k$ for fluctuations $\delta_{I}$ and $\delta_J$, as well as the dimensional cross-power spectrum as $\Delta_{I,J}\left( k\right) = \bar{I}_I\bar{I}_J \tilde{\Delta}_{I,J}\left( k\right) $ for mean intensities $\bar{I}_I$ and $\bar{I}_J$.  As a measure of how correlated or anticorrelated modes are, we also give the cross-correlation coefficient $\mathrm{CCC}$. For correlated modes, $0<\mathrm{CCC}<1$, and for anticorrelated modes, $-1<\mathrm{CCC}<0$; it is defined as
\begin{equation}
CCC_{I,J}\left(k \right) = \frac{\Delta_{I,J}\left(k \right)}{\sqrt{\Delta_{I}\left( k\right)\Delta_{J}\left( k\right)}} \,, \label{eq:CCC}
\end{equation}
with power spectra $\Delta_{I}$ and $\Delta_{J}$ of fluctuations $\delta_{I}$ and $\delta_J$, and the cross-power spectrum $\Delta_{I,J}$.

\begin{figure*}
\plottwo{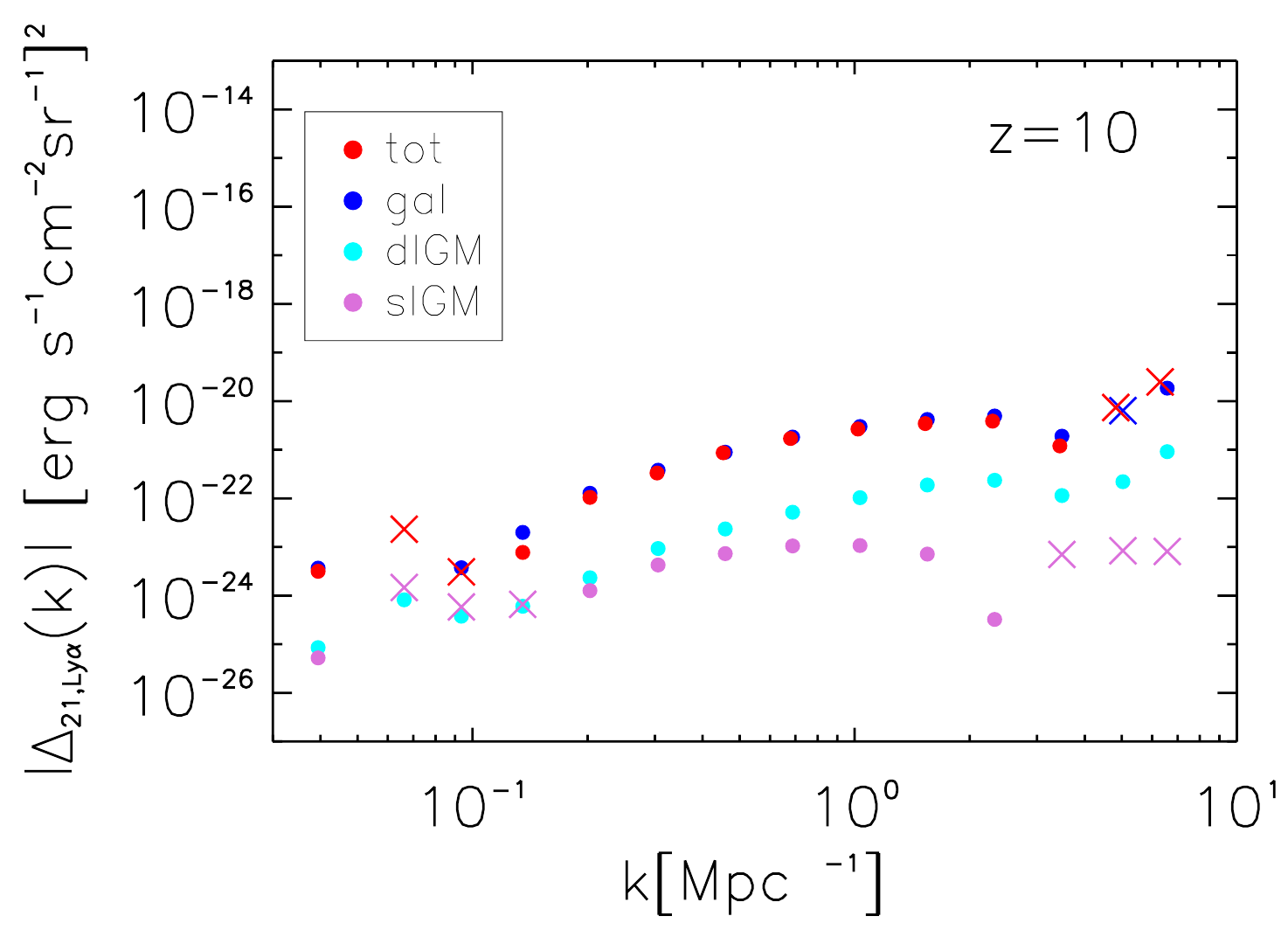}{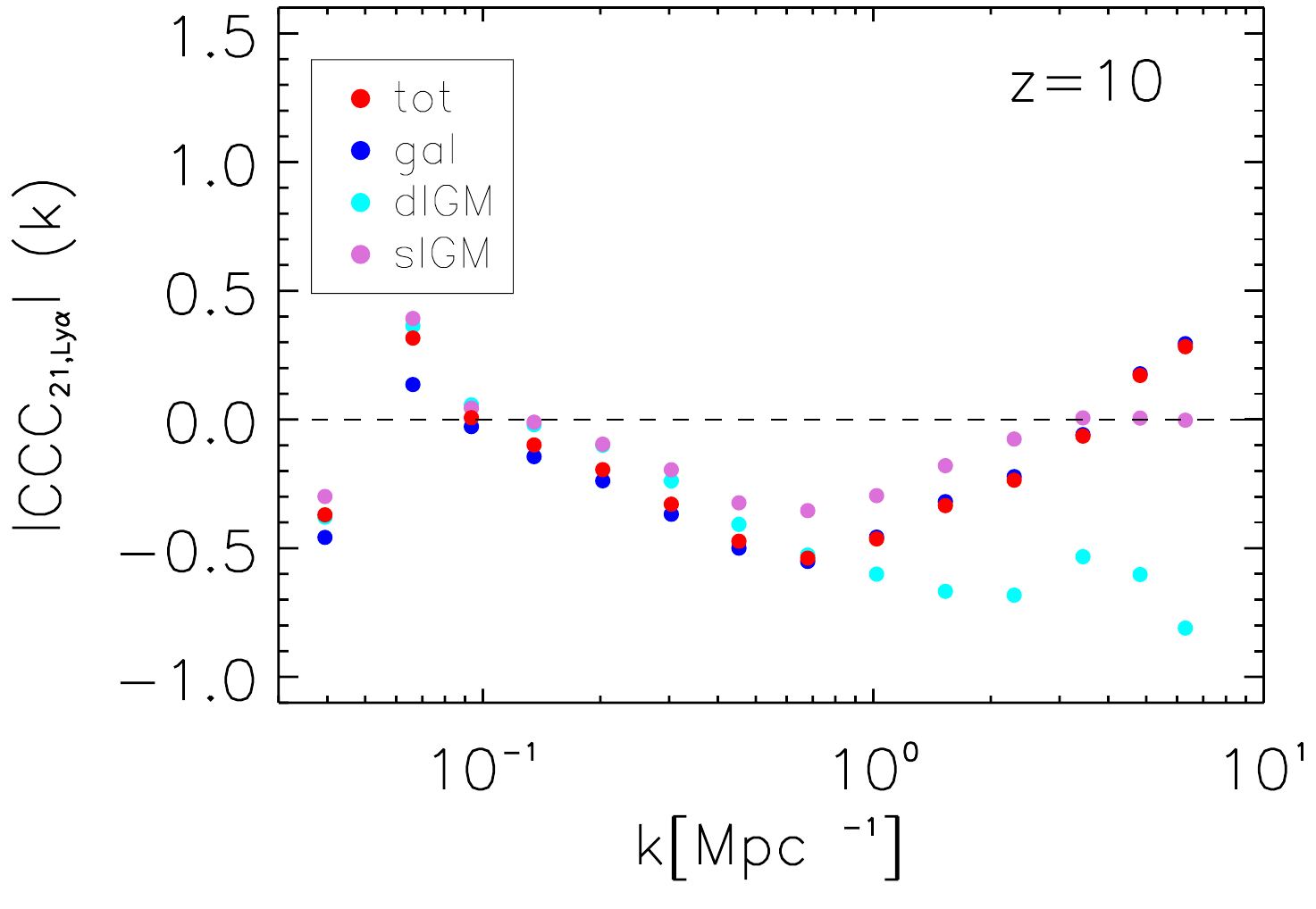} 
\vfill \vspace{0.1cm}
\plottwo{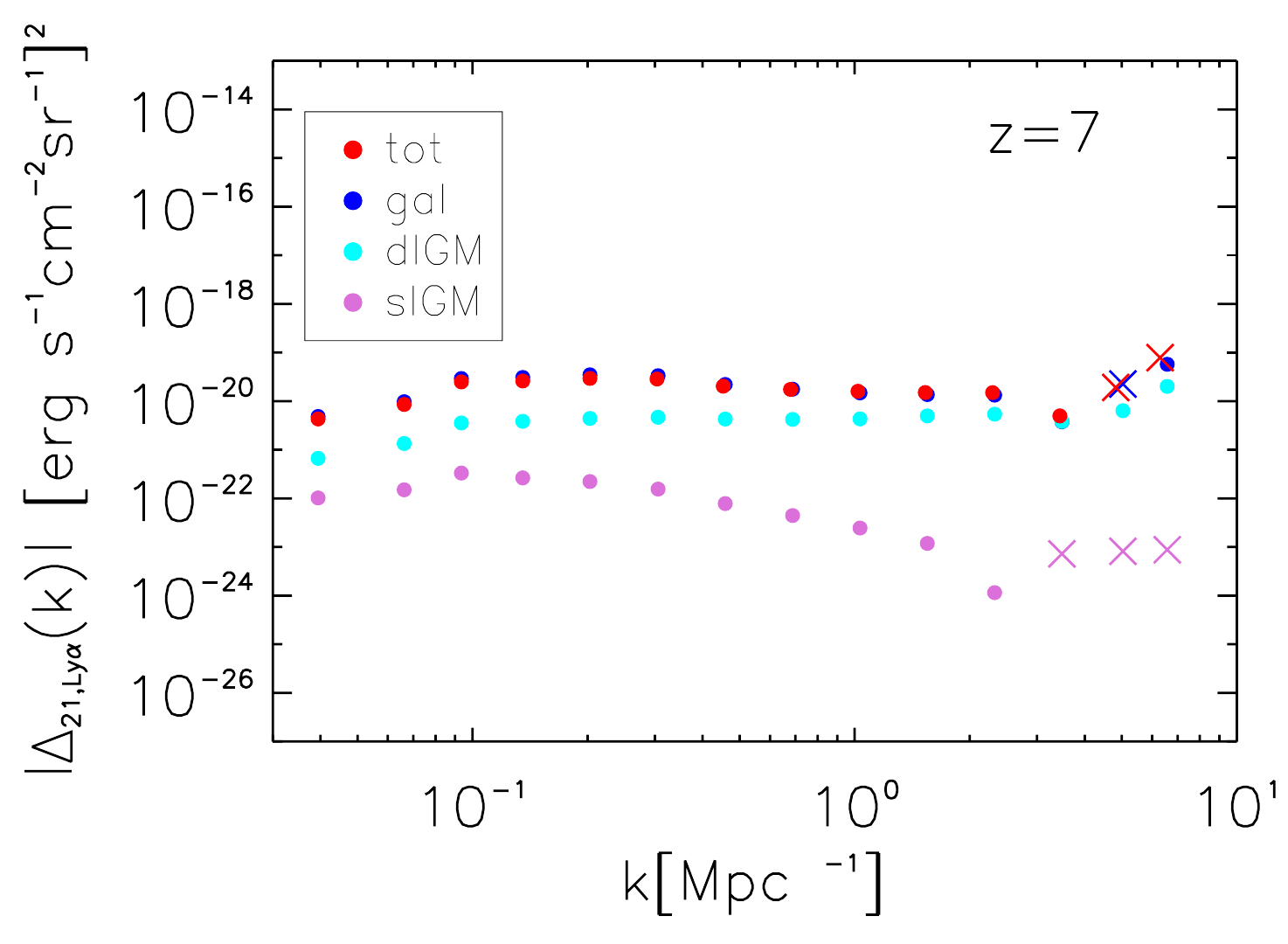}{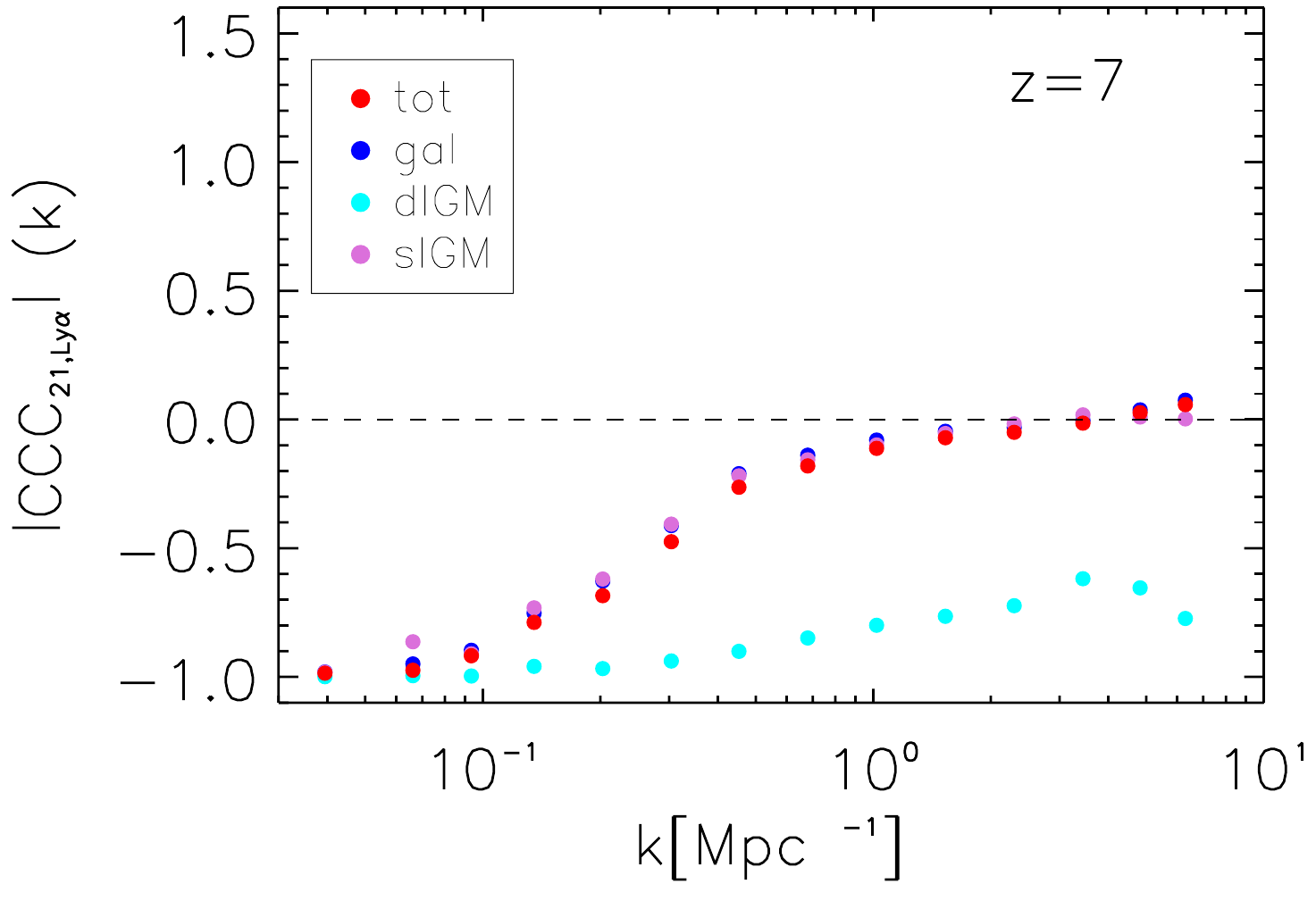}
\caption{Dimensional cross-power spectra (left) and cross-correlation coefficient CCC (right) of 21cm fluctuations and total Ly$\mathrm{\alpha}$ brightness fluctuations (tot, red), as well as three components of Ly$\mathrm{\alpha}$ emission, being galactic (gal, blue) and both diffuse IGM (dIGM, cyan) and scattered IGM (sIGM, orchid) at $z=10,\,\bar{x}_\mathrm{HI}=0.87$ (top panels) and $z=7,\,\bar{x}_\mathrm{HI}=0.27$ (bottom panels).}
 \label{fig:cross-comp}
\end{figure*}

\subsection{21 cm and Ly$\mathrm{\alpha}$ Fluctuations}\label{sec:cross21}

\subsubsection{Galactic, Diffuse IGM, and Scattered IGM} \label{sec:IGMcomp}
 The cross-correlation between fluctuations in 21cm and Ly$\mathrm{\alpha}$ brightness is useful in characterizing the IGM, as 21cm emission traces the neutral part of the IGM, and Ly$\mathrm{\alpha}$ emission is more closely connected to ionized regions. Ly$\mathrm{\alpha}$ emission is made up of galactic emission and emission in the diffuse ionized IGM, plus a subdominant contribution from scattering in the IGM. The cross-correlation with 21cm emission therefore is sensitive to the clustering and size of ionized regions. An anticorrelation between 21cm and Ly$\mathrm{\alpha}$ emission that is sensitive to the structure of the ionized medium during the EoR can be expected at large and intermediate scales, as well as a turnover to positive correlation at small scales (as both tracers follow the same underlying density field).

We cross-correlate 21cm fluctuations simulated as described in Section~\ref{sec:21cm} with the components of Ly$\mathrm{\alpha}$ fluctuations presented in Section~\ref{sec:lya}, that is, diffuse and scattered IGM components and the galactic emission component. Figure~\ref{fig:cross-comp} shows the breakdown of the dimensional cross-power spectrum (left) and the CCC (right) for the diffuse and scattered IGM components, as well as the galactic component of Ly$\mathrm{\alpha}$ fluctuations cross-correlated with 21cm fluctuations. Going from redshift $z=10$ (top) to $z=7$ (bottom) and therefore from a higher mean neutral fraction of $\bar{x}_\mathrm{HI}=0.87$ to $\bar{x}_\mathrm{HI}=0.27$, the morphology of the cross-correlation clearly shifts to a stronger anticorrelation at small {\it k} (larger scales). Note, for example, the interesting behavior of the CCC at $z=10$ (top right panel) with a characteristic peak of strongest anticorrelation, approximately corresponding to the typical distribution of sizes for ionized regions, where the Ly$\alpha$ and 21cm signals are strongly anticorrelated. At smaller {\it k} (larger scales), the medium still tends to be neutral, so the anticorrelation drops. The cross-correlation signal as shown in the dimensional cross-power spectrum (left panels) is dominated by galactic emission, and diffuse emission gains importance toward lower redshifts. The diffuse IGM component proves to be the strongest anticorrelated one of all components, with a CCC close to --1, tracing the extended ionized medium. 

Take, for example, the dimensional Ly$\alpha$ power spectra from Figure~\ref{fig:IGMgalz} at $z=7$: at a couple of Mpc$^{-1}$ the emission for the diffuse IGM is about four magnitudes smaller than the galactic emission, and the CCC in Figure~\ref{fig:cross-comp} (right) is two magnitudes higher for the diffuse IGM. This translates to a similar power for the dimensional cross-power spectrum of the diffuse IGM versus galactic emission at a couple of Mpc$^{-1}$ in the left panel of Figure~\ref{fig:cross-comp}, when comparing with Equation~(\ref{eq:CCC}). The scattered IGM displays a turnover from negative cross-correlation at intermediate {\it k} (larger scales) to positive cross-correlation at larger {\it k} (small scales). This turnover is shifted to larger scales with respect to the turnover for galactic emission, as one can anticipate already from the extension of emitting regions for different Ly$\alpha$ components in the simulation boxes shown above.

We also observe in our model, at lower redshift, smaller negative CCC for the turnover from negative to positive cross-correlation at $k\approx 4-5$ Mpc$^{-1}$, together with stronger anticorrelation at large scales, meaning the ionized bubbles extend to larger scales more frequently throughout the IGM when the universe is more ionized. The turnover scale around a few Mpc$^{-1}$ is somewhat sensitive to reionization history, as it gives an idea of the typical size of the smallest resolved ionized regions, whereas the morphology of the cross-correlation shows a clear dependence on reionization model parameters like the ionizing photon mean free path $R^\mathrm{UV}_\mathrm{mfp}$ (see, for example, Figure~\ref{fig:cross-Rmfp} in the following section). We leave the exact parameter dependence for the shift of the turnover scale for future studies, keeping the overall reionization history fixed throughout, except for a brief discussion in Section~\ref{sec:modelp}.

~\citet{2009ApJ...690..252L} noted that the cross-power spectrum with Ly$\alpha$ emitters turns positive on small scales around 1$\,$Mpc$^{-1}$. When the minimum detectable galaxy host mass is below the minimum host mass for ionizing sources, then a changed minimum detectable host mass leads to a shift in the turnover scale. For the relation between luminosity and halo mass chosen here, this shift seems to be negligible. Further studies with varied minimum host masses for galaxies and for ionizing sources, preferably at higher resolution, might be advisable. Also, in~\citet{Sobacchi:2016mhx}, a similar turnover seems possible above $\approx1\,$Mpc$^{-1}$ when cross-correlating 21cm fluctuations with Ly$\mathrm{\alpha}$ emitters. And~\citet{Silva12} find a turnover at high $k$, here at scales of the order of $\approx10\,$h$\,$Mpc$^{-1}$, when neglecting IGM emission and assuming Ly$\mathrm{\alpha}$ to be a biased tracer of the dark matter field, calculating the Ly$\mathrm{\alpha}$-galaxy/21cm cross-correlation via cross-correlation power spectra between the ionized field and matter density fluctuations and the matter power spectra themselves. This work suggests that when the fraction of ionized hydrogen becomes higher at lower redshift, the turnover scale is shifted to larger scales. Given differences in modeling and approximations made, for example, when defining ionized regions themselves, a similar behavior with scale is encouraging for future modeling efforts.

\begin{figure}
 \vfill \vspace{0.1cm}
\includegraphics[width=\columnwidth]{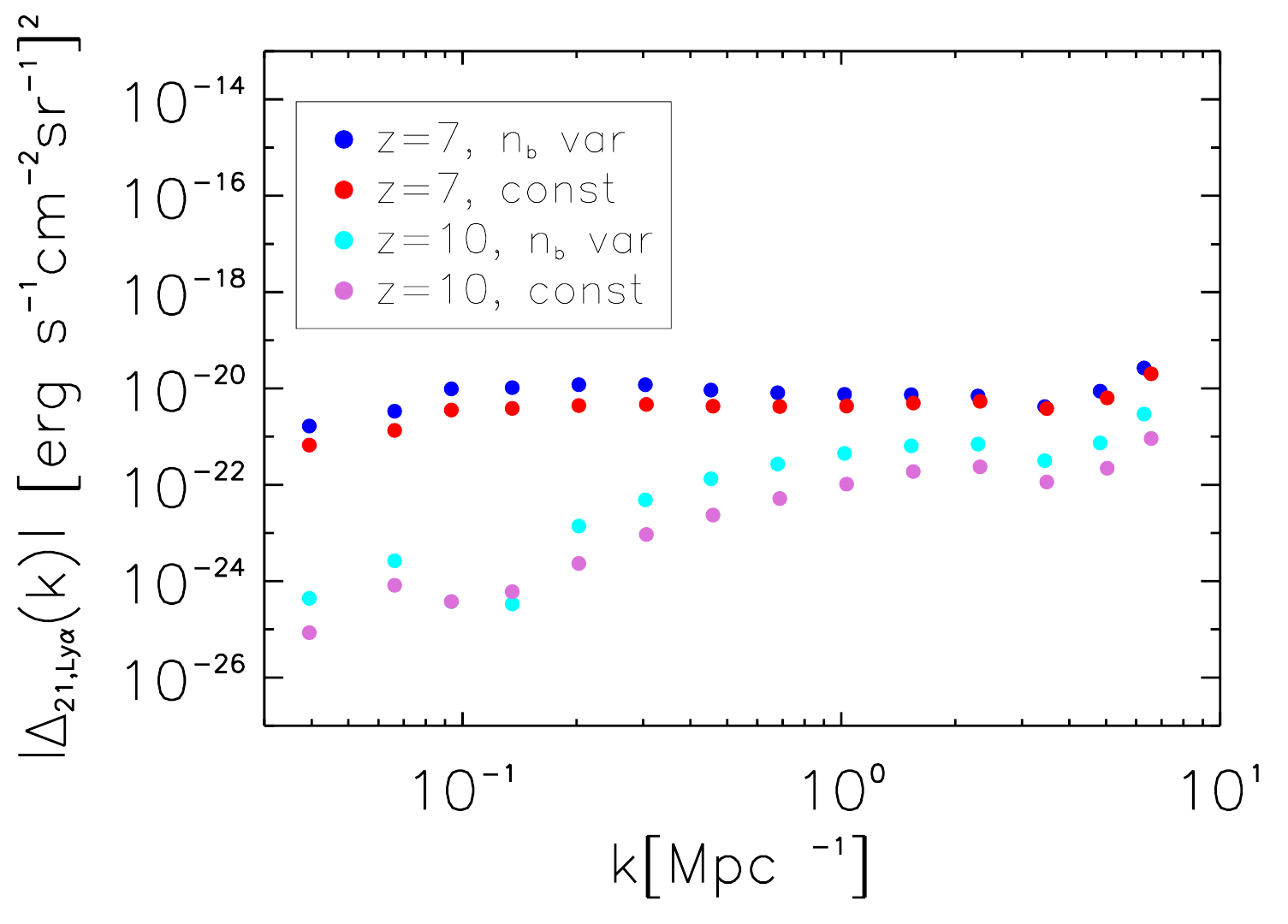}
\includegraphics[width=\columnwidth]{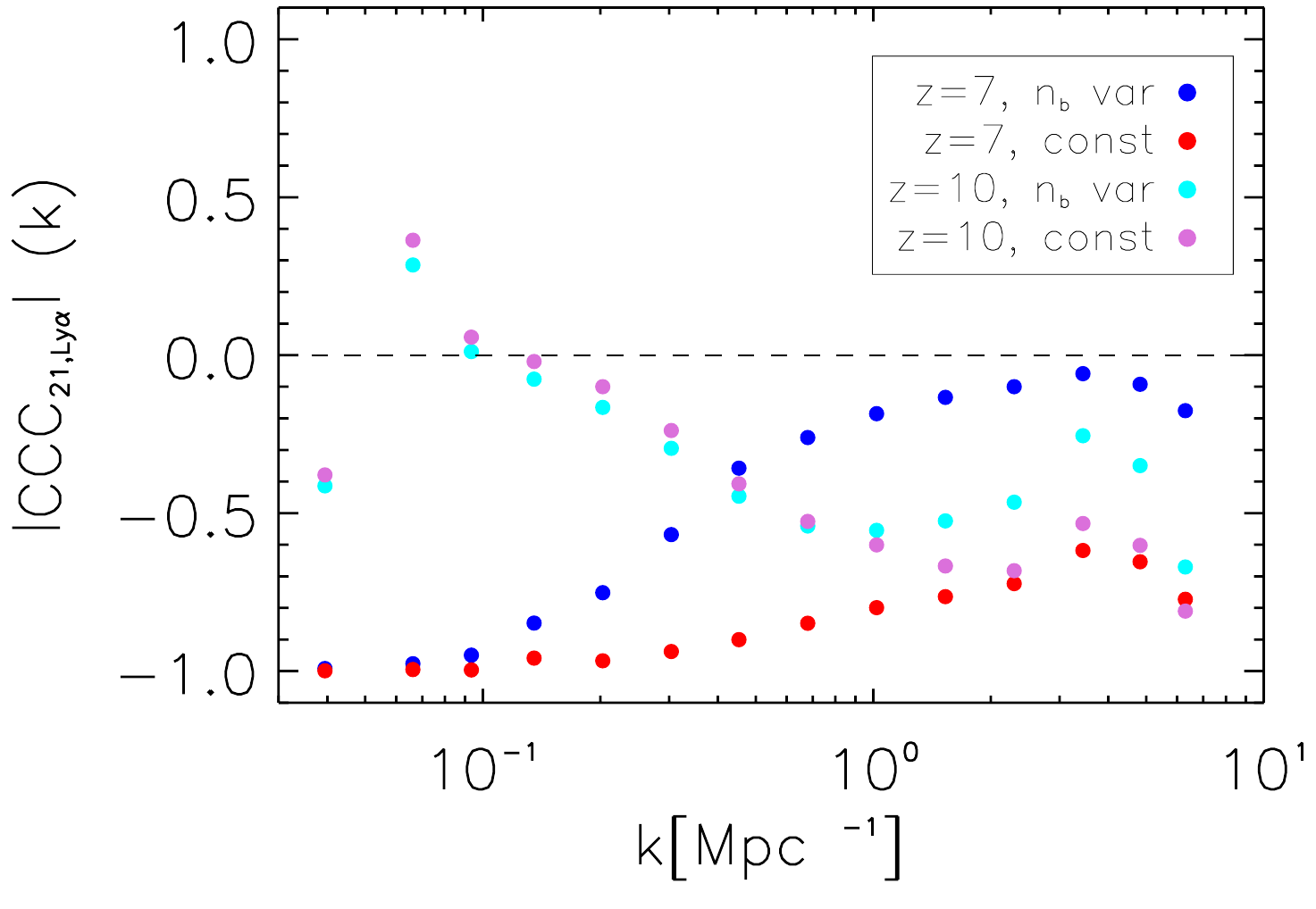}  
\caption{Dimensional cross-power spectra (left) and cross-correlation coefficient CCC (right) of 21cm fluctuations and the diffuse IGM component of Ly$\mathrm{\alpha}$ emission, taking into account fluctuations in the comoving baryonic density $n_\mathrm{b}$ (``$n_\mathrm{b}$ var") and for constant $n_\mathrm{b}$ (``const"), with constant $T_\mathrm{K}=10^4$K in ionized regions, at $z=10,\,\bar{x}_\mathrm{HI}=0.87$ and $z=7,\,\bar{x}_\mathrm{HI}=0.27$.}
 \label{fig:dIGMcross}
\end{figure}

In Figure~\ref{fig:dIGMcross} we illustrate the change of the dimensional cross-power spectra (top panel) and the CCC (bottom panel) for the diffuse IGM component of Ly$\mathrm{\alpha}$ emission at redshift $z=10$ and $z=7$, when neglecting fluctuations in comoving baryonic density $n_\mathrm{b}$, versus taking them into account, as discussed for simulation boxes and power spectra in Section~\ref{sec:lyasim}. The cross-correlation for constant gas temperature and comoving baryonic density sets a lower limit for the cross-correlation signal of diffuse IGM emission in Ly$\mathrm{\alpha}$. The characteristic shape is similar in both cases depicted at redshift $z=10$ and $z=7$.
 
\subsubsection{Some Parameter Studies}\label{sec:modelp}
\begin{figure}
\includegraphics[width=\columnwidth]{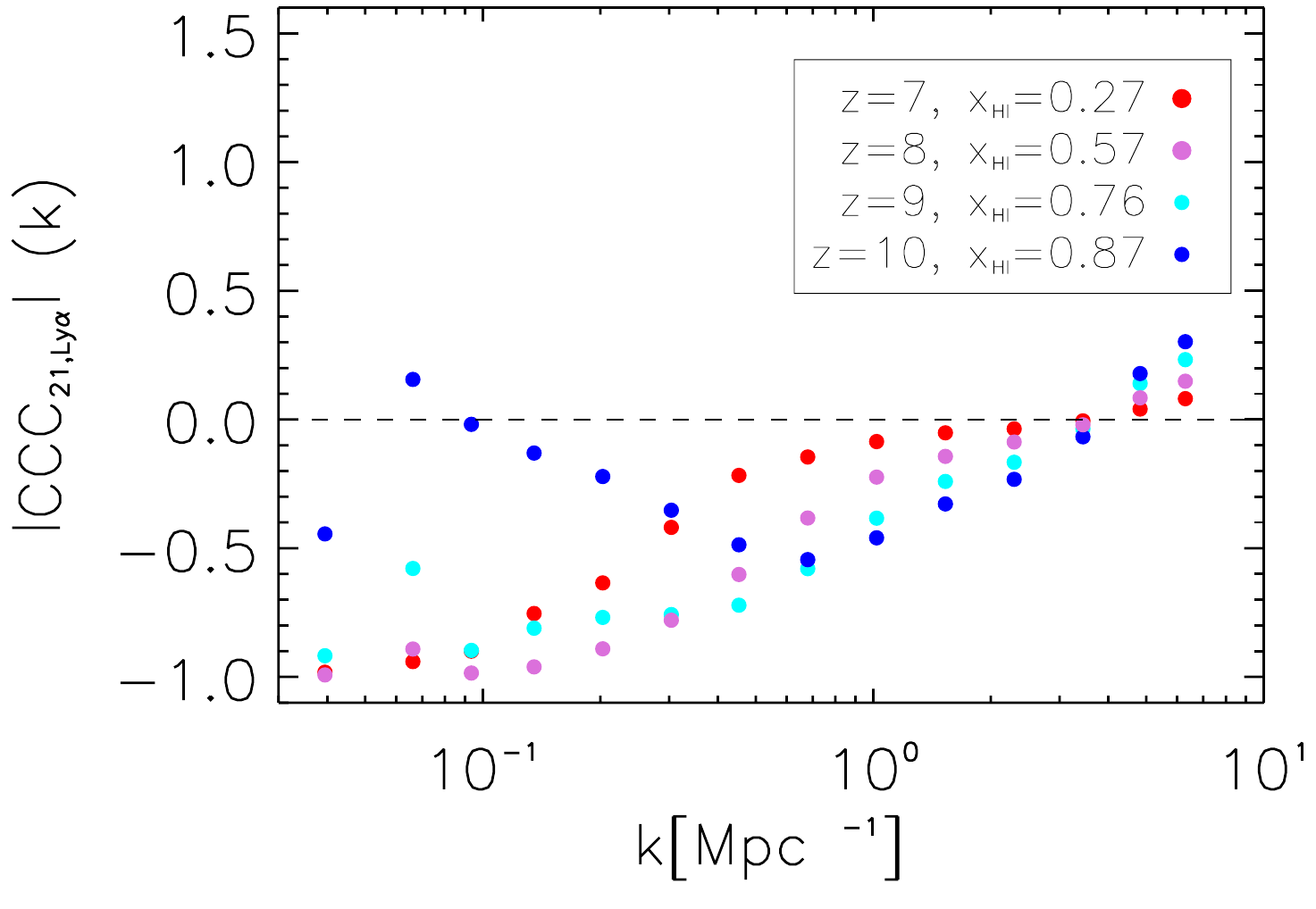}
\caption{Cross-correlation coefficient CCC of 21cm and galactic Ly$\mathrm{\alpha}$ fluctuations for redshift $z=\left\{ 7,8,9,10\right\}$ and corresponding $\bar{x}_\mathrm{HI}=\left\{0.27,0.56,0.76,0.87 \right\}$.}
 \label{fig:cross-zvar}
\end{figure}

 Here we show the impact of varying selected model parameters on the cross-correlation signal between 21cm and Ly$\mathrm{\alpha}$ brightness fluctuations. The parameters that we vary, while keeping the overall reionization history fixed, are the duty cycle $f_\mathrm{duty}$, which determines the halo-occupying fraction for Ly$\mathrm{\alpha}$-emitting galaxies as introduced in Section~\ref{sec:lyasim}, and the escape fraction $f_\mathrm{esc}$ of Ly$\mathrm{\alpha}$ photons from Ly$\mathrm{\alpha}$-emitting galaxies. We also vary the mean free path of ionizing radiation $R_\mathrm{mfp}^\mathrm{UV}$, which will affect the reionization history. In addition, the cross-correlation signal for a range of mean ionized fractions $\bar{x}_\mathrm{HI}$ is displayed in Figure~(\ref{fig:cross-zvar}), following the redshift evolution of our fiducial model.
 
We note that the variation of parameters like the escape fraction $f_\mathrm{esc}$ will also alter the reionization history, when, instead of the usual ionizing efficiency $\zeta$ as an effective parameter for the amount of ionizing radiation released, the equilibrium between ionizing and recombination rate is used to define ionized regions, as was done in~\citet{Silva12}. Studying the impact on the cross-correlation of the definition applied for ionized regions might be an interesting future avenue. 

\begin{figure}
\includegraphics[width=\columnwidth]{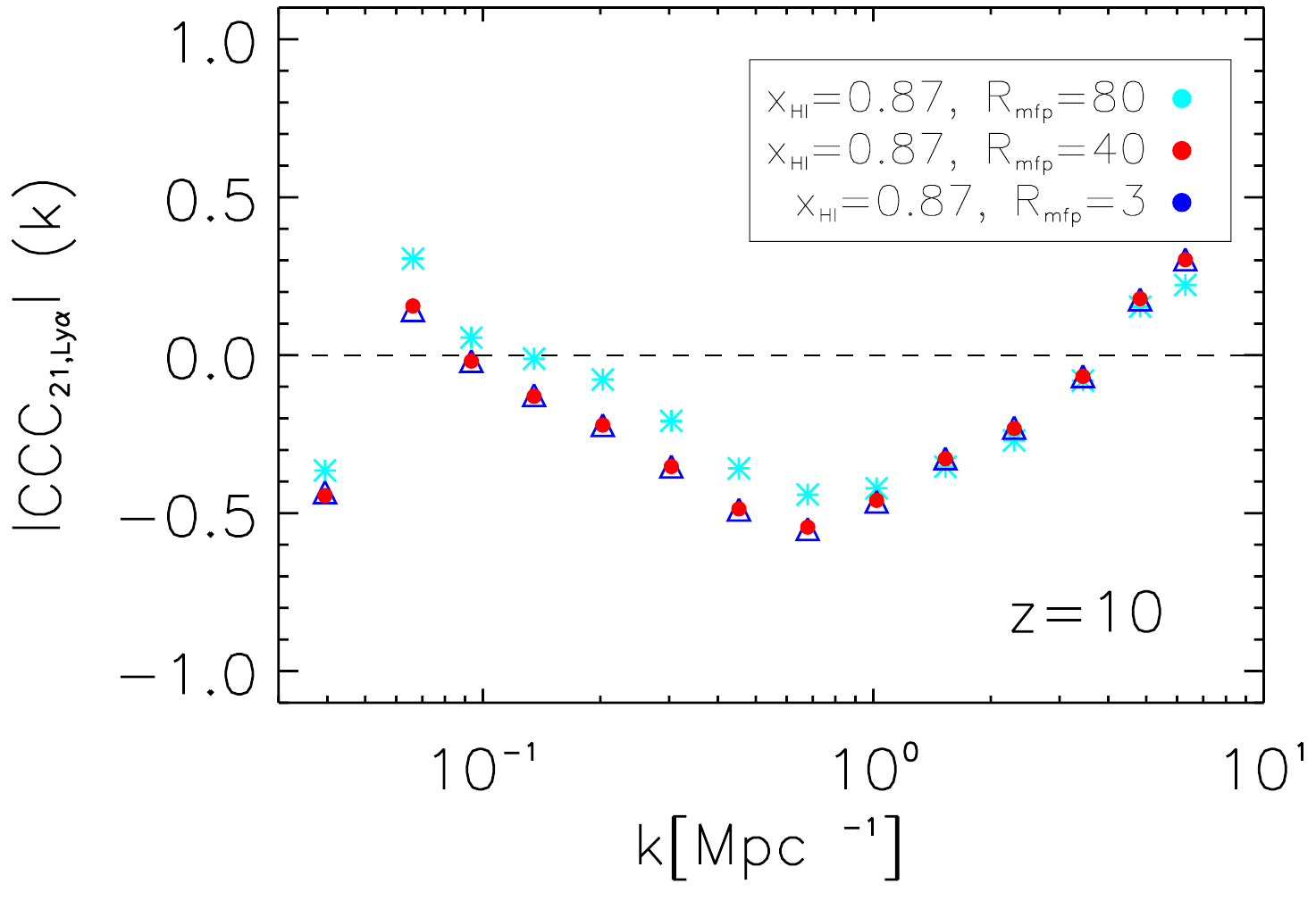}
\vspace{0.1cm}
\vfill
\includegraphics[width=\columnwidth]{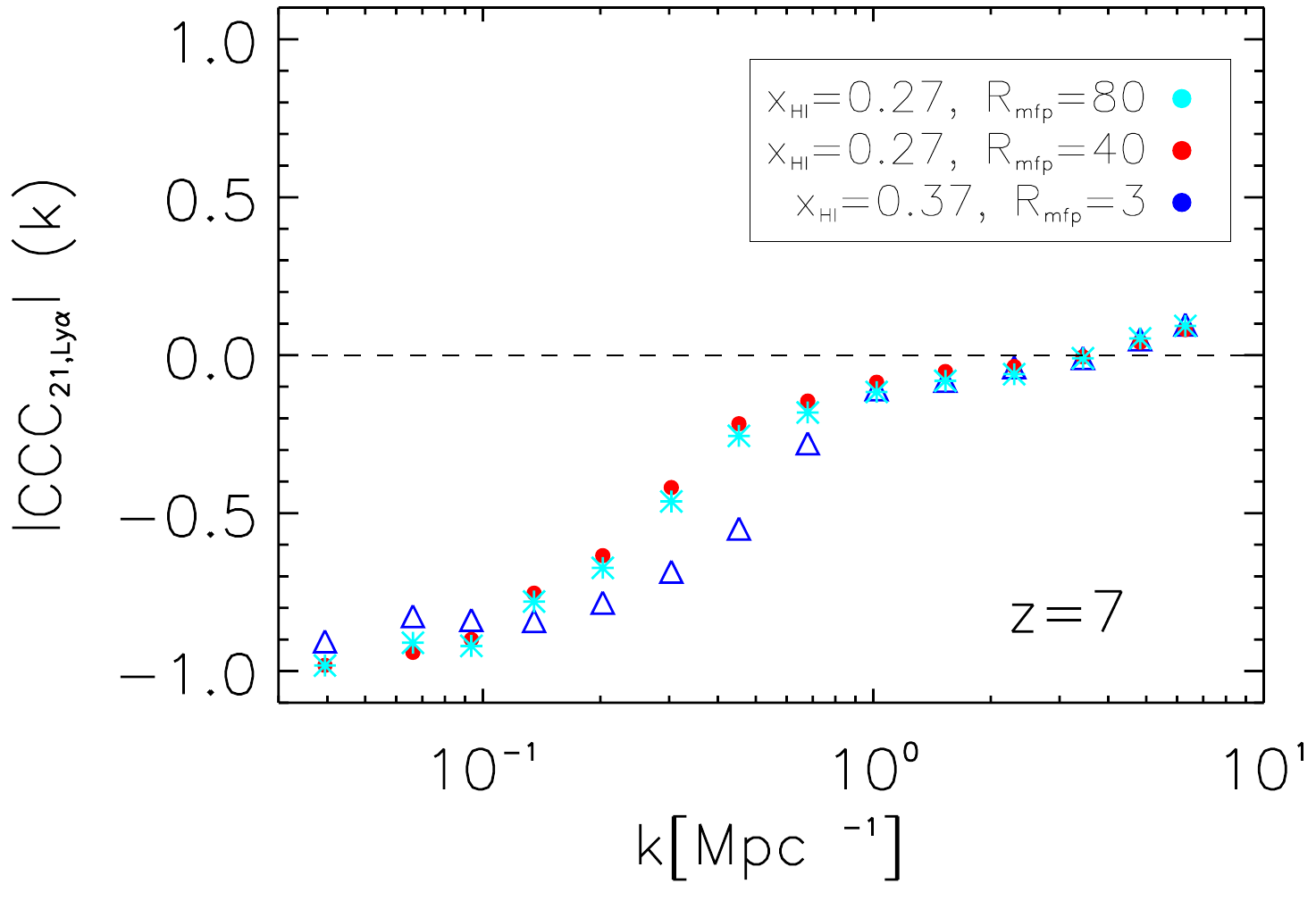}
\caption{Cross-correlation coefficient CCC of 21cm and galactic contribution to Ly$\mathrm{\alpha}$ fluctuations for mean free path of ionizing radiation $R_\mathrm{mfp}=$ $\left\{ 80,40,3\right\}$Mpc with $\bar{x}_\mathrm{HI}=$ $\left\{ 0.27,0.27,0.37\right\}$ (asterisks, points, triangles) at redshift $z=10$ (top) and $z=7$ (bottom).}
 \label{fig:cross-Rmfp}
\end{figure}

In Figure~\ref{fig:cross-Rmfp} the CCCs for a mean free path of ionizing radiation $R_\mathrm{mfp}= \left\{3,40,80 \right\}\,$Mpc are compared. At redshift $z=10$, the CCC shows a very similar behavior, with ionized regions of mean size $\approx 1.5\,$Mpc for all three values of $R_\mathrm{mfp}$. For the mean sizes, we trace through our simulation from each halo center along a line of sight (LOS), chosen to be the z axis, until we cross the first phase transition from ionized to neutral, and we calculate the mean of the distances obtained. Until redshift $z=7$, a stronger dependence on $R_\mathrm{mfp}$ becomes apparent. The case of highest mean free path $R_\mathrm{mfp}= 80\,$Mpc displays a lower neutral fraction of $\bar{x}_\mathrm{HI}= 0.27$ as well as larger ionized regions of $\approx 12.7\,$Mpc on average, $R_\mathrm{mfp}=40\,$Mpc has $\bar{x}_\mathrm{HI}=0.27$ and average size $\approx 12.8\,$Mpc of ionized regions, and $R_\mathrm{mfp}=3\,$Mpc leads to $\bar{x}_\mathrm{HI}=0.37$ and average size $\approx 6.5\,$Mpc. Note the slightly higher mean bubble size at $R_{mfp}=40$ as compared to $R_{mfp}=80$. This might be due to the effect of bubble sizes saturating at higher mean free paths (as radiation of a certain energy is only able to penetrate the medium up to a certain distance), while at the same time a slight scatter is introduced  by the variance of different density field realizations. We stress that this effect is only present in this parameter study section, as we used the same density field realization for the other sections that analyze results for our fiducial model. Note as well that, as for $R_\mathrm{mfp}$, the variation of ionizing efficiency $\zeta$ and virial temperature $T_\mathrm{vir}$ will also have the effect of altering the reionization history. 

\begin{figure}
\includegraphics[width=\columnwidth]{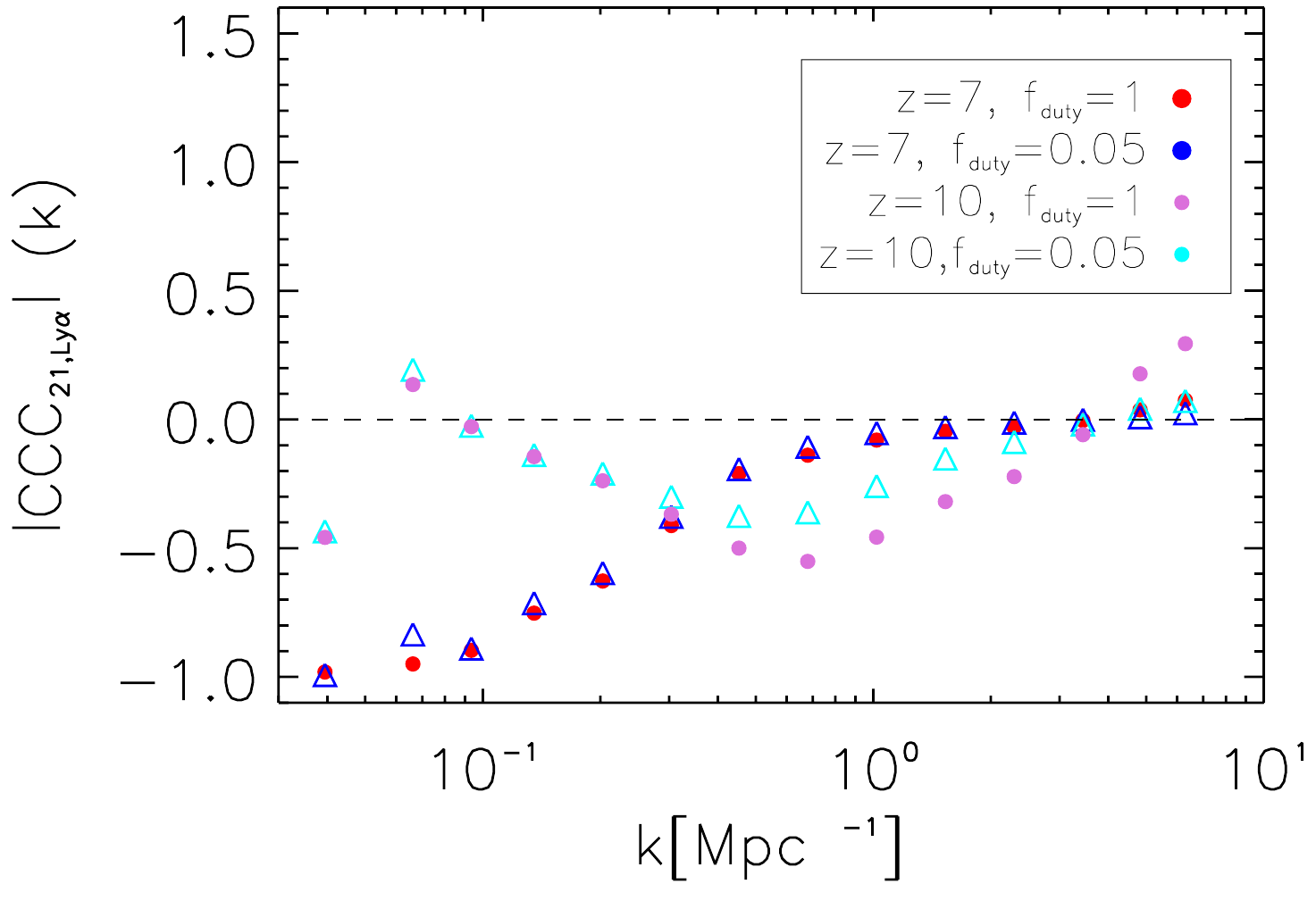}
\caption{Cross-correlation coefficient CCC of 21cm and galactic Ly$\mathrm{\alpha}$ fluctuations for duty cycles $f_\mathrm{duty}=1$ and $f_\mathrm{duty}=0.05$.}
 \label{fig:cross-wedge}
\end{figure}

\begin{figure}
\includegraphics[width=\columnwidth]{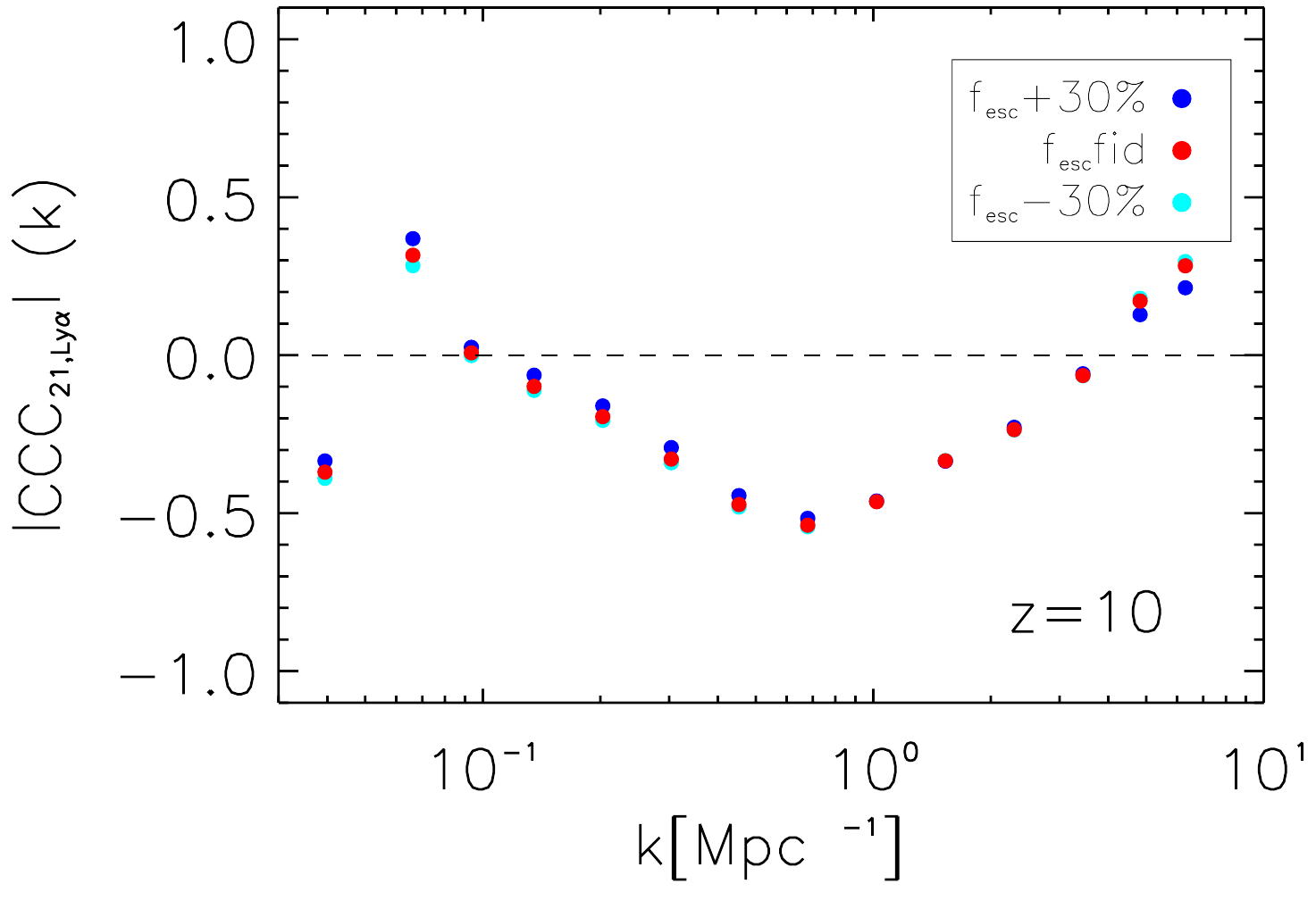}
\vspace{0.1cm}
\vfill
\includegraphics[width=\columnwidth]{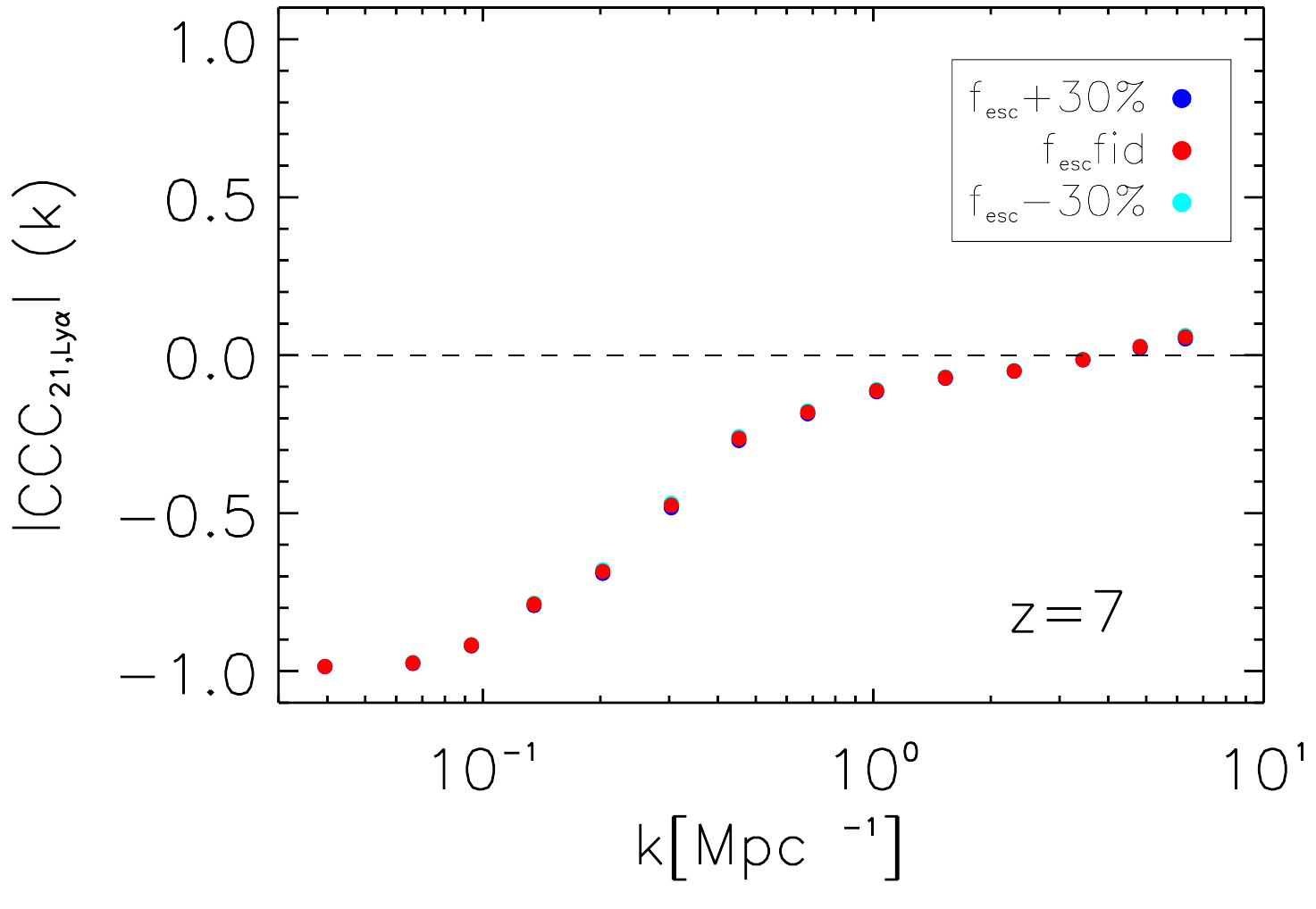}
\caption{Cross-correlation coefficient CCC of 21cm and total Ly$\mathrm{\alpha}$ fluctuations for $30\%$ higher and lower escape fraction $f_\mathrm{esc}$ as compared to the fiducial values from~\cite{Razoumov:2010} at redshift $z=10$ (top) and $z=7$ (bottom).}
 \label{fig:cross-fesc}
\end{figure}

Figure~\ref{fig:cross-wedge} shows the CCC for two assumed duty cycles $f_\mathrm{duty}=1$ and $f_\mathrm{duty}=0.05$ at redshift $z=10$ and $z=7$ and tests the impact on the cross-correlation signal of reducing the fraction of halos occupied with Ly$\mathrm{\alpha}$-emitting galaxies, where halos above a minimum mass $M_\mathrm{min}$ that corresponds to a virial temperature of $T_\mathrm{vir}=10^4\,$K were randomly populated. As expected, a reduction of the fraction of halos that host a Ly$\mathrm{\alpha}$-emitting galaxy also reduces the power of our cross-correlation signal. We also test the impact of varying the Ly$\mathrm{\alpha}$ escape fraction $f_\mathrm{esc}$ in Figure~\ref{fig:cross-fesc} for redshift $z=10$ (top panel) and $z=7$ (bottom panel). The two cases of increasing and decreasing the escape fraction by $30\%$ are shown together with the fiducial case that follows~\cite{Razoumov:2010}. Increasing the escape fraction $f_\mathrm{esc}$ has a slight tendency to decrease the cross-correlation signal at some scales, while decreasing $f_\mathrm{esc}$ can slightly increase the signal. It needs to be noted again, though, that varying both $f_\mathrm{duty}$ and $f_\mathrm{esc}$ will have an effect on the reionization history, when defining ionized regions not by mean collapse fraction but by radiation equilibrium within the ionized regions.

For comparison, Figure~\ref{fig:cross-zvar} shows the change in CCC with redshift and therefore $\bar{x}_\mathrm{HI}$ for our fiducial model. Here it becomes obvious, for example, how the peak in negative CCC shifts to smaller {\it k}, or larger scales, when reionization progresses and the mean neutral fraction decreases.

To sum up, the cross-correlation signal of 21cm and Ly$\mathrm{\alpha}$ fluctuations during the EoR is sensitive to parameters that change the reionization history or the clustering properties of emitting galaxies.

\subsection{Ly$\mathrm{\alpha}$ Damping Tail} \label{sec:damp}
 In order to more realistically simulate the observed galactic Ly$\mathrm{\alpha}$ emission, IGM attenuation due to the damping tail of Ly$\mathrm{\alpha}$ needs to be taken into account. We relate the intrinsic luminosity in Ly$\mathrm{\alpha}$ assigned to halos as in Equation~(\ref{eq:Lintr}) to the observed luminosity via optical depth $\tau_\mathrm{Ly\alpha}$ for Ly$\mathrm{\alpha}$. This gives for the observed galactic Ly$\mathrm{\alpha}$ luminosity
\begin{equation}
L^\mathrm{gal}_\mathrm{obs} = L^\mathrm{gal}e^{-\tau_\mathrm{Lya}} \label{eq:Lobs} \,. 
\end{equation}
The optical depth at Ly$\mathrm{\alpha}$ line resonance in neutral hydrogen, which makes up the not yet ionized part of the IGM, can under the assumption of uniform gas distribution be approximated at high redshift by~\citep{1965GP,Barkana:2000fd}
\begin{equation}
\tau_\mathrm{s} \approx 6.45\times 10^5 \left( \frac{\Omega_\mathrm{b} h}{0.03}\right) \left( \frac{\Omega_\mathrm{m}}{0.3}\right)^{-0.5} \left( \frac{1+z_\mathrm{s}}{10}\right)^{1.5},
\end{equation}
with source redshift $z_\mathrm{s}$, and present-day density parameters of matter $\Omega_\mathrm{m}$ and of baryons $\Omega_\mathrm{b}$.

The Ly$\mathrm{\alpha}$ radiation is redshifted between the emitting source sitting in an ionized bubble and the edge of the neutral medium around the bubble, and therefore gets shifted from the line core in resonance to the line wings of lower optical depth on the way to the observer. For Ly$\mathrm{\alpha}$ emission at source redshift $z_\mathrm{s}$, which redshifts by $z_\mathrm{s}-z_\mathrm{obs}$ before reaching the edge of the neutral IGM fully ionized at $z_\mathrm{reion}$,~\citet{Escude98} finds for the optical depth $\tau_\mathrm{Ly\alpha}$ of Ly$\mathrm{\alpha}$ emission the analytical result 
\begin{align}
\tau_\mathrm{Ly\alpha} \left( z_\mathrm{obs}\right) = &  \tau_\mathrm{s} \bar{x}_\mathrm{HI}  \left( \frac{2.02\times 10^{-8} }{\pi}\right) \left(\frac{1+z_\mathrm{s} }{1+z_\mathrm{obs}}\right)^{1.5} \nonumber \\ &  \vspace{0.2cm}\times \left[ I\left( \frac{1+z_\mathrm{s}}{1+z_\mathrm{obs}}\right) -  I\left( \frac{1+z_\mathrm{reion}}{1+z_\mathrm{obs}}\right)\right] , \label{eq:tauLya}
\end{align} 
with average neutral hydrogen fraction $\bar{x}_\mathrm{HI}$ and helper function $I\left( x\right)$ defined as
\begin{align}
I\left( x\right) = & \frac{x^{4.5}}{1-x} + \frac{9}{7}x^{3.5} + \frac{9}{5}x^{2.5} + 3x^{1.5} +9x^{0.5 } \nonumber \\ & - 4.5\ln\left( \frac{1+x^{0.5}}{1-x^{0.5}}\right) . 
\end{align}
The approach taken in Equation~(\ref{eq:tauLya}) to calculate $\tau_\mathrm{Ly\alpha}$ assumes that the sum over neutral patches can be replaced by an average neutral fraction, where the Ly$\alpha$ damping wing averages over a sufficiently long path length. Alternatively, one can sum the contribution to $\tau_\mathrm{Ly\alpha}$ of neutral patches along the LOS~\citep{Mesinger:2007kd}. This yields for $\tau_\mathrm{Ly\alpha}$, summing over each neutral patch that extends from $z_{ai}$ to $z_{ei}$ with $z_{ai}>z_{ei}$,
\begin{align}
\tau_\mathrm{Ly\alpha} \left( z_\mathrm{obs}\right) = &  \tau_\mathrm{s} \sum_i x_{\mathrm{HI},i}  \left( \frac{2.02\times 10^{-8} }{\pi}\right) \left(\frac{1+z_{\mathrm{a}i} }{1+z_\mathrm{obs}}\right)^{1.5} \nonumber \\ &  \vspace{0.2cm}\times \left[ I\left( \frac{1+z_{\mathrm{a}i}}{1+z_\mathrm{obs}}\right) -  I\left( \frac{1+z_{\mathrm{e}i}}{1+z_\mathrm{obs}}\right)\right] . \label{eq:tauLya2}
\end{align} 

\begin{figure*}
\plottwo{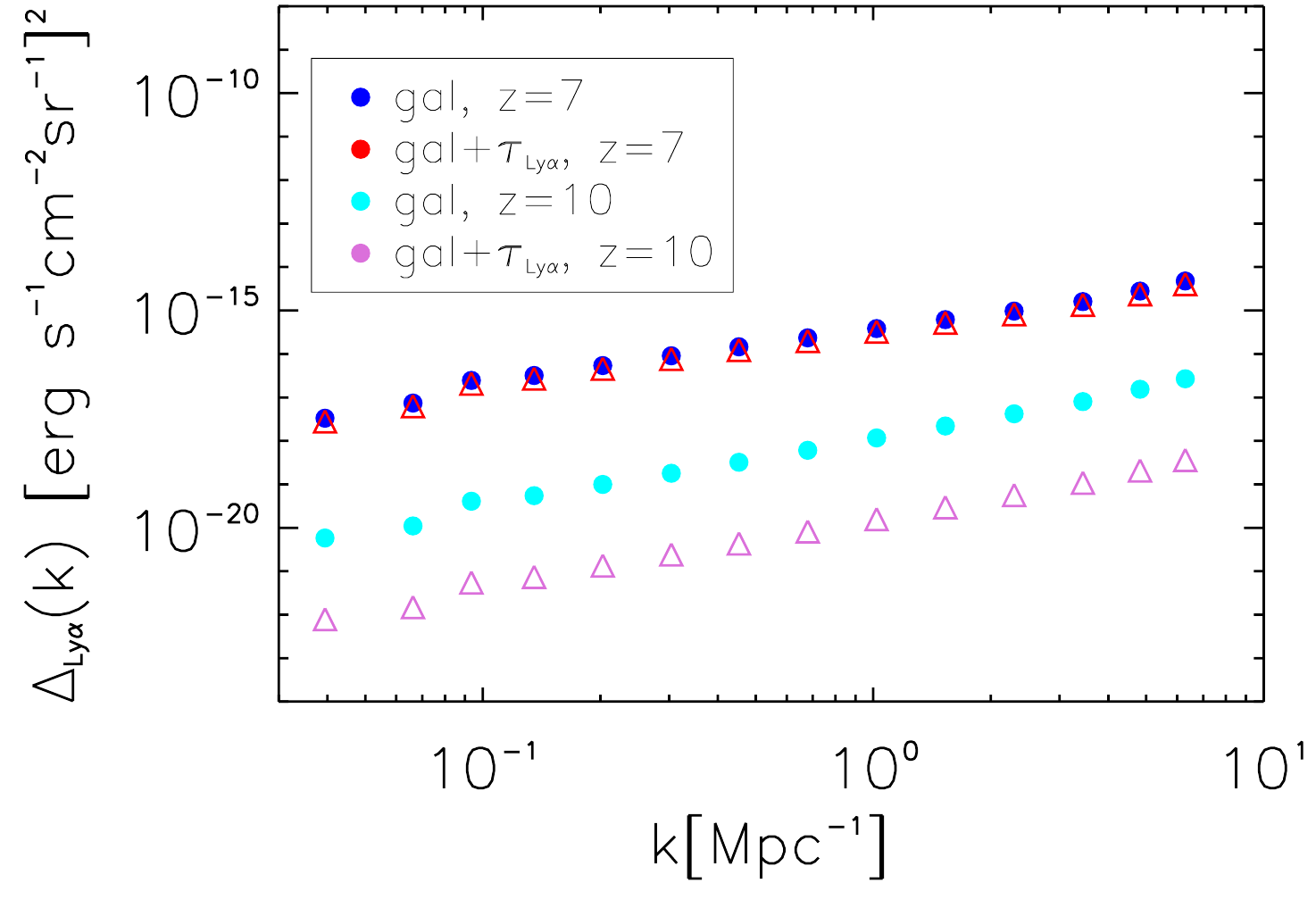}{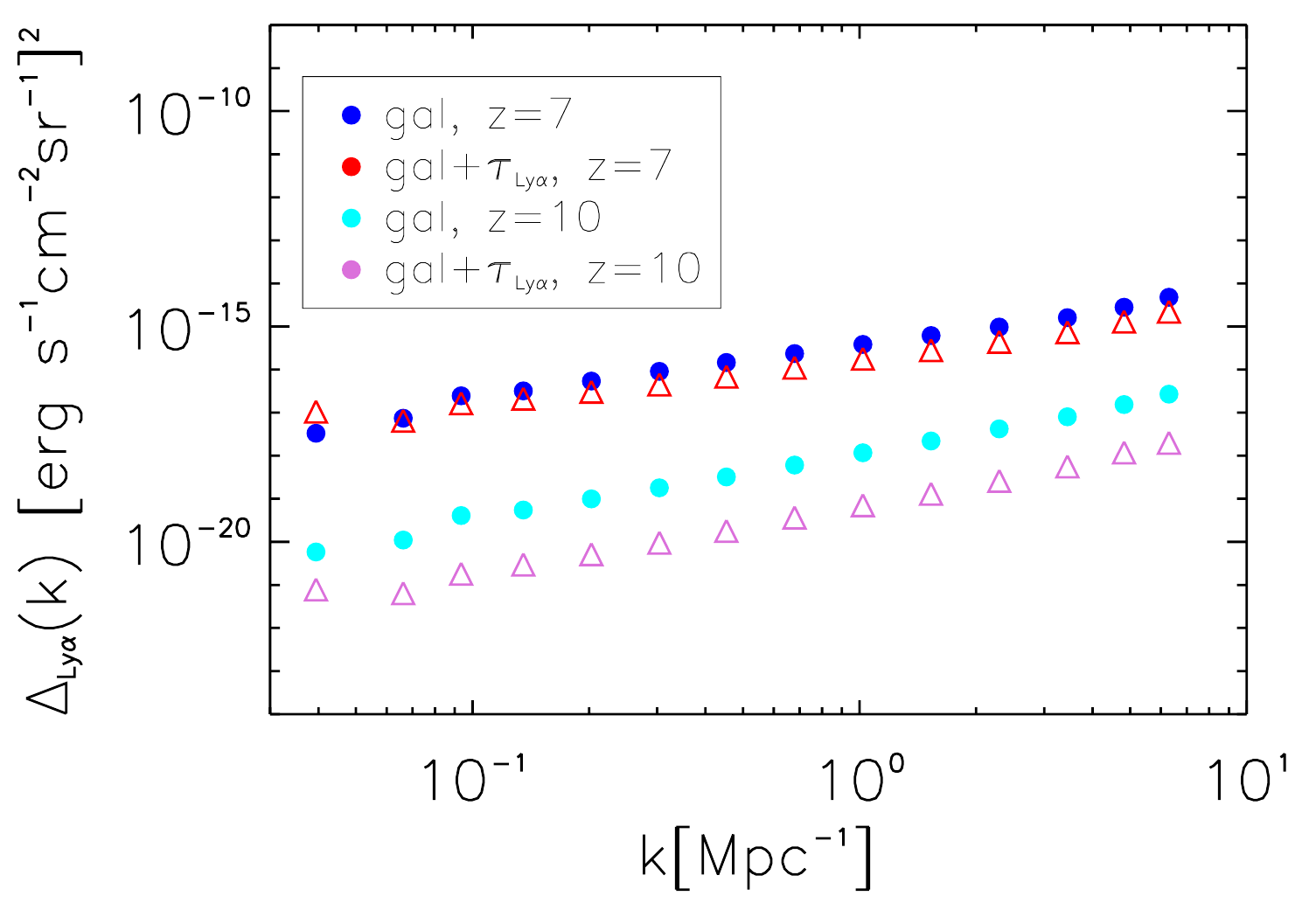}
\vfill \vspace{0.1cm} 
\plottwo{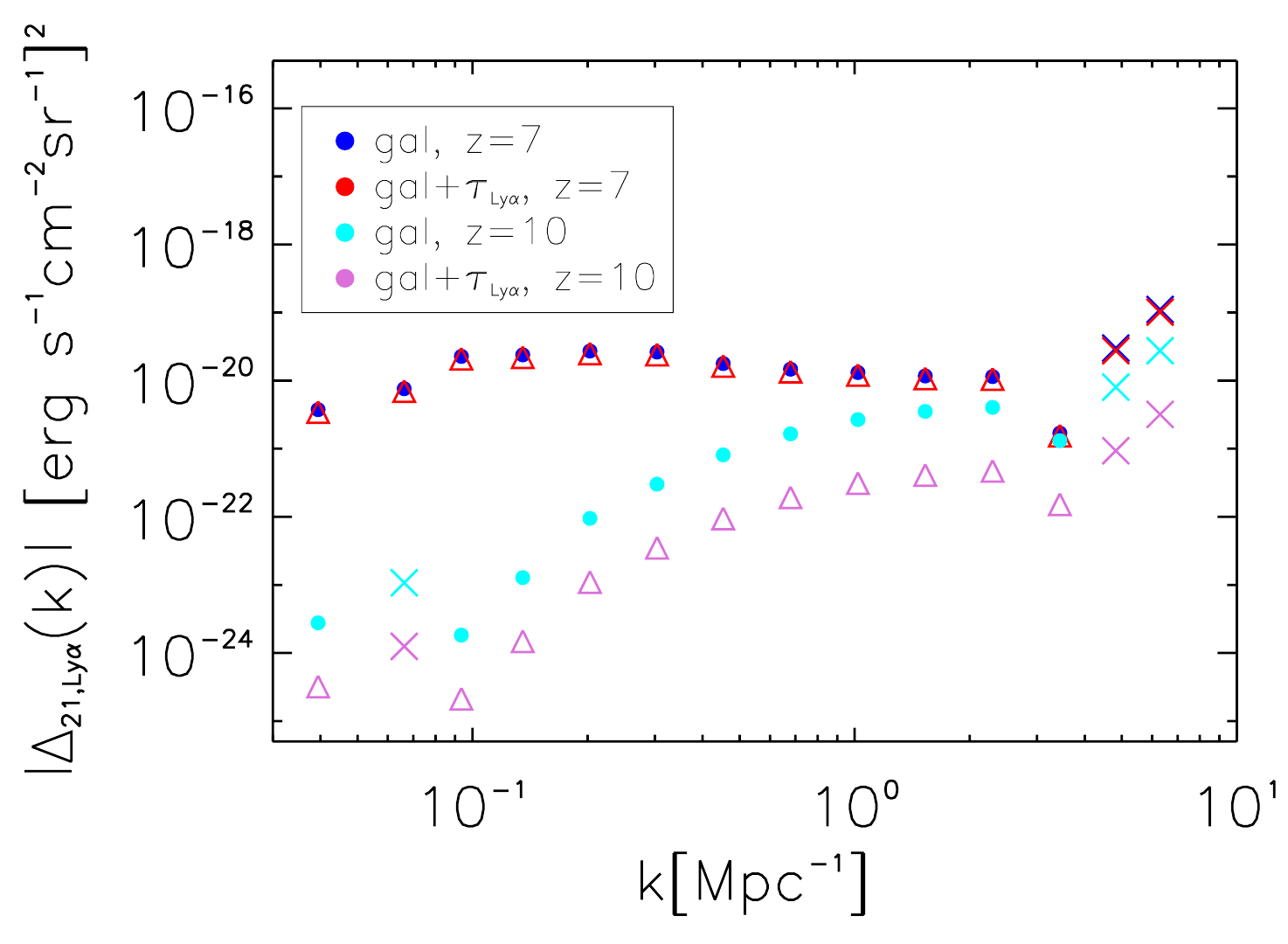}{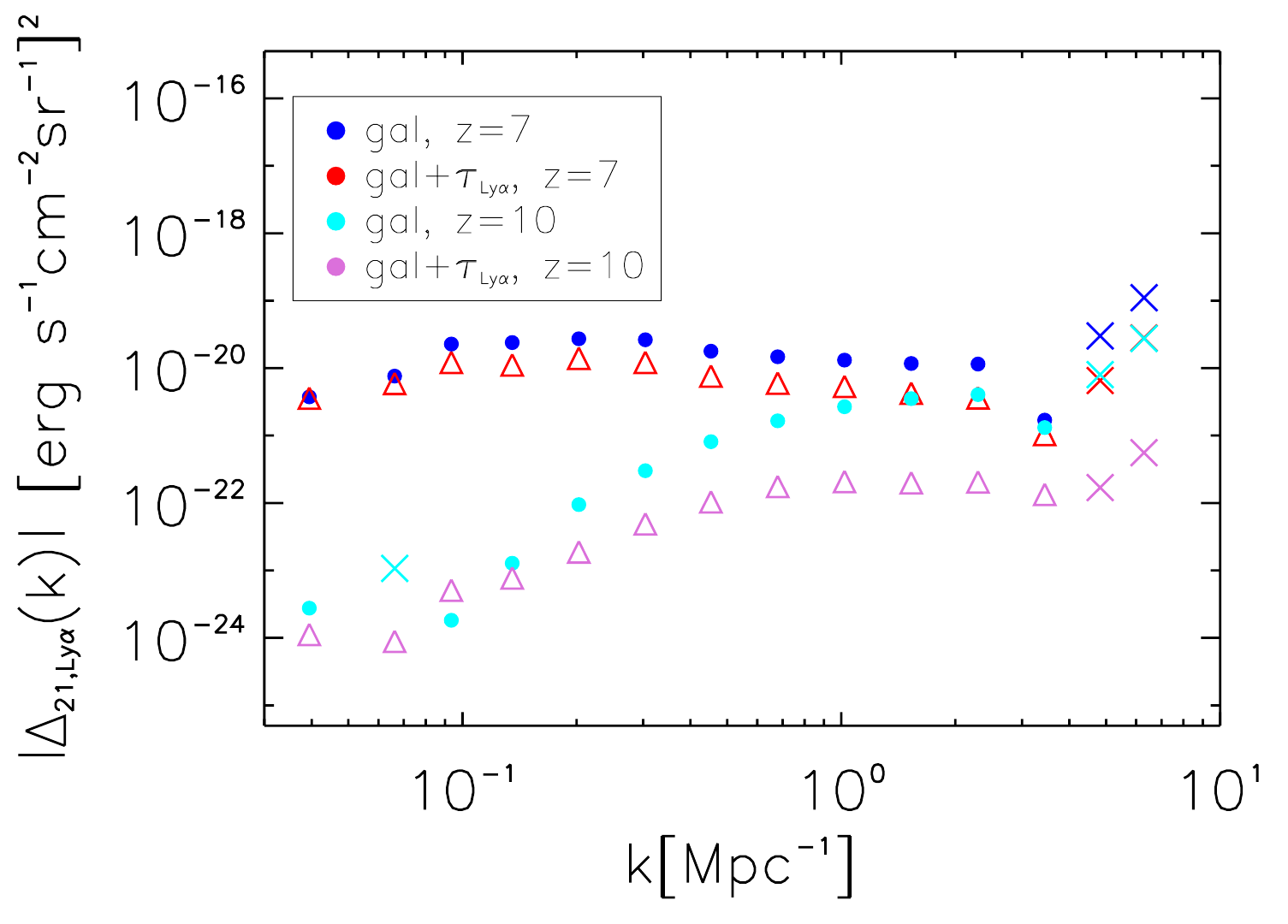}
\vfill \vspace{0.2cm} 
\plottwo{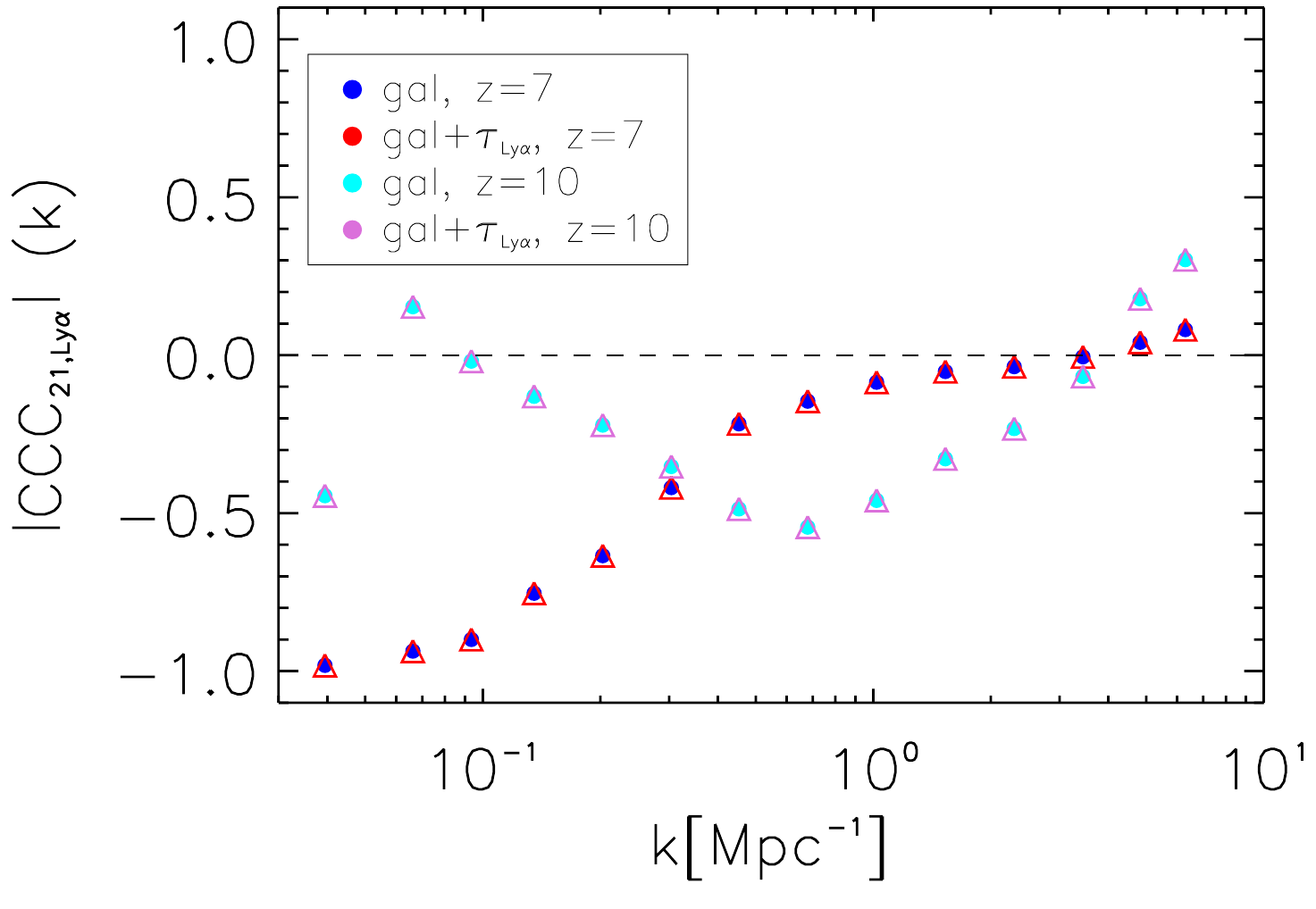}{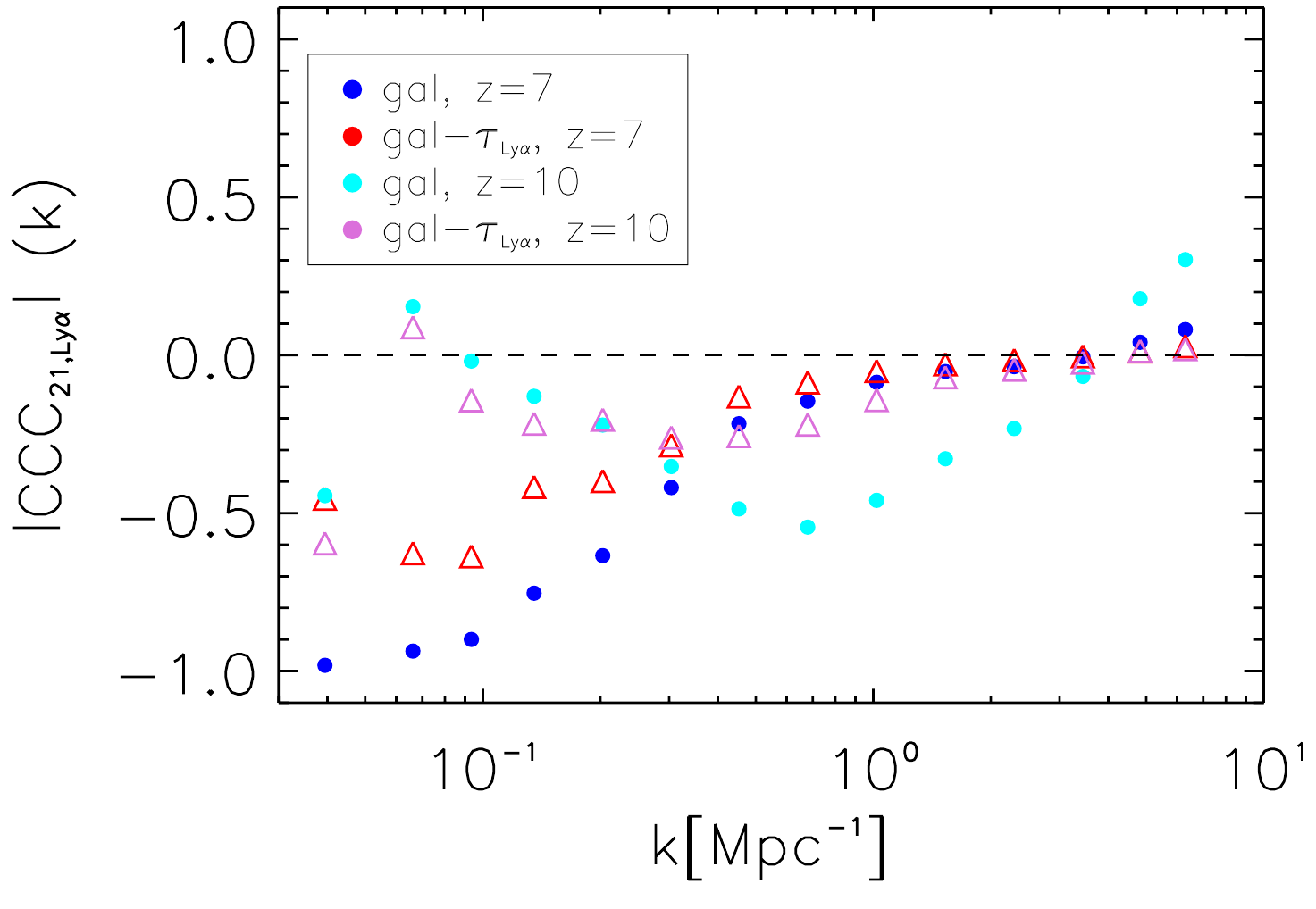}
\caption{Left panels: dimensional Ly$\mathrm{\alpha}$ power spectra (top), dimensional cross-power spectra (middle), and cross-correlation coefficient CCC$_\mathrm{21,Ly\alpha}$ (bottom) for the galactic contribution to the Ly$\mathrm{\alpha}$ emission with (triangles) and without (points) Ly$\mathrm{\alpha}$ damping at redshift $z=10$ (cyan, orchid) and $z=7$ (blue, red), assuming the commonest filter scale as the typical size of an ionized region. Right panels: same as left panels, but Ly$\mathrm{\alpha}$ damping calculated for tracing of neutral (damping) regions through the simulation along the z-axis LOS.} \label{fig:damp}
\end{figure*}

In order to calculate the redshift offset for the patches of neutral IGM, we need to trace phase transitions from ionized to neutral, and vice versa, along the LOS, starting from the center of each ionized halo. To each phase transition the corresponding redshift offset is assigned. We thereby match our halo catalog at given redshift to corresponding ionized regions, assuming for now each galaxy to be in the center of the halo it is assigned to. The optical depth is then used to correct intrinsic luminosities and calculate observed luminosities for each halo that includes Ly$\mathrm{\alpha}$ damping following Equation~(\ref{eq:Lobs}). 

For the sizes of the ionized regions surrounding each halo, we compare two approximations. The first simple approach consists of taking the commonest filter scale as the typical size of an ionized bubble, which is similar for most halos at a given redshift and corresponds to about 4$\,$Mpc at $z=10$, and about 20$\,$Mpc at $z=7$ for our fiducial model. In the approach of Equation~(\ref{eq:tauLya}) we trace through our simulation box along an LOS, chosen to be from each halo center along the z axis here, until we cross the phase transition from ionized to neutral. Mean sizes of ionized regions are $\approx1.5\,$Mpc at $z=10$ and $\approx12.8\,$Mpc at $z=7$, therefore about a factor of two smaller than in our first simple approach, leading to a generally stronger damping effect. Tracing through the simulation and summing the optical depth for each neutral patch as in Equation~(\ref{eq:tauLya2}) results in similar damping as compared to the use of Equation~(\ref{eq:tauLya}) at larger {\it k}. But for smaller {\it k} (larger scales,) the power spectra are up to an order of magnitude less damped at $z=10$, and up to about $30\%$ at $z=7$.
  
In Figure~\ref{fig:damp} we show the uncorrected dimensional power spectra (top), cross-power spectra (middle), and cross-correlation coefficient CCC (bottom) for redshift $z=10$ and $z=7$ alongside the the corrected power spectra for galactic emission in Ly$\mathrm{\alpha}$, in the left panels for the first simple approach of assuming the commonest filter scale as the typical size of an ionized bubble and calculating the optical depth as in Equation~(\ref{eq:tauLya}), and in the right panels for the sizes of ionized bubbles via tracing through the simulation and summing the damping effect for each neutral patch as in Equation~(\ref{eq:tauLya2}). As at a given redshift the typical bubble sizes are fairly similar, we observe a rather uniform decrease in power with scale, with a stronger decrease for high {\it k} in the case of tracing neutral patches along the LOS. Also, at higher redshift, the ionized bubbles are significantly smaller, the redshifting is away from the line core until the bubble edge is smaller, the extent of neutral patches is larger, and therefore the damping effect is bigger (up to one order of magnitude) at redshift $z=10$ as compared to $z=7$, where the effect is at the level of 10$\%$--20$\%$ for the commonest filter scale and up to a factor of two for tracing along the LOS.  For the cross-correlation power spectra (middle panels), as well as the cross-correlation coefficient CCC (bottom panels), taking into account Ly$\mathrm{\alpha}$ damping in the (more accurate) approach of tracing the neutral patches in the simulation, instead of using the mean filtering scales as a rough approximation, displays a mostly stronger and more scale-dependent damping effect.

\subsection{Cross-correlation of Ly$\mathrm{\alpha}$ and H$\mathrm{\alpha}$} \label{sec:crossHa}
 Different line fluctuations trace galactic and intergalactic emission in differing ways. For example, H$\mathrm{\alpha}$ fluctuations mostly stem from galactic emission, whereas Ly$\mathrm{\alpha}$ fluctuations stem from both galactic emission and a contribution from the IGM. We therefore cross-correlate H$\mathrm{\alpha}$ (galactic plus very subdominant diffuse IGM contribution) and Ly$\mathrm{\alpha}$ fluctuations in order to pick out the IGM contribution of Ly$\mathrm{\alpha}$ emission from the total Ly$\mathrm{\alpha}$ emission. 
 
\begin{figure}
\includegraphics[width=0.9\columnwidth]{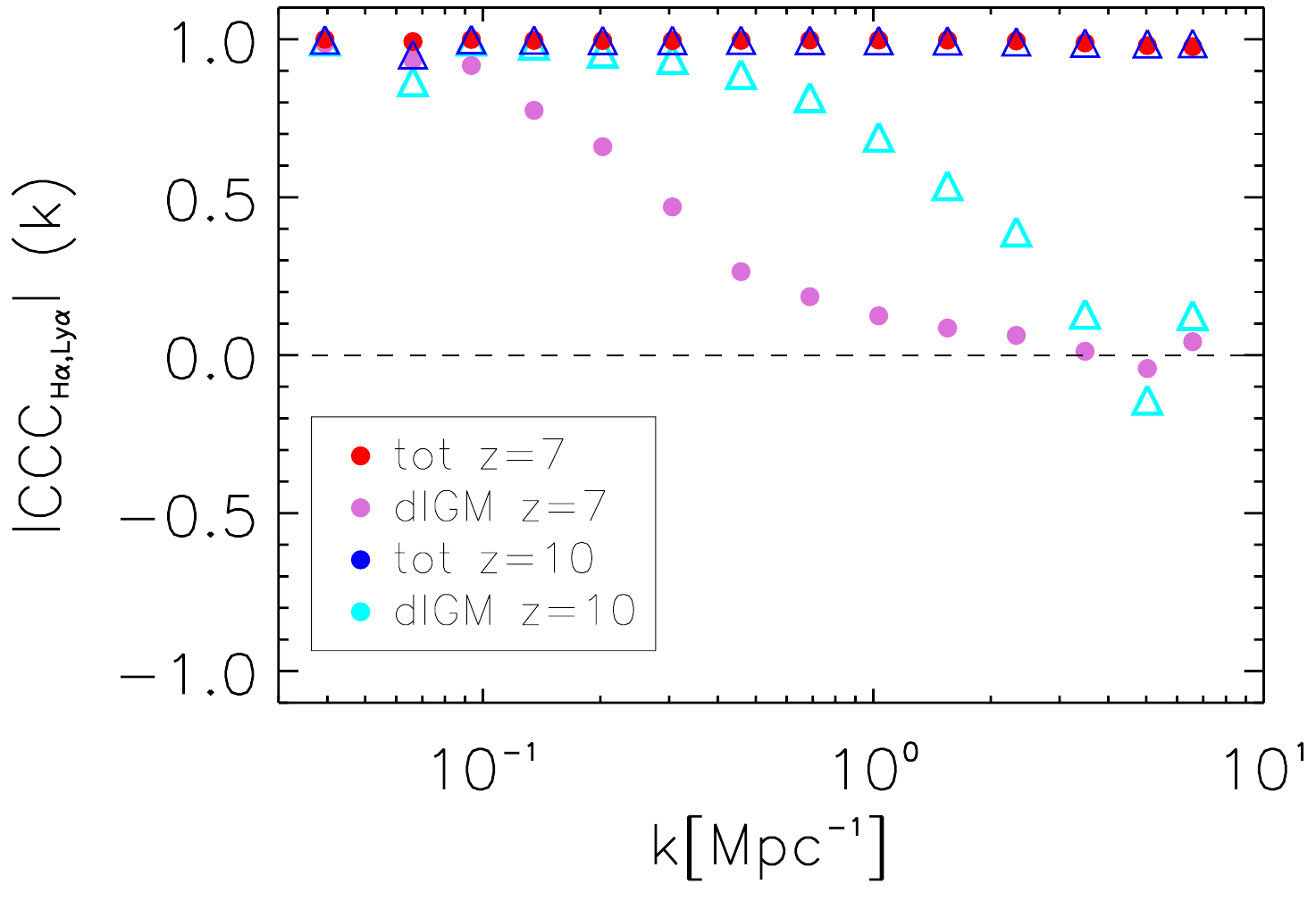}
\vfill
\vspace{0.2cm}
\includegraphics[width=0.9\columnwidth]{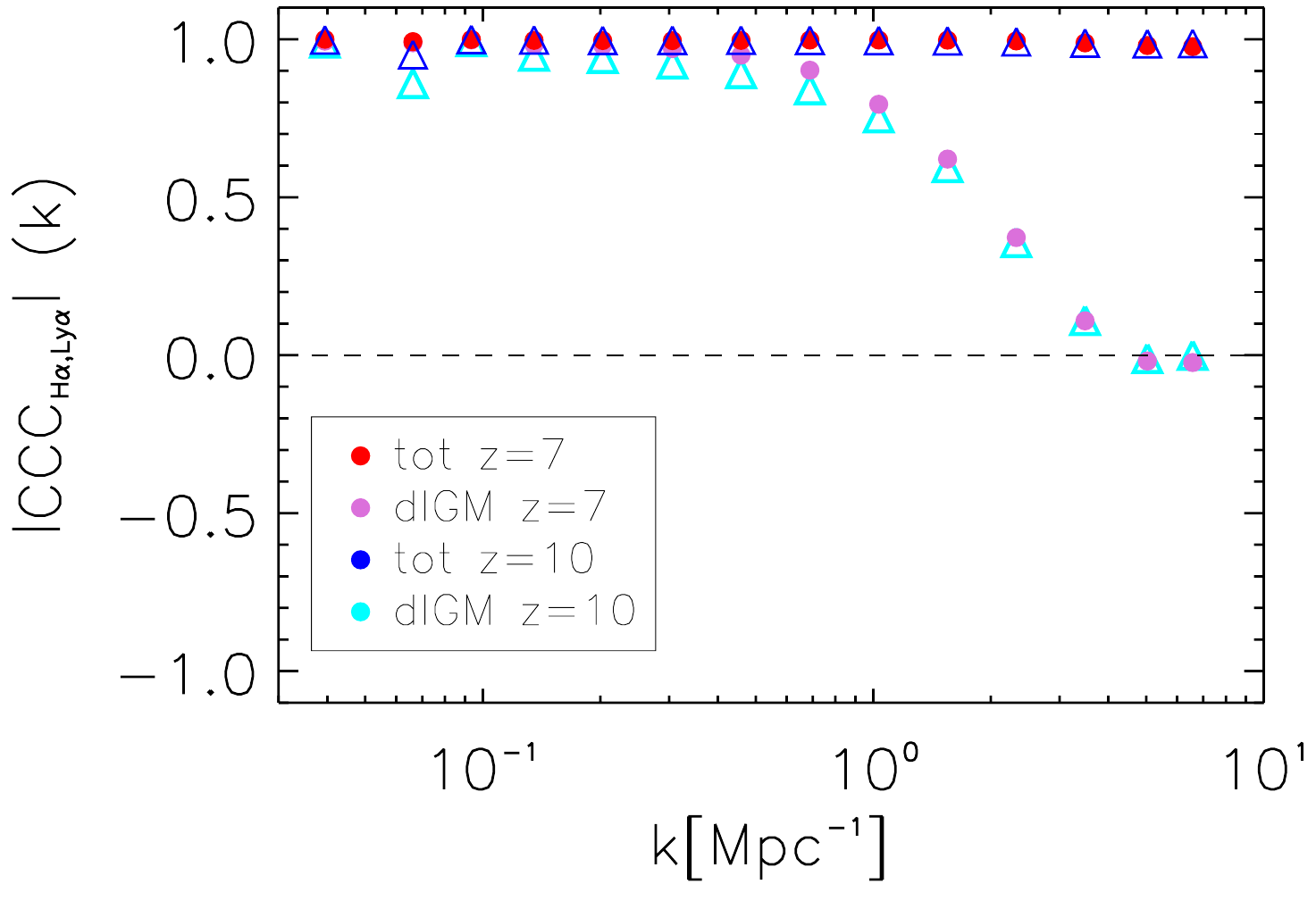}
\caption{H$\mathrm{\alpha}$ to Ly$\mathrm{\alpha}$ cross-correlation coefficient $\mathrm{CCC}_{\mathrm{H_{\alpha}},\mathrm{Ly\alpha}}$ of brightness fluctuations at redshift $z=10$ and $z=7$. Shown is the cross-correlation of the sum of galactic and diffuse IGM fluctuations in H$\alpha$ with total Ly$\mathrm{\alpha}$ fluctuations ``Ly$\mathrm{\alpha}$-tot'' and with the diffuse IGM contribution ``Ly$\mathrm{\alpha}$-dIGM'' (top), as well as the scattered IGM contribution ``Ly$\mathrm{\alpha}$-sIGM'' (bottom).} \label{fig:Ha-cross}
\end{figure}

 The resulting cross-correlation coefficient is shown in Figure~\ref{fig:Ha-cross}; it is defined as $\mathrm{CCC}_\mathrm{H\alpha,Ly\alpha}=\Delta_\mathrm{H\alpha,Ly\alpha}/\sqrt{\Delta_\mathrm{H\alpha}\Delta_\mathrm{Ly\alpha}}$ (see Equation~(\ref{eq:CCC})) and is equal to one if the two variables are perfectly correlated with each other. When cross-correlating H$\mathrm{\alpha}$ emission with total Ly$\mathrm{\alpha}$ emission, ``Ly$\mathrm{\alpha}$-tot'' in both panels of Figure~\ref{fig:Ha-cross}, the CCC is close to one both at both redshifts $z=10$ and $z=7$, with a slight decrease toward higher {\it k}. When cross-correlating H$\mathrm{\alpha}$ emission with the diffuse (top panel) and the scattered (bottom panel) IGM component of Ly$\mathrm{\alpha}$ emission, the CCC sharply decreases toward smaller scales (higher {\it k}). There even is a turnover from positive cross-correlation at lower $k$ to negative cross-correlation at high $k$, at both redshifts $z=10$ and $z=7$. The most prominent decrease of the CCC with $k$ is visible for the diffuse IGM at redshift $z=7$ (top panel, orchid dots). Interestingly, the redshift behavior of the CCC for diffuse IGM versus scattered IGM is different. 
 
 The different behaviors for components of Ly$\mathrm{\alpha}$ emission when cross-correlated with H$\mathrm{\alpha}$ emission mostly tracing galactic emission was shown in this section. This can be used to single out the IGM contribution to the total Ly$\mathrm{\alpha}$ emission and distinguish galactic and IGM components of Ly$\mathrm{\alpha}$ emission.

%%%%%%%%

\section{Signal-to-noise Ratio Calculation} \label{sec:StoN}
 Now that we have simulated 21cm and Ly$\mathrm{\alpha}$ emission in order to calculate their respective auto and cross-power spectra, as well as investigated parameter effects, we turn to estimating the detectability of these spectra by future probes of the EoR. We first discuss the 21cm and Ly$\mathrm{\alpha}$ noise auto spectra and then their noise cross-power spectra in the following sections.

\subsection{21 cm Noise Auto Spectrum and Foreground Wedge}\label{sec:21noise}
 In this section, we consider the noise power spectrum of 21cm emission, with our signal-to-noise ratio (S/N) calculation including cosmic variance and thermal and instrumental noise. We proceed to integrate the so-called 21cm foreground wedge in our S/N calculations.  Instrument specifications are taken to match the SKA stage 1~\citep{2015PritchardSKA} for line intensity mapping of the 21cm brightness temperature during the EoR. 
\newline \indent The variance for a (dimensional) 21cm power spectrum estimate for mode $k$ and angle $\mu$ between the line of sight and $k$~\citep{2006ApJ...653..815M,2008ApJ...680..962L}, when neglecting systematic effects such as imperfect foreground removal, reads as
\begin{equation}
\sigma^2_{21} \left( k,\mu\right) = \left[  P_{21}\left( k,\mu\right) + \frac{T_\mathrm{sys}^2 V_\mathrm{sur}}{B\,t_\mathrm{int}n\left( k_{\perp}\right)}W_{21}\left(k,\mu \right) \right] , \label{eq:sigma21}
\end{equation}
 where the first term is due to cosmic variance, the second term describes the thermal noise of the instrument, and the window function $W_{21} \left( k,\mu \right)$ includes the limited spectral and spatial instrumental resolution.
As we want to consider SKA stage 1, we take $B=8\,$MHz for the survey bandwidth, a total observing time time of $t_\mathrm{int}=1000$ hr, an instrument system temperature of $T_\mathrm{sys}=400\,$K,  and an effective survey volume of $V_\mathrm{sur}=\chi^2\Delta\chi\left( \lambda_{21}\left( z\right)^2/A_\mathrm{e}\right)^2$, with redshifted 21cm wavelength $\lambda_{21}\left(z \right)$, effective area per antenna $A_\mathrm{e}=925$m$^2$ $\left( z=8\right)$, and comoving distance and survey depth $\chi$ and $\Delta \chi$. The antenna distribution enters via the number density of baselines $n\left( k_{\perp}\right)=0.8$ that observe transverse wavenumber $k_{\perp}$, which we (simplistically) assume to be constant as in~\citet{Chang:2015era}. 
 The window function $W_{21}\left( k,\mu \right)$ reads, as in~\citet{2011ApJ...741...70L}, as
\begin{equation}
W_{21}\left( k,\mu \right) = e^{ \left( k_{\parallel}/k_{\parallel,\mathrm{res}}\right)^2 + \left( k_{\perp}/k_{\perp,\mathrm{res}}\right)^2}, \label{eq:window}
\end{equation}
with parallel modes $k_{\parallel} = \mu k$ along the line of sight and perpendicular modes $k_{\perp}=\left( 1- \mu^2\right)^{1/2} k$. The spectral  and spatial instrumental resolution in parallel and perpendicular modes is given by
\begin{equation}
k_{\parallel,\mathrm{res}} = \frac{R_\mathrm{res}H\left( z\right) }{c\left(1+z\right)} = \frac{1}{\Delta x_\mathrm{\parallel,res}}
 \label{eq:korres}
\end{equation} 
and  
\begin{equation}
k_{\perp, \mathrm{res}} = \frac{1}{\chi\left( z\right)\theta_\mathrm{min}} = \frac{1}{\Delta x_\mathrm{\perp, res}} \,, \label{eq:kparres}
\end{equation}
with comoving resolution elements $\Delta x_\mathrm{\parallel, res}$ and $\Delta x_\mathrm{\perp, res}$, comoving distance $\chi\left( z\right)$, and angular beam (or spatial pixel) size in radians $\theta_\mathrm{min}=\left(x_\mathrm{pix}/60\right) \left(\pi/180\right)$. 
The instrumental resolution for a  radio telescope is determined by $R_\mathrm{res} = \nu_{21}\left(z \right)/ \nu_\mathrm{res}$, with frequency resolution $\nu_\mathrm{res}=3.9\times10^3\,$kHz for a SKA stage 1 type survey, and angular resolution $x_\mathrm{pix} =\left( \lambda_{21}\left( z \right)/l_\mathrm{max}\right)\left(\pi/180\right)/60$, with maximum baseline $l_\mathrm{max}=10^5$ cm.
For example, at redshift $z=7$, we have $k_{\parallel, \mathrm{res}}\left( z=7\right) \approx 16$ Mpc$^{-1}$ and $k_{\perp, \mathrm{res}} \left( z=7\right) \approx 242$ Mpc$^{-1}$.
The total variance $\sigma^2\left(k \right)$ for the full spherically averaged power spectrum is the binned sum over all angles $\mu$, or equivalently all modes $k^2 = k^2_{\parallel}+k^2_{\perp}$, divided by the respective number of modes per bin; it is given by 
\begin{equation}
\frac{1}{\sigma^2\left(k \right)} = \sum_{\mu} \frac{N_\mathrm{m}}{\sigma^2\left(k,\mu \right)} ,  \label{eq:sigma-tot}
\end{equation}
with number of modes $N_\mathrm{m}=\Delta k \Delta \mu k^2 V_\mathrm{sur}/\left(4\pi^2\right)$ for binning logarithmically in $k$, survey volume $V_\mathrm{sur}$, and mode as well as angle bin sizes $\Delta k$ and $\Delta \mu$. In our S/N calculation, we explicitly counted the number of modes $N_\mathrm{m}$ in each bin.
The sum over angles $\mu$ is restricted by minimal and maximal allowed values $\mu_\mathrm{min}^2 = \max \left( 0,1-k_{\perp,\mathrm{max}}^2/k^2\right)$ and $\mu_\mathrm{max} = \min\left( 1,k/k_{\parallel,\mathrm{min}}\right)$~\citep{2006ApJ...653..815M} that are determined by minimum mode $k_{\parallel,\mathrm{min}}=2\pi/r_\mathrm{pix}$ due to survey depth and maximum mode $k_{\perp,\mathrm{max}}=k_{\perp,\mathrm{res}}$ spatially resolvable by the survey.

\begin{figure*}
\plottwo{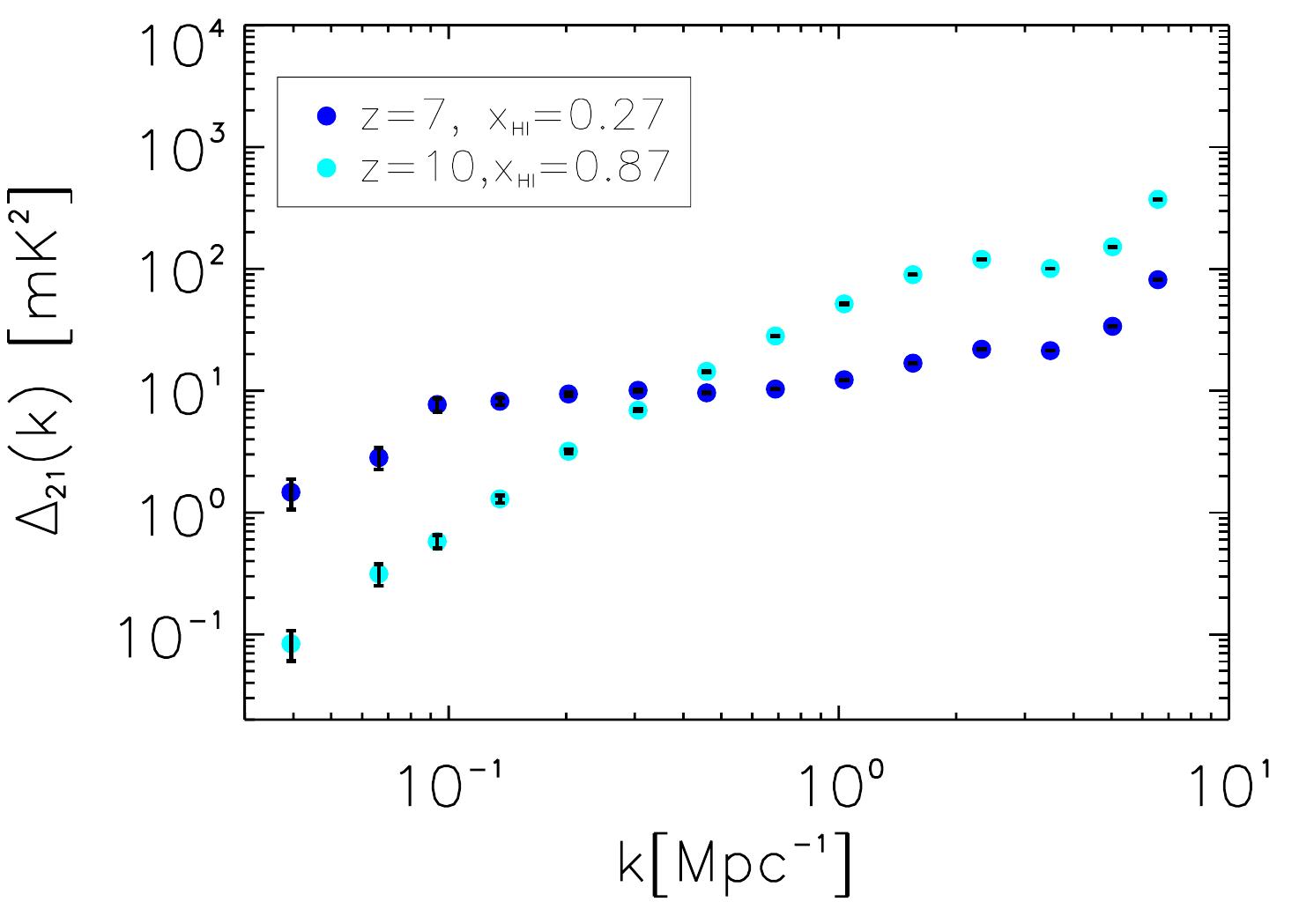}{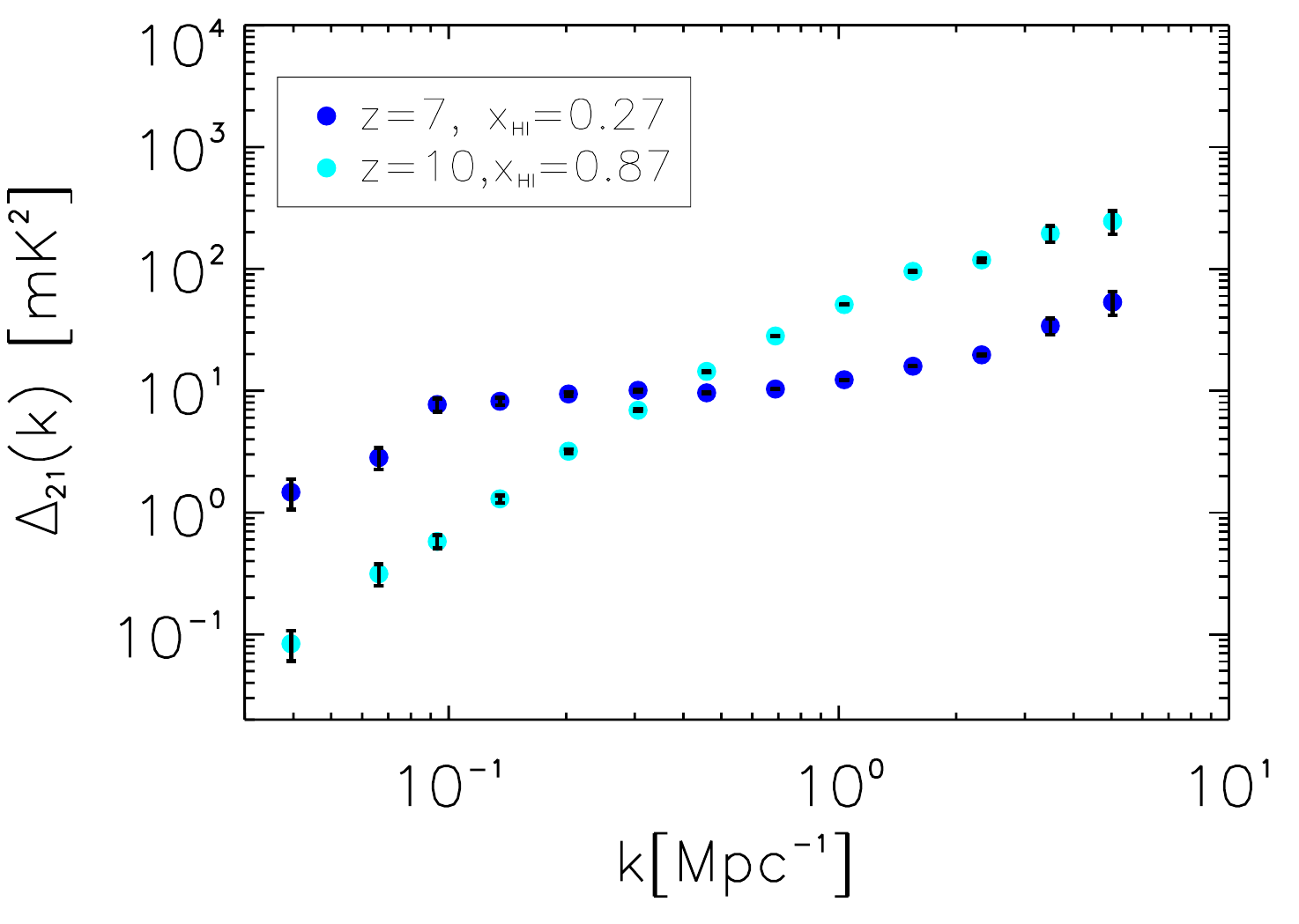}
\caption{Left: 21cm noise power spectrum (spherically averaged), including cosmic variance and thermal and instrumental noise for an SKA stage 1 type survey. Right: 21cm noise power spectrum after removal of the foreground wedge defined in Equation~(\ref{eq:wedge}), for survey characteristic angle $\theta_0=15^{\circ}$, where scales roughly left of the vertical red dashed line might be lost to removal of smooth foregrounds, again including cosmic variance and thermal and instrumental noise; see Table~\ref{tab:exp} for instrument specifications; redshift $z=7$ and mean neutral fraction $\bar{x}_\mathrm{HI}=0.27$ in blue, $z=10$ and $\bar{x}_\mathrm{HI}=0.87$ in cyan.} \label{fig:21noise}
\end{figure*}

Besides thermal and instrumental noise and cosmic variance, we want to incorporate the so-called 21cm foreground wedge in our S/N calculation, in order to restrict ourselves to an EoR window where foreground model errors do not contaminate the signal.
This 21cm foreground wedge stems from a combination of foregrounds and instrument systematics due to leakage in the 21cm radio window. 
By subtraction of the foreground wedge, we mask, that is, avoid, a significant amount of foreground. The wedge is defined for the cylindrically averaged 2D power spectrum via a relation between mode $k_{\perp}$ perpendicular and mode $k_{\parallel}$ parallel to the line of sight. This relation reads~\citep{MoralesWedge12,LiuWedge14} as
\begin{equation}
k_{\parallel} \leq \frac{\chi\left( z\right) E\left(z\right) \theta_0}{d_\mathrm{H}\left( 1+z \right)} k_{\perp} \,, \label{eq:wedge}
\end{equation}
with characteristic angle $\theta_0$, comoving distance $\chi\left( z\right)$, Hubble distance $d_\mathrm{H}$, and Hubble function $E\left( z\right)=H\left( z\right)/H_0$, which determine the slope of the wedge. 
The most pessimistic assumption for the characteristic angle $\theta_0$ would be to include contamination from sources on the horizon, i.e., $\theta_0 = \pi/2$. But contaminations from residual sources are band limited by the instrument field of view, so that it is possible to avoid contamination from sources outside the primary beam, which would make the EoR window significantly larger~\citep{Pober:2013jna,Jensen:2015xua} and $\theta_0$ significantly smaller, of the order of $10$ degrees. In addition, modes with low $k_\parallel$, below roughly $k_\mathrm{\parallel, min}\sim 0.05$\,Mpc$^{-1}h$~\citep{Dillon:2013rfa,Dillon:2014jla}, are affected by spectrally smooth foregrounds. We include this region in our power spectrum calculation and indicate it by a vertical red dashed line, for example in Figure~\ref{fig:21noise}. We keep this region as, when removed, only the points within that region are lowered significantly (one point in our case), as working within the wedge might be possible for the cross-power spectrum that is less sensitive to foreground contamination, and as the exact horizontal cutoff with its model and redshift dependency is unknown.

\begin{figure*}
\plottwo{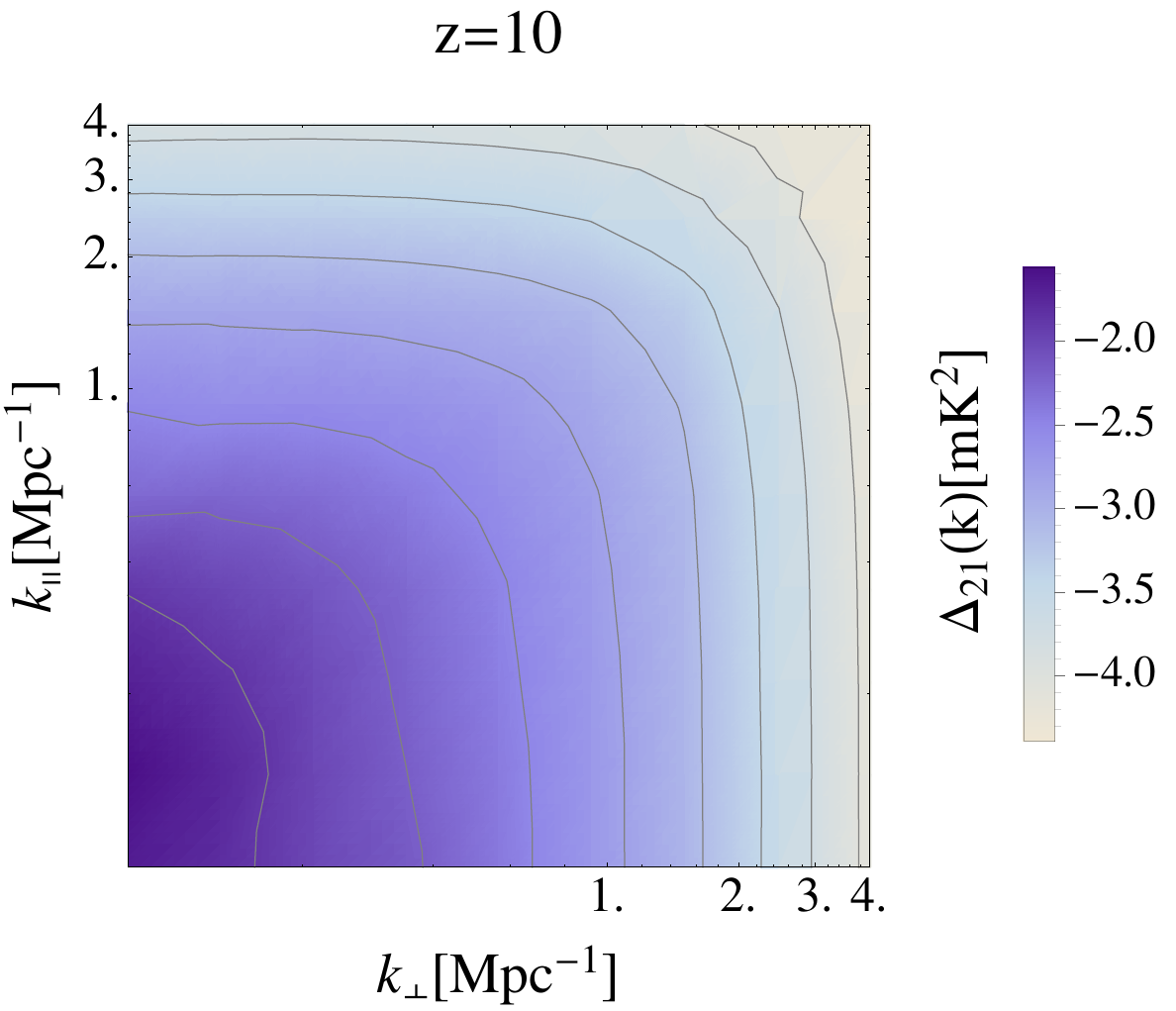}{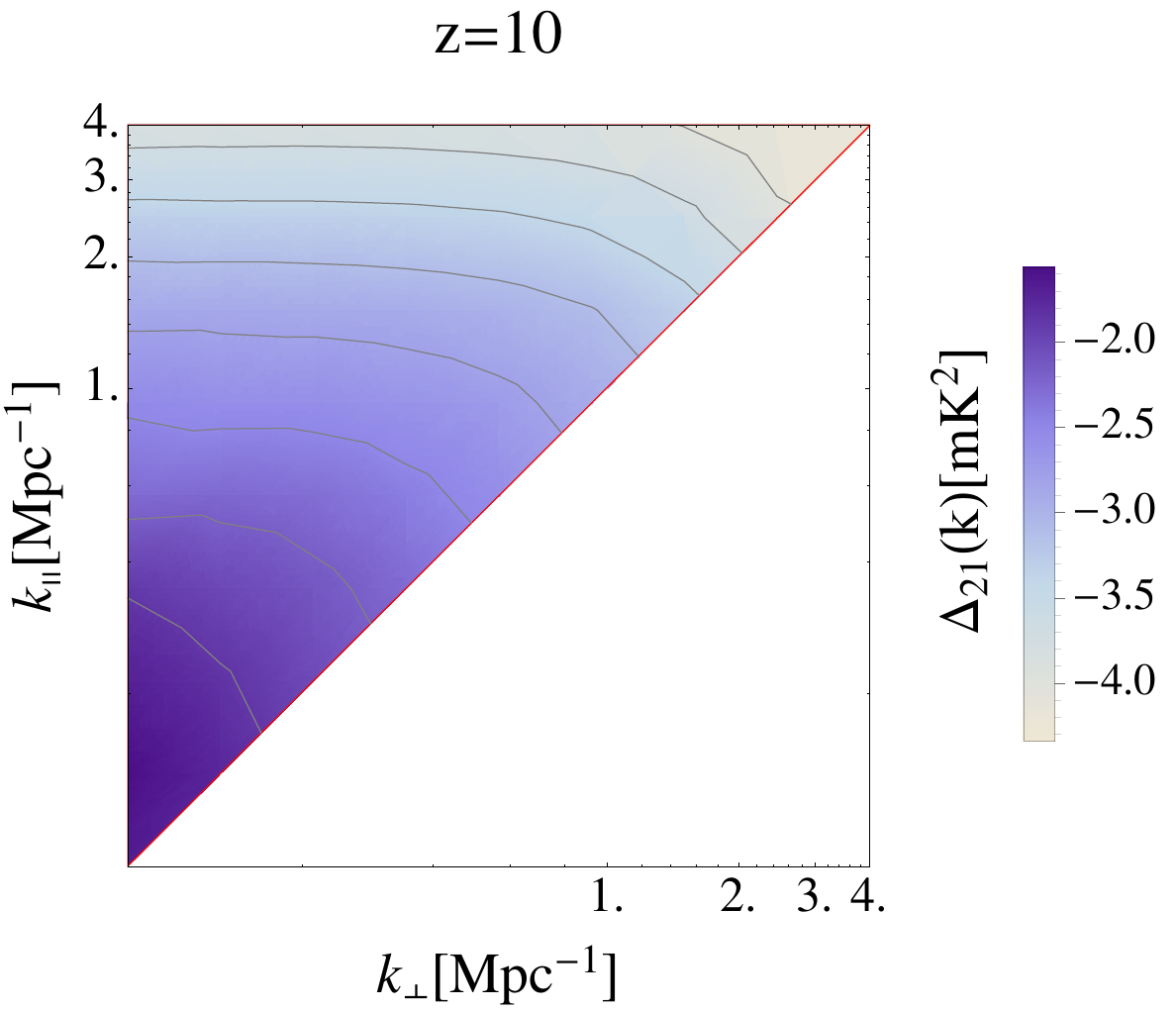}
\vfill \vspace{0.3cm}
\plottwo{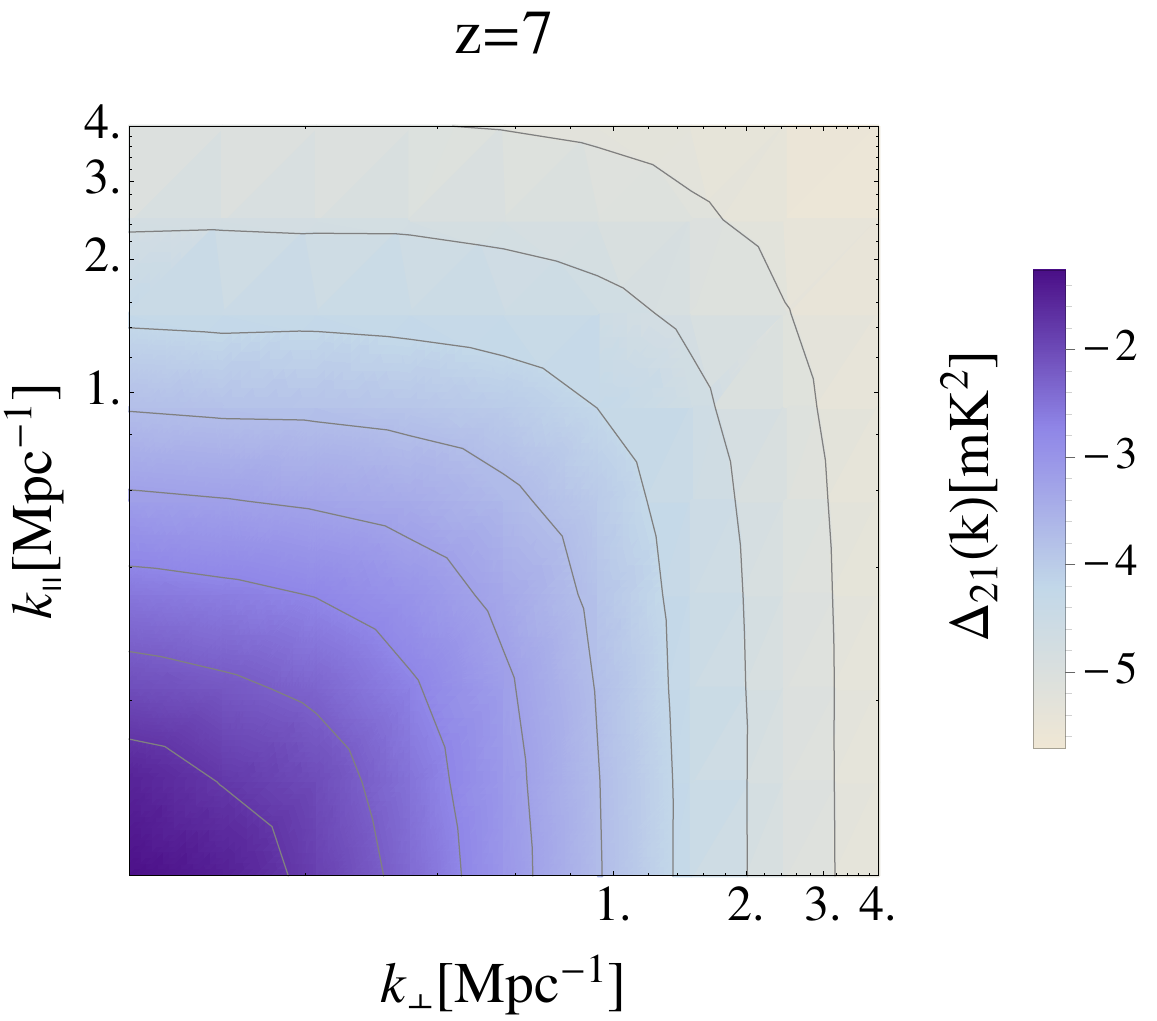}{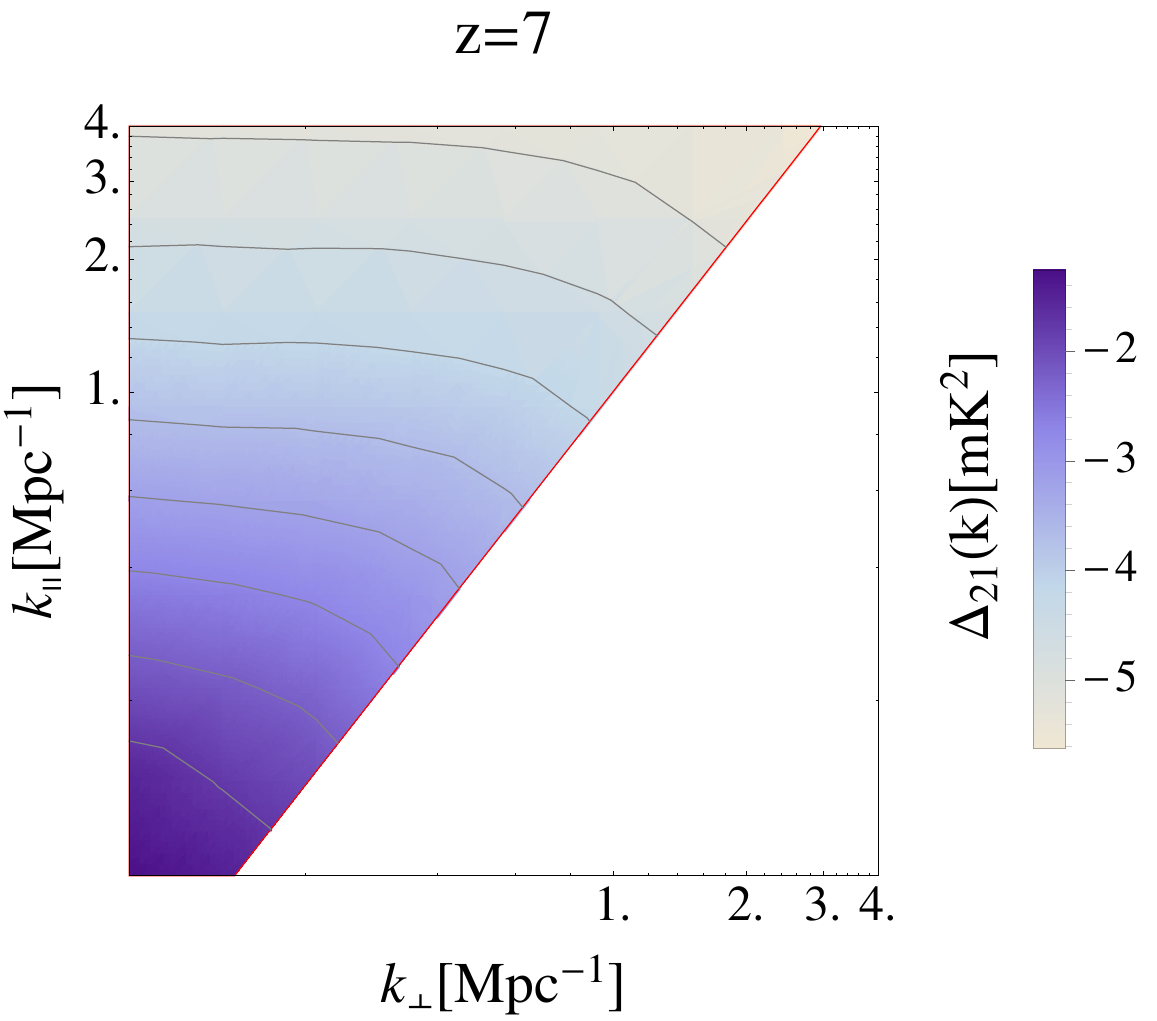}
\caption{Cylindrically averaged 21cm power spectra at $z=10$, $\bar{x}_\mathrm{HI}=0.87$ (top) and $z=7$, $\bar{x}_\mathrm{HI}=0.27$ (bottom), displayed for $k_\mathrm{\parallel}, k_\mathrm{\perp}>0.08$ Mpc$^{-1}$. Left: no foreground removal, full power spectra extracted from the simulation boxes with 200 Mpc box length as shown in Figure~\ref{fig:21cm} (middle). Right:  cylindrically averaged 21cm power spectra where the foreground wedge defined in Equation~(\ref{eq:wedge}) for survey characteristic angle $\theta_0 \approx15^{\circ}$ is removed.} \label{fig:Pk2D}
\end{figure*}

Figure~\ref{fig:Pk2D} shows the cylindrically averaged 21cm power spectrum both with and without foreground wedge subtraction for a survey with characteristic angle $\theta_0 \approx 15^{\circ}$ for redshift $z=10$ (top panels) and $z=7$ (bottom panels). The same characteristic angle was used for the 21cm spherically averaged noise power spectrum with foreground avoidance shown in Figure~\ref{fig:21noise} (right panel). The subtraction of the foreground wedge leads to loss in power and S/N for larger $k$ modes as compared to the 21cm noise power spectrum without the wedge removed (left panel); in both panels, error bars account for cosmic noise, thermal noise, and instrumental resolution. Encouragingly, the loss in power for the spherically averaged power spectrum is restricted to higher $k$ modes, and a reconstruction of the full power spectrum from data might be possible. As we can see here, the detection of the power spectrum of 21cm fluctuations over around two decades in spatial scale is feasible with future 21cm experiments, making the detection range of the Ly$\mathrm{\alpha}$ power spectrum the limiting factor for the cross-correlation of 21cm and Ly$\mathrm{\alpha}$ fluctuations.

\begin{table*}
\centering
\caption{Instrument Specifications for 21cm Survey: SKA Stage 1}
\centering
\begin{tabular}{P{2.0cm} P{2.0cm} P{2.0cm} P{2.0cm} P{2.5cm} P{2.5cm} P{2.0cm}}
\hline \hline
 $\nu_\mathrm{res}$  &  $l_\mathrm{max}$ & $T_\mathrm{sys}$ & $t_\mathrm{int}$ & B (z=8) & $A_\mathrm{e}$ (z=8) & $n_{\perp}$  \\
   (kHz) & (cm) & (K) & (hr) & (MHz) & (m$^2$) & \\
\hline
 3.9 &  $10^5$  & 400 & 1000 & 8  & 925 & 0.8 \\ 
 \hline
\multicolumn{7}{c}{Notes. See Section \ref{sec:21noise} for details on error calculations; specifications taken from~\citet{2015PritchardSKA,Chang:2015era}.}
\end{tabular}
\label{tab:exp}
\end{table*}

\begin{table*}
\centering
\caption{Instrument Specifications for Ly$\mathrm{\alpha}$ Experiments}
\begin{center}
\begin{tabular}{ P{4cm} P{2.5cm}P{3cm}P{4cm} P{2.0cm}}
 \hline \hline
 Experiment & $x_\mathrm{pix}$ &  $R_\mathrm{res}$  & $\sigma_\mathrm{N}$  & $V_\mathrm{vox}$ at $z=7$  \\
  & ('') & & (erg$\,$s$^{-1}$cm$^{-2}$Hz$^{-1}$sr$^{-1}$) & (Mpc$^3$)  \\
\hline
SPHEREx & 6.2  &  41.5  & $3\times10^{-20}$ & 0.3 \\ 
 CDIM & 1  &  300  & $1.5\times10^{-21}$ & $1.3\times 10^{-3}$ \\ 
 \hline
\multicolumn{5}{c}{Notes. See Section \ref{sec:LyaNoise} for details on error calculations; specifications taken from~\citet{2014spherex} and~\citet{CDIM}.}
\end{tabular}
\end{center}
\label{tab:expL}
\end{table*}

\begin{figure*}
\plottwo{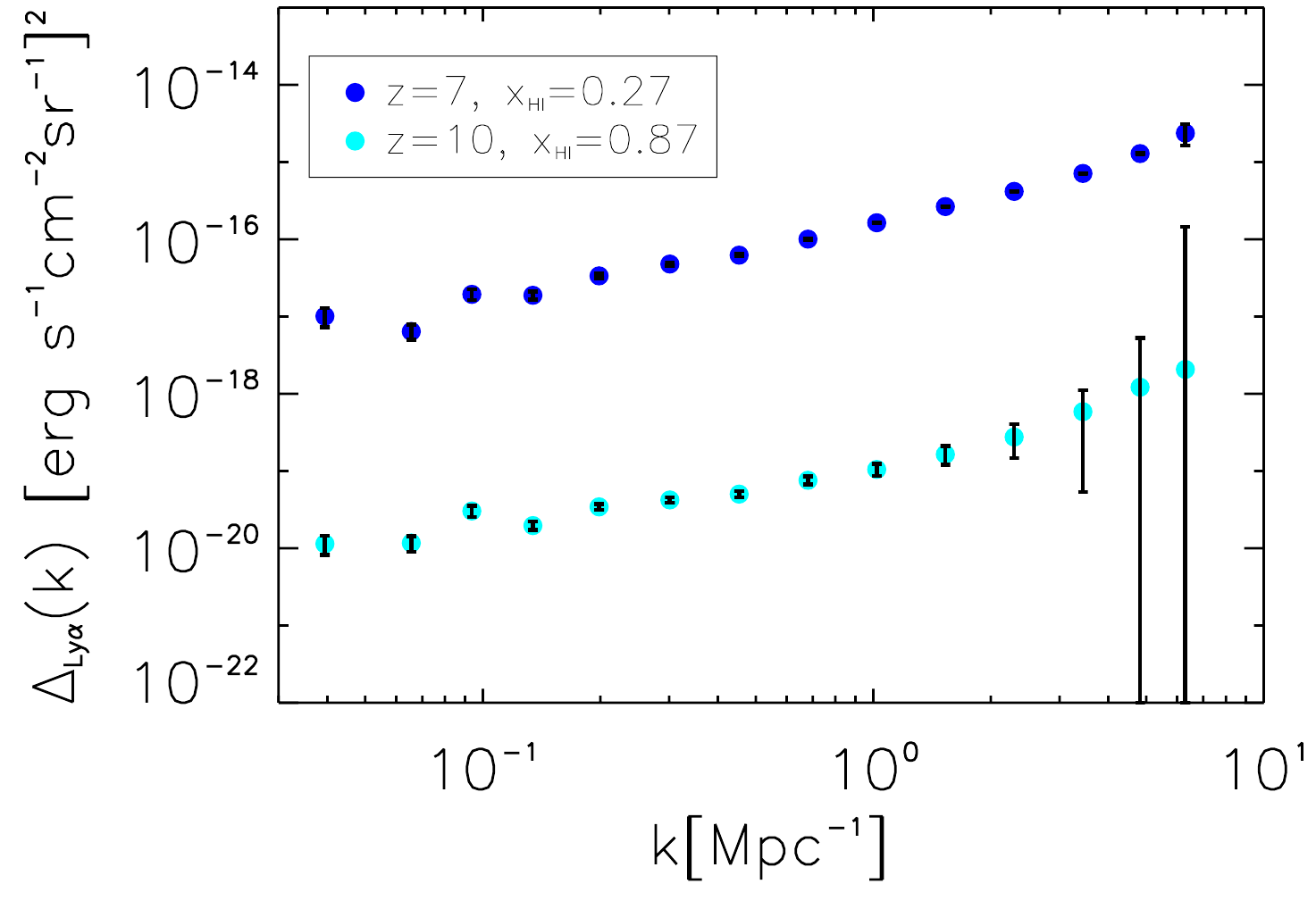}{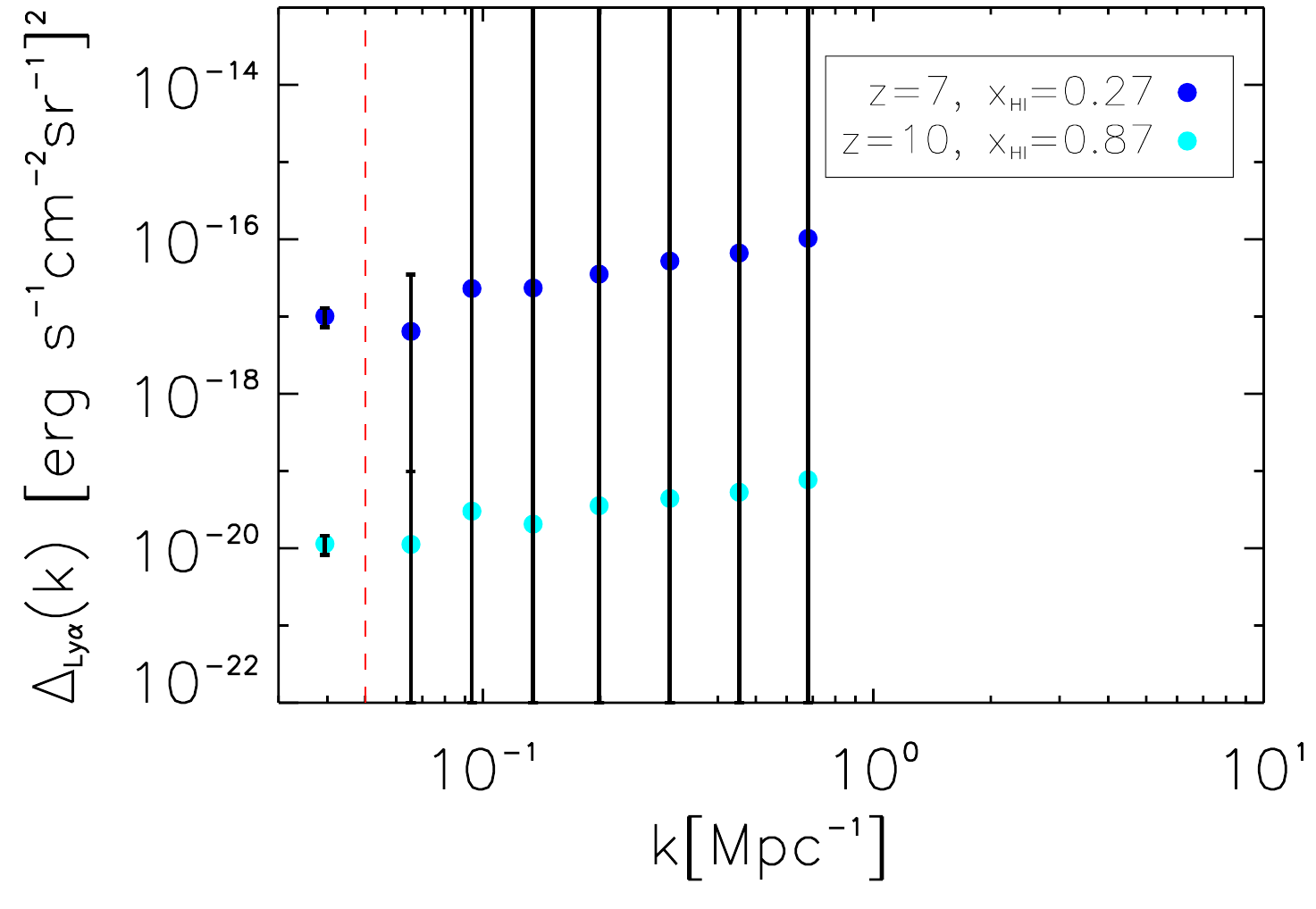}
\plottwo{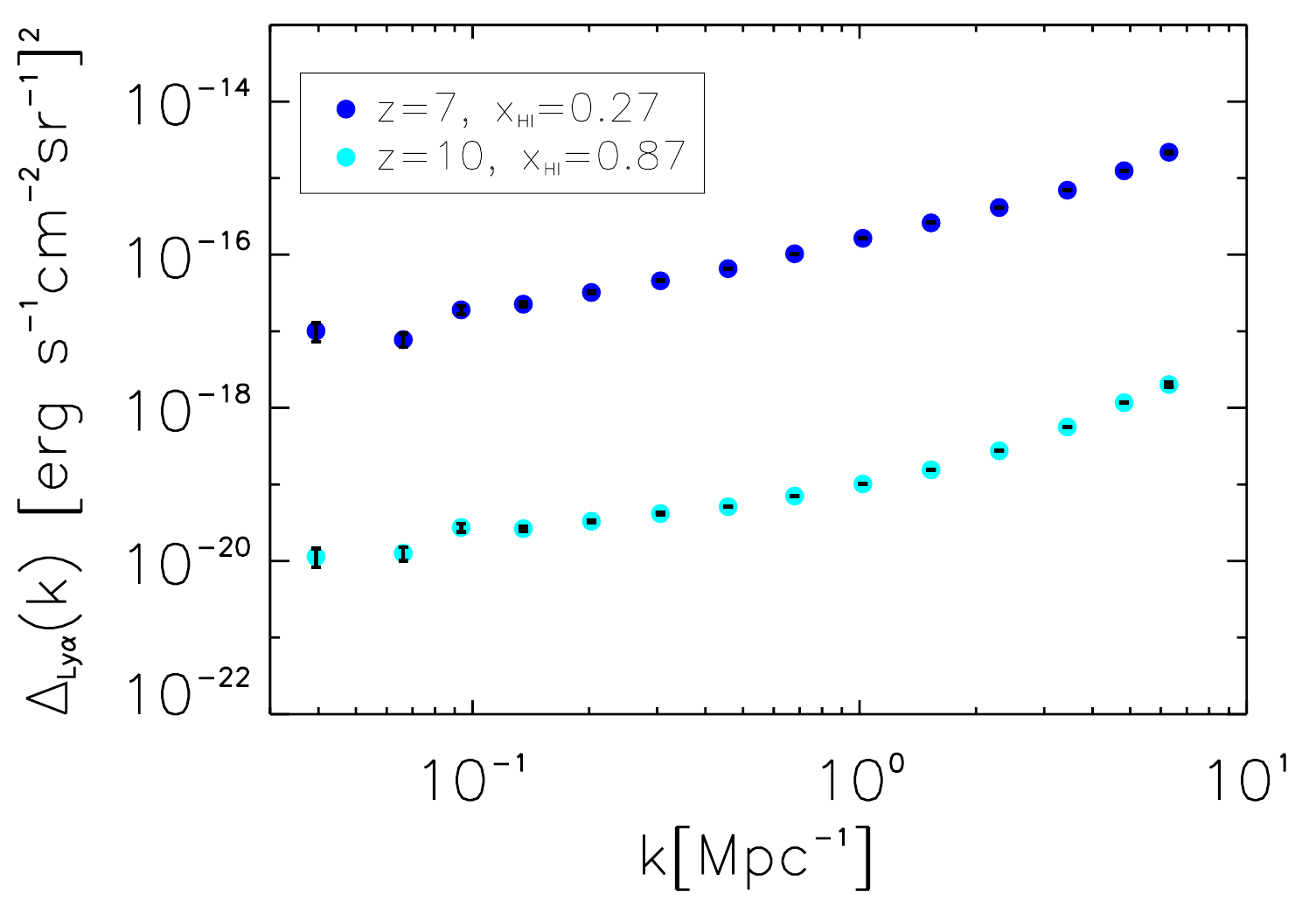}{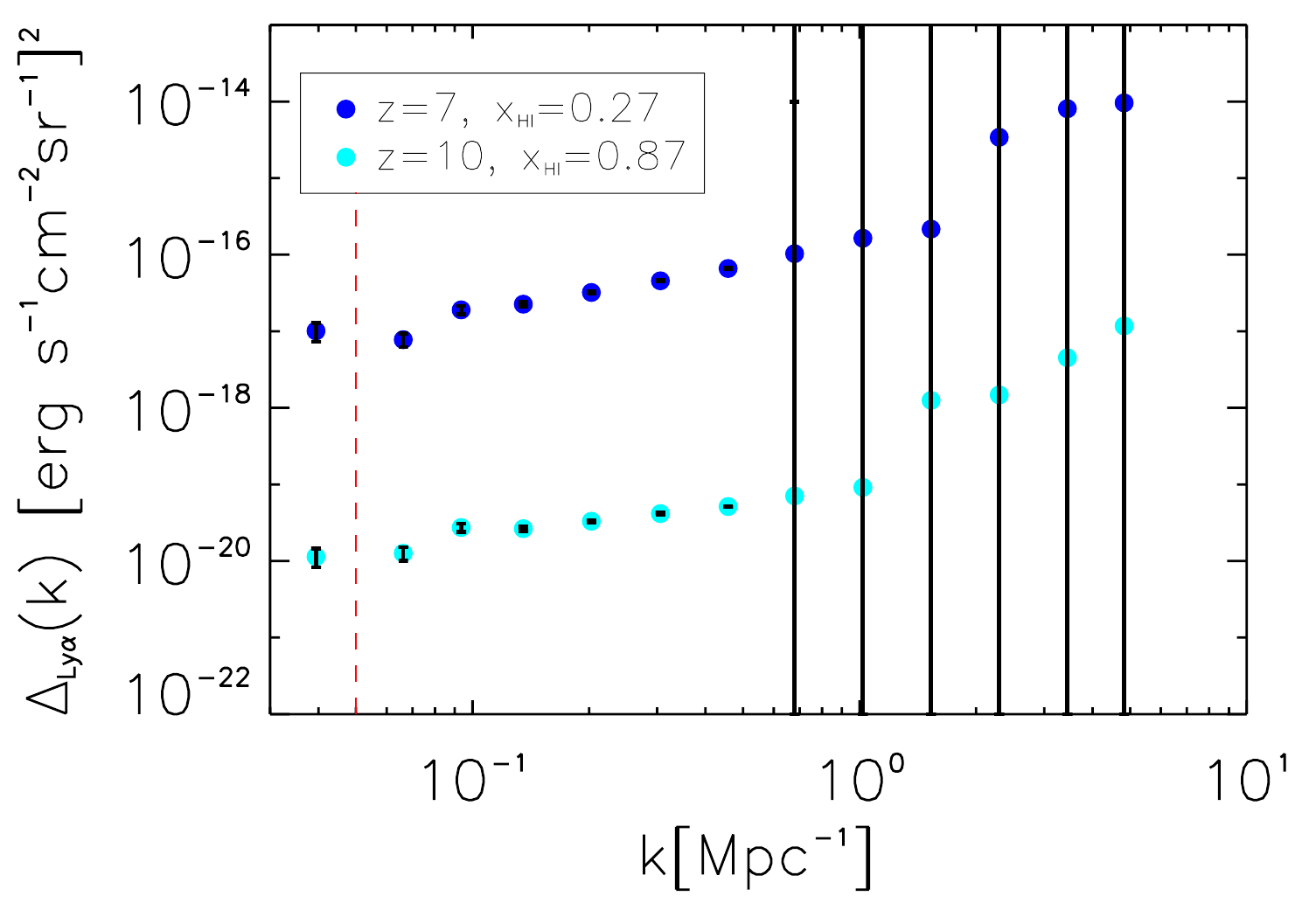}
\caption{Left: Ly$\mathrm{\alpha}$ noise power spectrum for a SPHEREx (top panel) and a CDIM (bottom panel) type of survey, including cosmic variance and thermal and instrumental noise with a $k_{\parallel}>0.5\,$ Mpc$^{-1}$ cut for CDIM and a $k_{\parallel}>0.06\,$ Mpc$^{-1}$ cut for SPHEREx (for the choice of this cut, see discussion in Section~\ref{sec:LyaNoise} and Appendix~\ref{app:SN}). Right: Ly$\mathrm{\alpha}$ noise power spectrum after removal of the foreground wedge defined in Equation~(\ref{eq:wedge}) for survey characteristic angle $\theta_0\approx15^{\circ}$, where scales roughly left of the vertical red dashed line might be lost to removal of smooth foregrounds, again including cosmic variance and thermal and instrumental noise with a $k_{\parallel}>0.06\,$ Mpc$^{-1}$ cut and a $k_{\parallel}>0.5\,$ Mpc$^{-1}$ cut, for SPHEREx (top panel) and CDIM (bottom panel), respectively, type of surveys (see Table~\ref{tab:expL} for instrument specifications) with redshift $z=7$ and neutral fraction $\bar{x}_\mathrm{HI}=0.27$ in blue, $z=10$ and $\bar{x}_\mathrm{HI}=0.87$ in cyan; all power spectra include Ly$\alpha$ damping for tracing through the simulation along the z-axis LOS.} \label{fig:lyasim-noise}
\end{figure*}

\subsection{Ly$\mathrm{\alpha}$ Noise Auto Spectrum} \label{sec:LyaNoise}
Here we consider the noise power spectrum of total Ly$\mathrm{\alpha}$ emission, composed of galactic, diffuse, and scattered IGM contributions. In the S/N calculation, we include cosmic variance, as well as thermal and instrumental noise, while also taking  Ly$\mathrm{\alpha}$ damping into account (see Section~\ref{sec:damp}). In the following, we use instrument specifications of the proposed all-sky near-infrared survey satellites SPHEREx~\citep{2014spherex} and the CDIM~\citep{CDIM} for line intensity mapping at high redshifts, as summarized in Table~\ref{tab:expL}. 
For the thermal noise variance of SPHEREx, we take $\sigma_\mathrm{N} \approx 3$ kJy$\,$sr$^{-1}$, corresponding to $\sigma_\mathrm{N} \approx 3\times 10^{-20}$ erg$\,$s$^{-1}$cm$^{-2}$Hz$^{-1}$sr$^{-1}$, which is consistent with sensitivity at 5$\sigma$ given in~\citet{Dore:2016tfs} of 18--19 in AB magnitude for relevant bands.\footnote{Magnitude to flux density converter: \newline http://ssc.spitzer.caltech.edu/warmmission/propkit/pet/magtojy/} For CDIM, we have for the thermal noise variance $\sigma_\mathrm{N} \approx 0.15$ kJy$\,$sr$^{-1}$, corresponding to $\sigma_\mathrm{N} \approx 1.5\times 10^{-21}$ erg$\,$s$^{-1}$cm$^{-2}$Hz$^{-1}$sr$^{-1}$.
%flux to flux/sr: 4*10^(-28)*4*pi/5 (5 because of 5 sigma)

Assuming a pure white-noise spectrum, the thermal noise power spectrum reads as
\begin{equation}
P_\mathrm{N,Ly\mathrm{\alpha}} = \sigma_\mathrm{N}^2 V_\mathrm{vox} \,.
\end{equation}
The comoving pixel volume corresponds to $V_\mathrm{vox}=A_\mathrm{pix}\, r_\mathrm{pix} \approx 0.3$ Mpc$^3$ for SPHEREx and $V_{vox}\approx 1.3\times10^{-3}$ for CDIM, both at $z=7$, the product of the pixel area $A_\mathrm{pix}=x_\mathrm{pix} \times x_\mathrm{pix}$ in comoving Mpc and comoving pixel depth $r_\mathrm{pix}=\chi\left( R_\mathrm{res}\right)$, which corresponds to the comoving length at frequency resolution $R_\mathrm{res}$.  
The frequency resolution is $R_\mathrm{res}=41.5$ for SPHEREx and $R_\mathrm{res}=300$ for CDIM, in the frequency range of interest for Ly$\mathrm{\alpha}$ emission during reionization. 
The variance, as a function of $k$ mode and angle $\mu$ between the line of sight and mode $k$, reads as
\begin{equation}
\sigma^2_\mathrm{Ly\mathrm{\alpha}} \left( k,\mu\right) = \left[  P_\mathrm{Ly\mathrm{\alpha}}\left( k,\mu\right) + \sigma_\mathrm{N}^2\,V_\mathrm{vox}\,W_\mathrm{Ly\mathrm{\alpha}}\left(k,\mu \right) \right] \,. \label{eq:sigmaLya}
\end{equation}
 The first term is due to cosmic variance, $\sigma_\mathrm{N}$ includes thermal noise, and the window function $W_\mathrm{Ly\mathrm{\alpha}} \left( k,\mu \right)$ accounts for limited spatial and spectral instrumental resolution and is defined analogous to Equation~(\ref{eq:window}). For example, at redshift $z=7$, Equations~(\ref{eq:korres}) and (\ref{eq:kparres}) give an angular resolution of $k_{\parallel, \mathrm{res}}\left( z=7\right) \approx 0.02\,$ Mpc$^{-1}$ and a spectral resolution of  $k_{\perp, \mathrm{res}} \left( z=7\right) \approx 3.8\,$ Mpc$^{-1}$ for the characteristics of the SPHEREx satellite, as well as $k_{\parallel, \mathrm{res}}\left( z=7\right) \approx 0.1\,$ Mpc$^{-1}$ and $k_{\perp, \mathrm{res}} \left( z=7\right) \approx 23.4\,$ Mpc$^{-1}$ for a CDIM-like experiment. The total variance $\sigma^2_\mathrm{Ly\mathrm{\alpha}} \left(k \right)$ for the full spherically averaged power spectrum again is the sum over the upper-half plane of angles $\mu$, or equivalently $k$ modes with $k^2 = k^2_{\parallel}+k^2_{\perp}$, divided by the respective number of modes per bin as defined in Equation~(\ref{eq:sigma-tot}). We explicitly counted the number of modes $N_\mathrm{m}$ in each bin.

Figure~\ref{fig:lyasim-noise} shows the noise power spectrum of Ly$\mathrm{\alpha}$ fluctuations at $z=10$ and $z=7$. The error bars account for cosmic noise and thermal noise, as well as instrumental noise. A cut in parallel modes of $k_{\parallel}>0.06\,$ Mpc$^{-1}$ for SPHEREx (top panels) and $k_{\parallel}>0.5\,$ Mpc$^{-1}$ for CDIM (bottom panels) was applied, as the instrumental noise in parallel modes, that is, the limitation due to spectral resolution, dominates over the signal at higher modes. As shown in Appendix~\ref{app:SN}, this cut roughly corresponds to the $k$ mode where the S/N drops below one. Of course, this presents a trade-off between a loss of power and a gain of precision.
A high-significance Ly$\mathrm{\alpha}$ power spectrum measurement is possible for CDIM across more than a decade in spatial scale, even when the 21cm foreground wedge is removed (right panels), which is encouraging for cross-correlation studies with 21cm emission. Also for SPHEREx, a detection of the Ly$\mathrm{\alpha}$ power spectrum can be achieved around a scale of $k\approx 0.06\,$Mpc$^{-1}$. In the following, we will employ for the cross-correlation signal the Ly$\mathrm{\alpha}$ power spectrum measurements with CDIM specifications.

\newcommand*{\figuretitlel}[1]{%
    {%   <--------  will only affect the title because of the grouping (by the
    \textbf{#1}%              braces before \centering and behind \medskip). If you remove
    \par\medskip}%            these braces the whole body of a {figure} env will be centered.
}

\begin{figure*}[!htb]
\plottwo{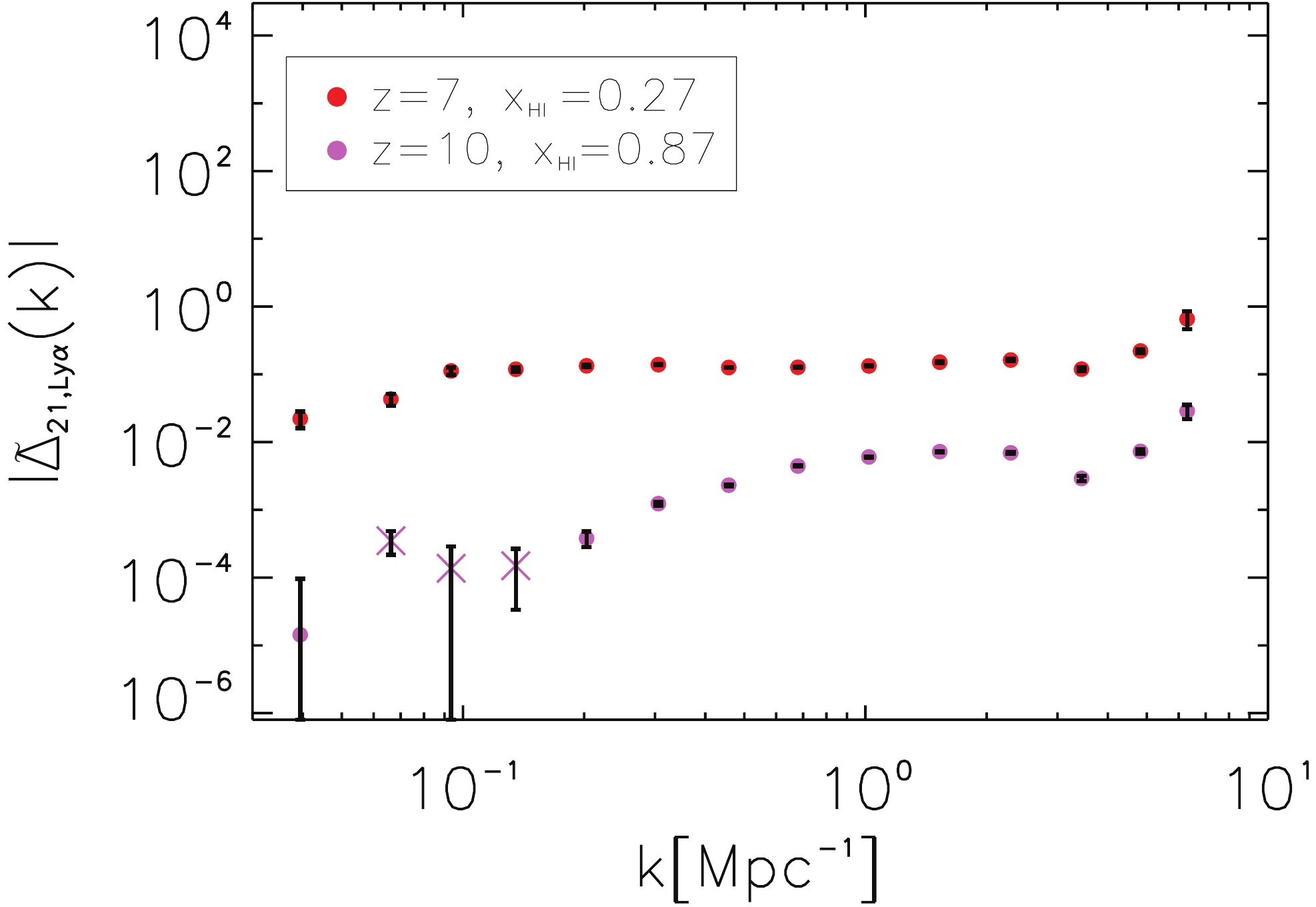}{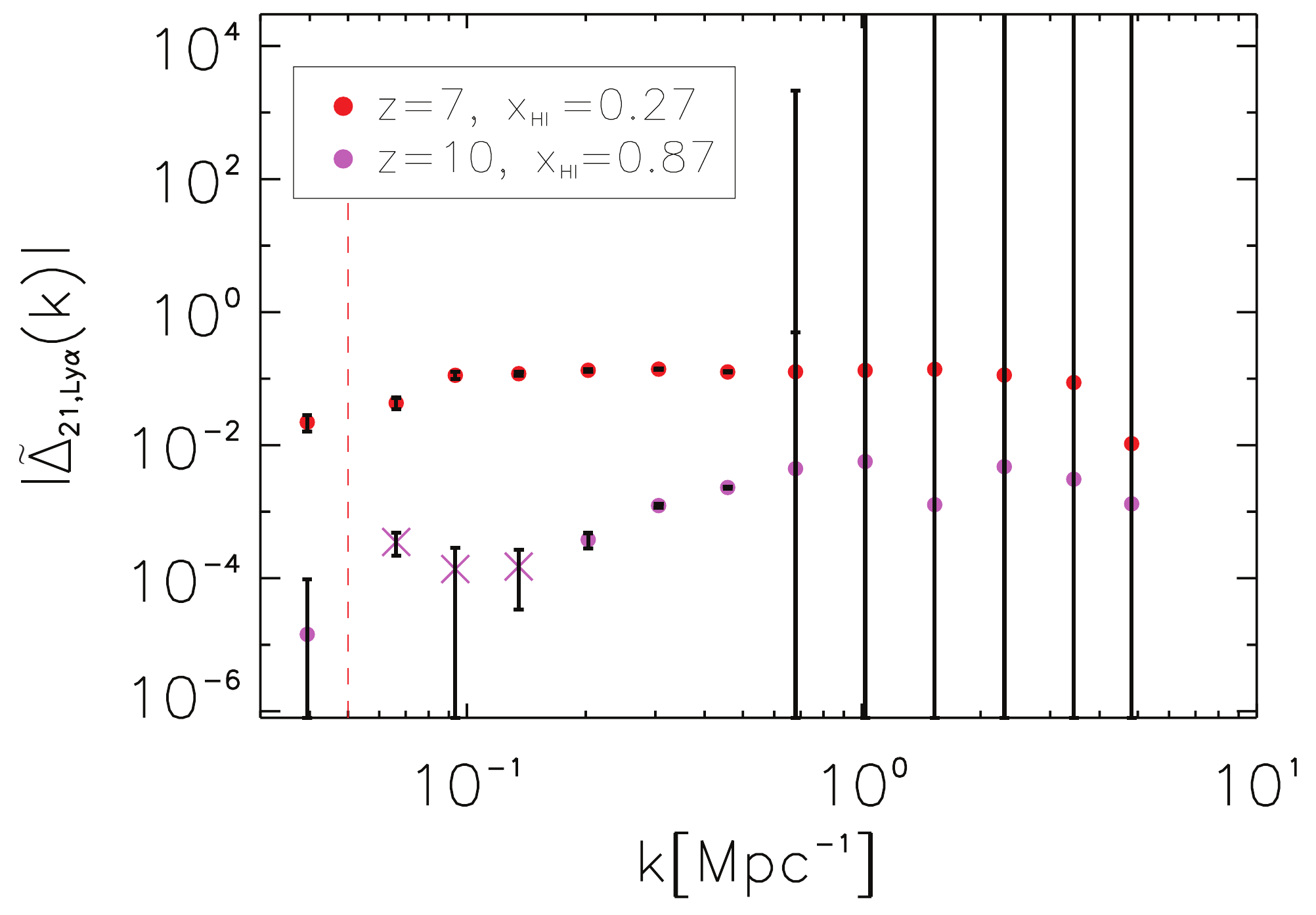}
\plottwo{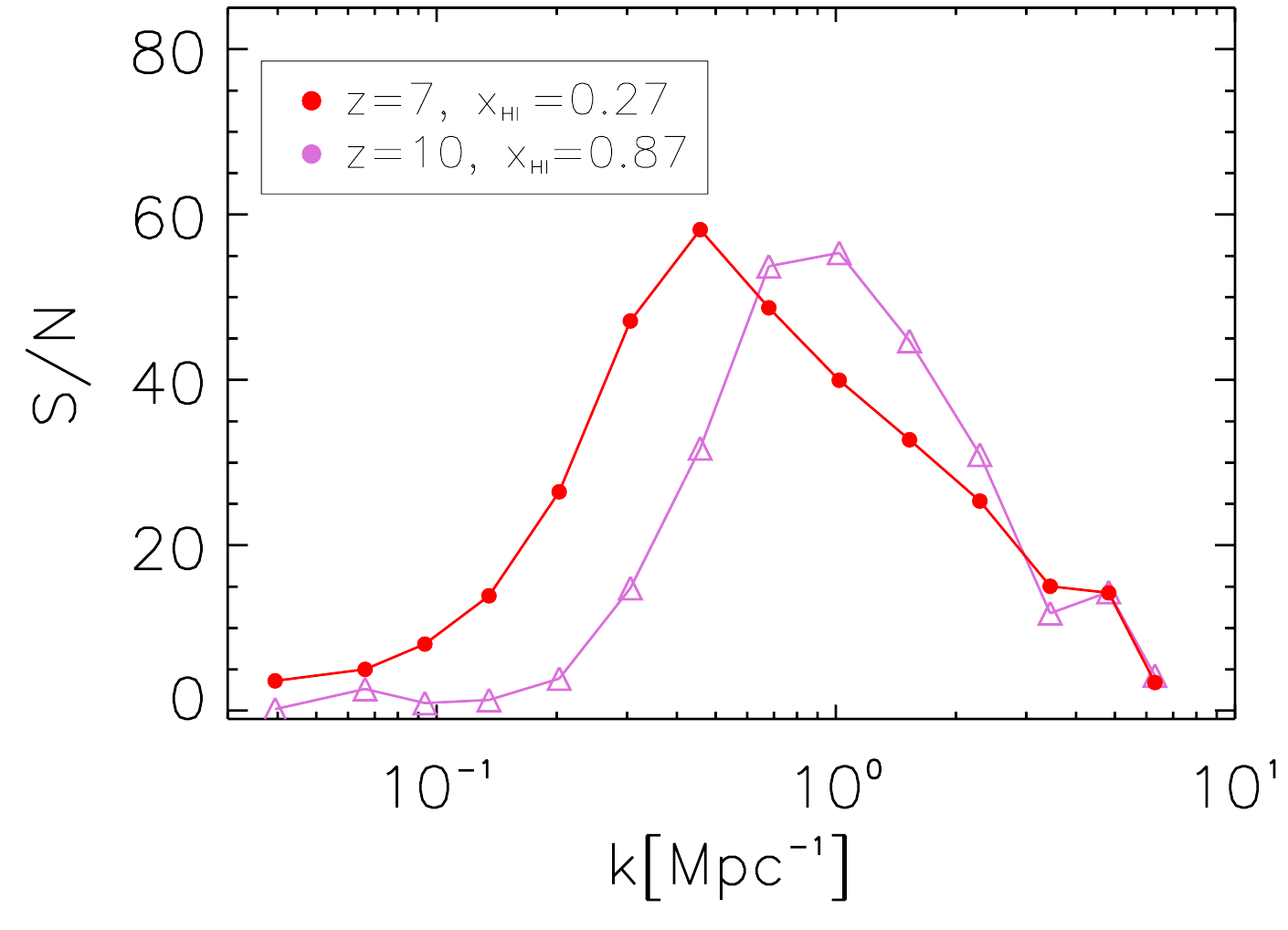}{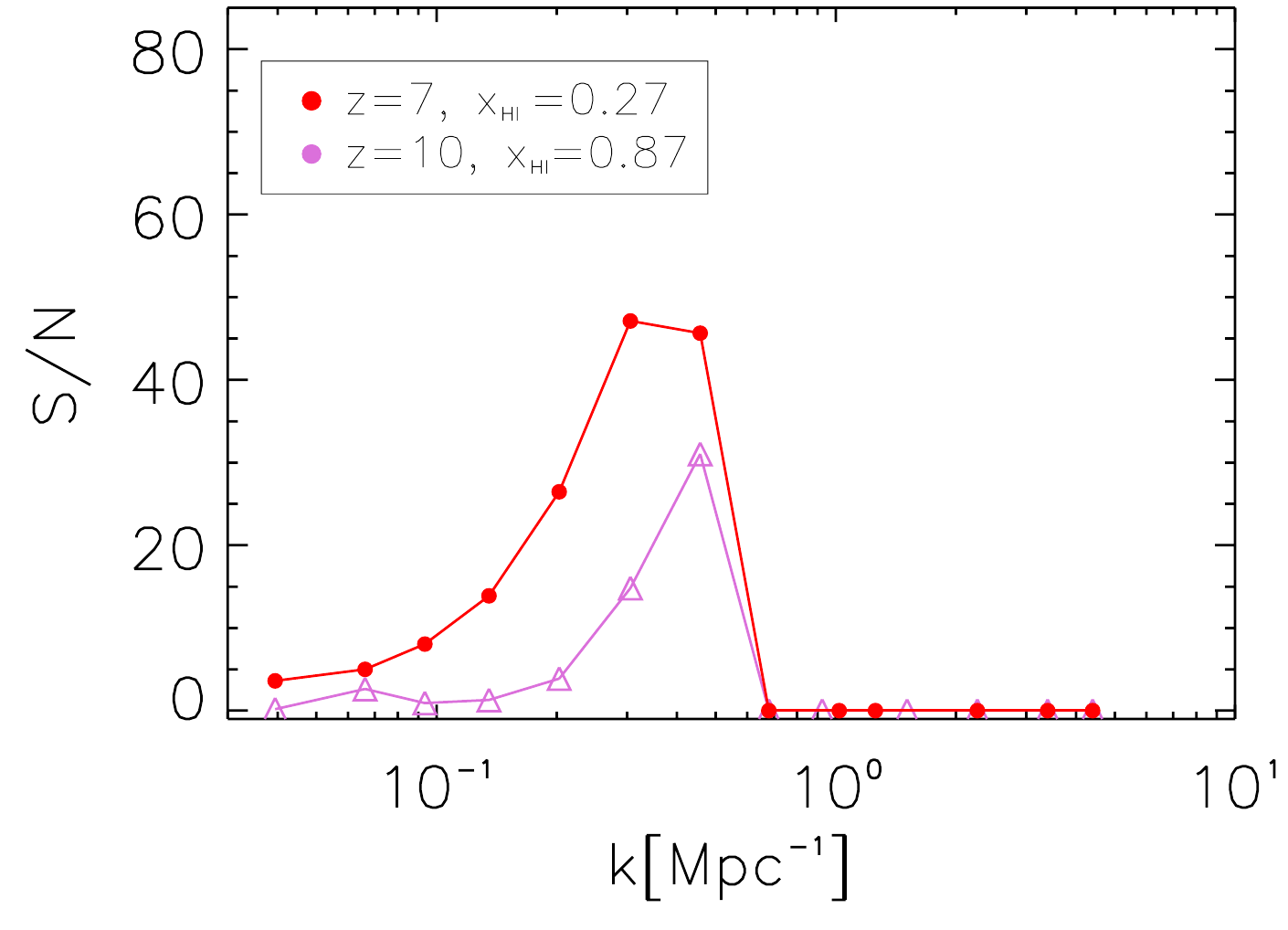}
\plottwo{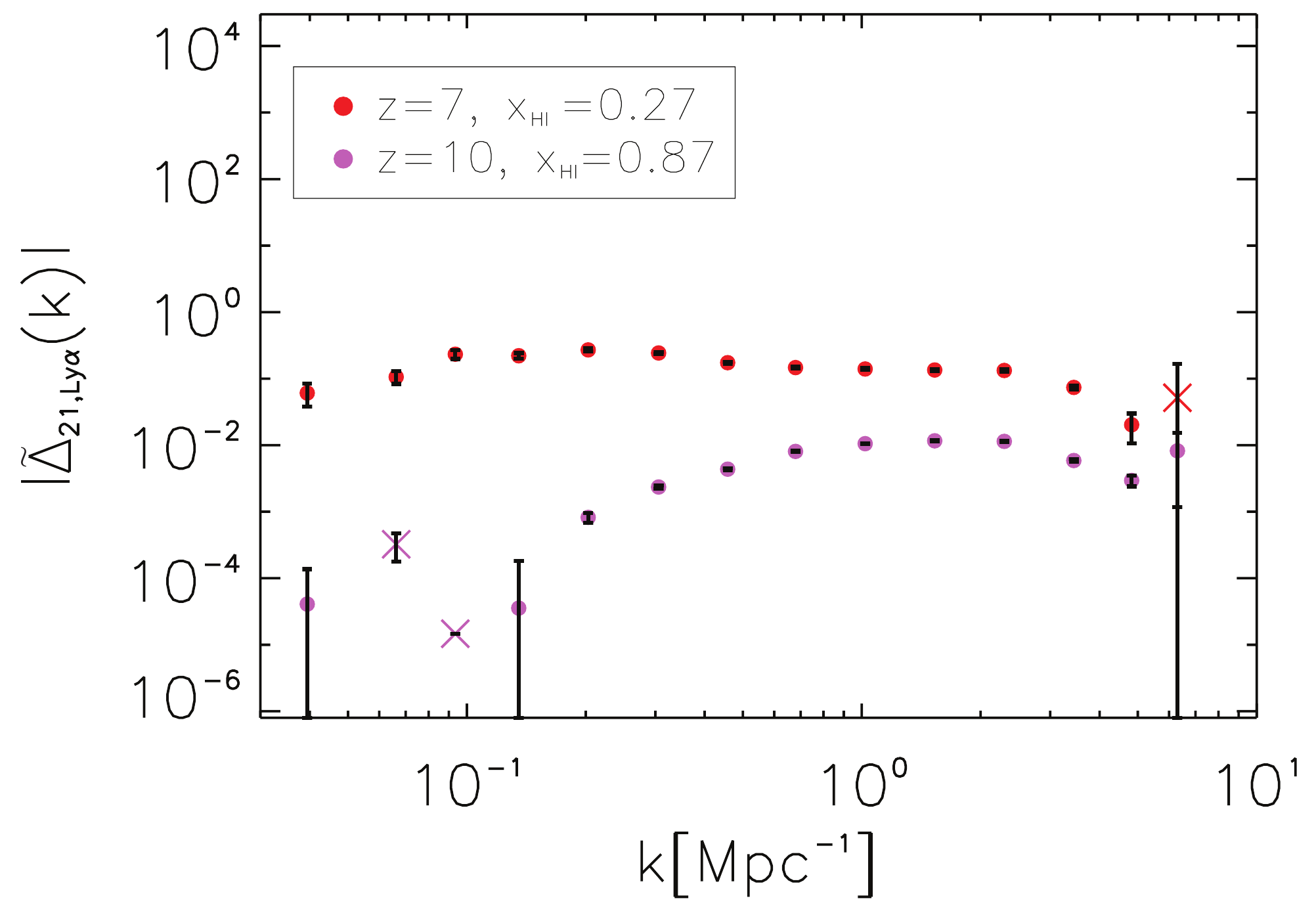}{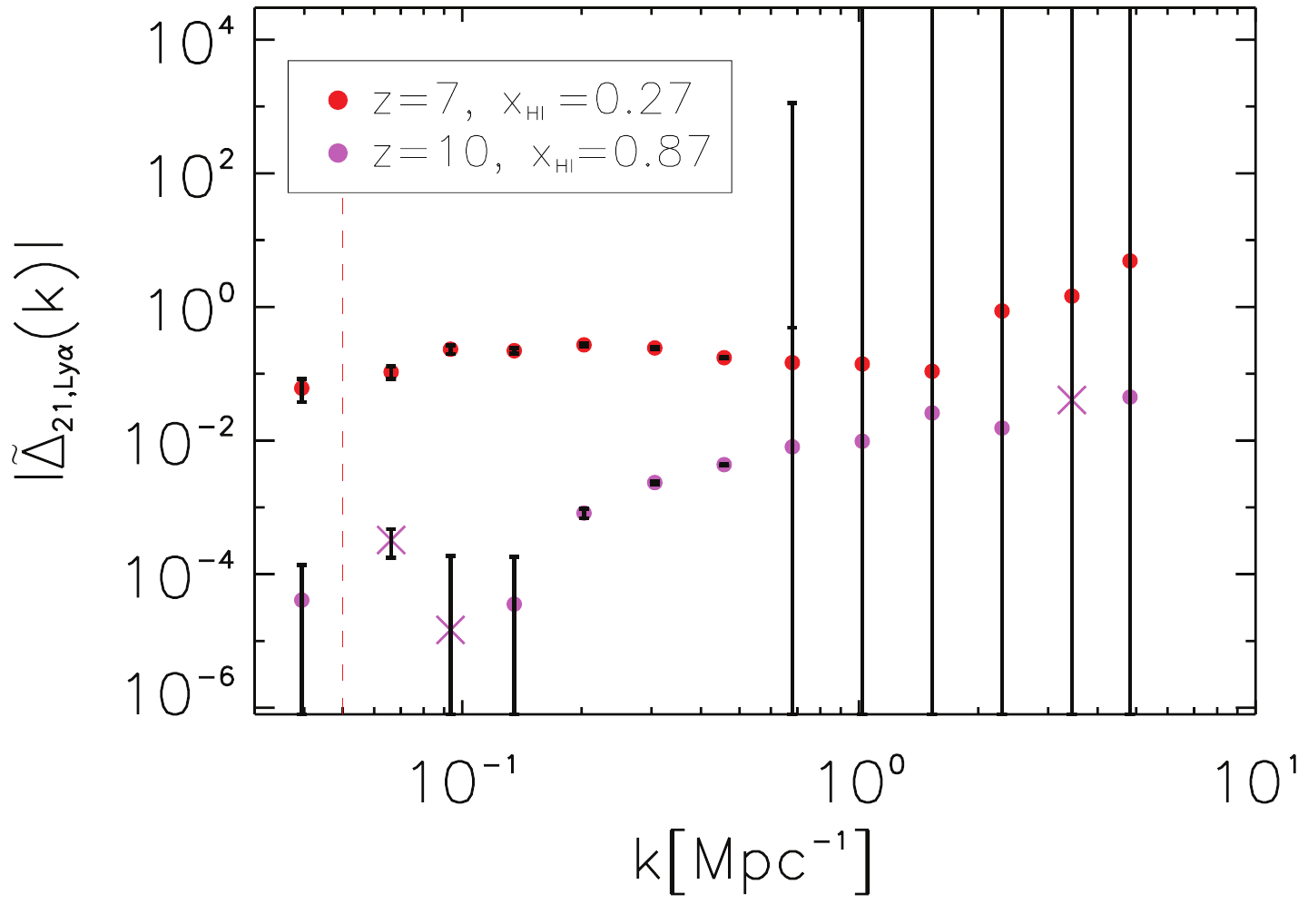}
\plottwo{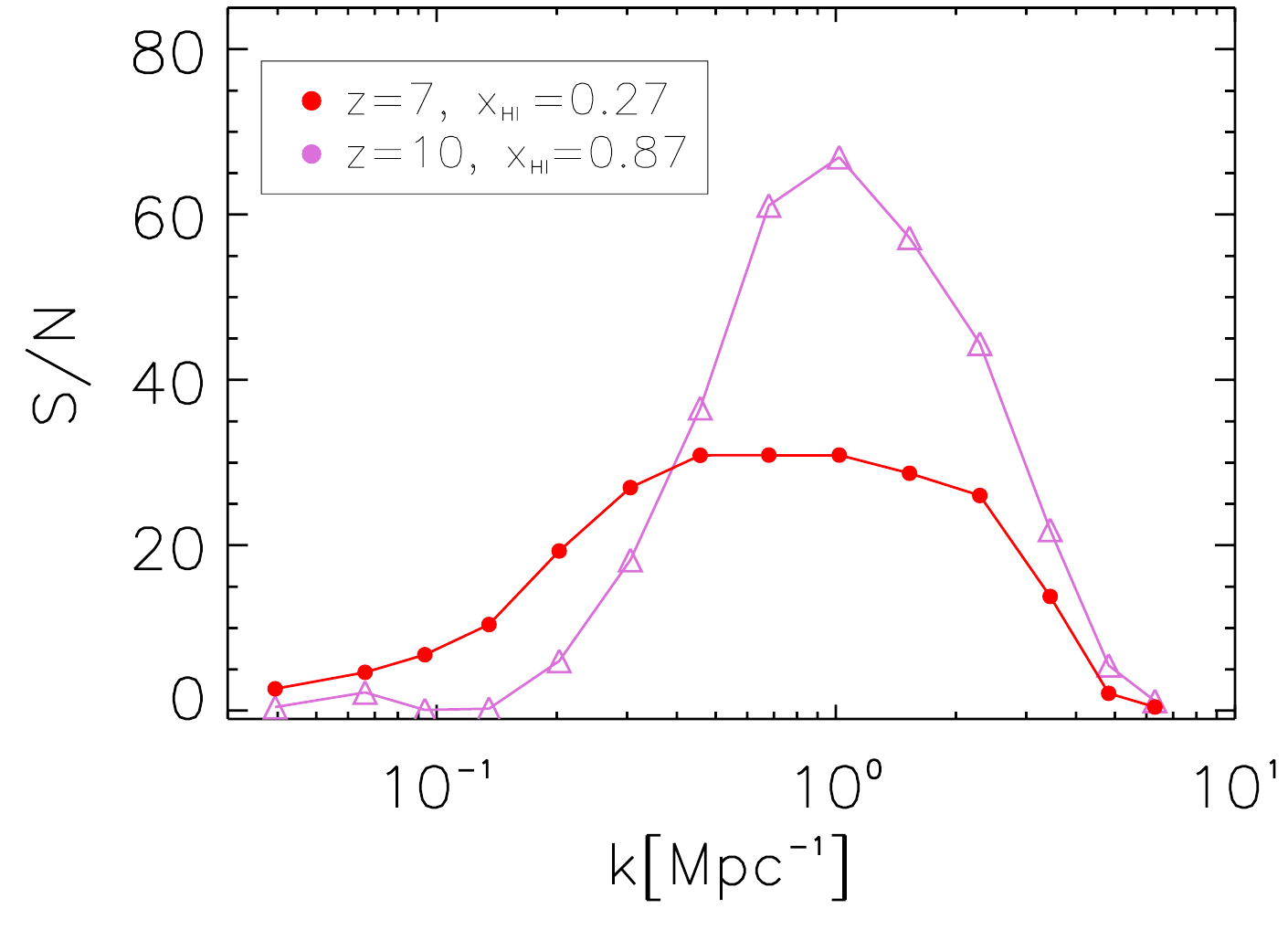}{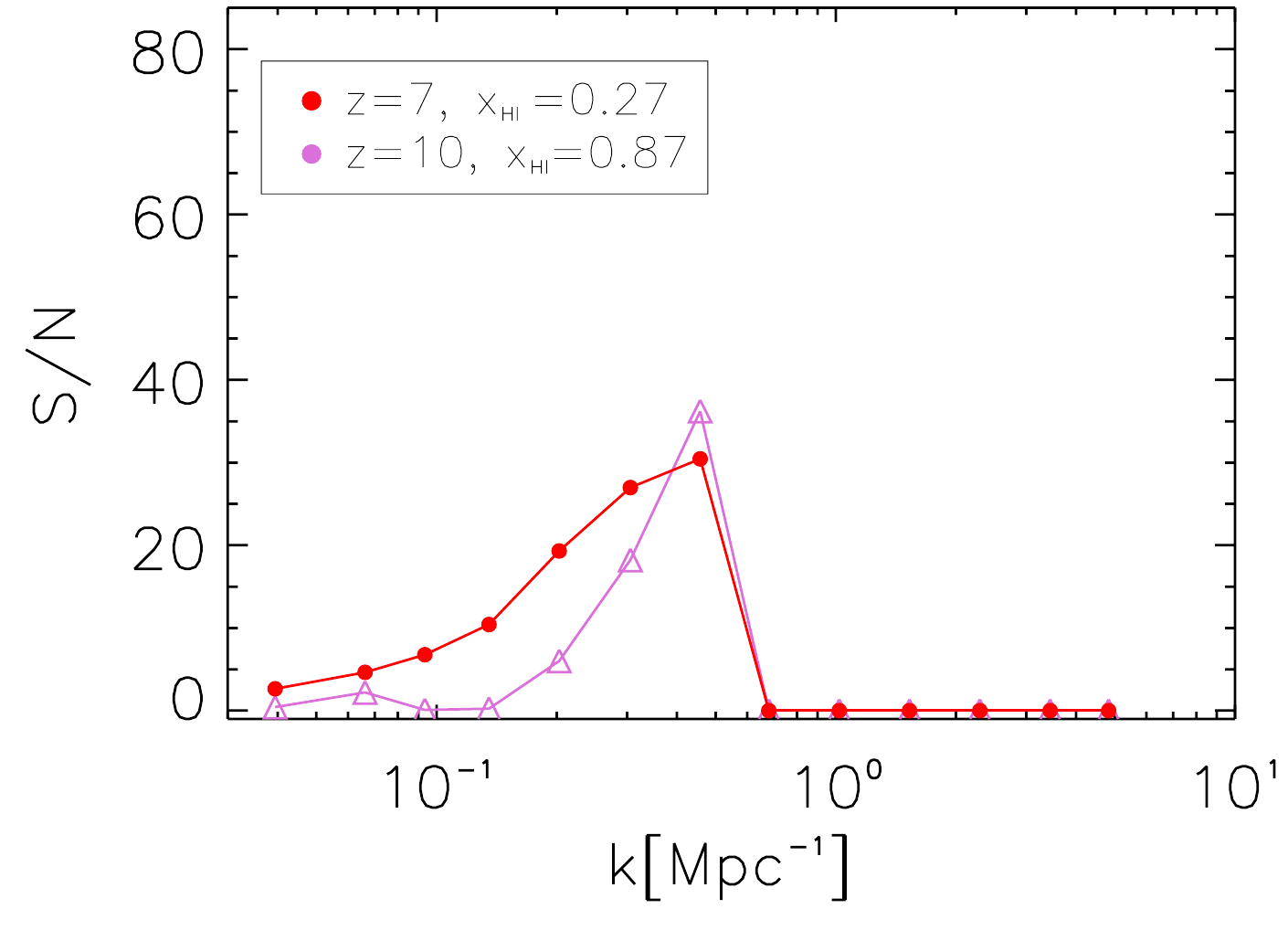}
\caption{Top two rows: dimensionless cross-correlation power spectra (top) and signal-to-noise ratio (bottom) of 21cm and total Ly$\mathrm{\alpha}$ fluctuations with error calculations including cosmic variance and thermal and instrumental noise for a survey of 21cm emission, type SKA stage 1, and a survey of Ly$\mathrm{\alpha}$ emission, type CDIM. For experiment characteristics, see Tables~\ref{tab:exp} and~\ref{tab:expL}; points denote negative and crosses positive cross-correlation. Left: cut of $k_{\parallel}>0.5\,$ Mpc$^{-1}$ (see discussion in Section~\ref{sec:LyaNoise} and Appendix~\ref{app:SN}). Right: cut of $k_{\parallel}>0.5\,$ Mpc$^{-1}$ and removal of the foreground wedge defined in Equation~(\ref{eq:wedge}) for survey characteristic angle $\theta_0\approx15^{\circ}$; scales roughly left of the vertical red dashed line might be lost to removal of smooth foregrounds; redshift $z=7$ and neutral fraction $\bar{x}_\mathrm{HI}=0.27$ in red, $z=10$ and $\bar{x}_\mathrm{HI}=0.87$ in orchid. All spectra include Ly$\alpha$ damping assuming the commonest filter scale as the typical size of an ionized region; see Section~\ref{sec:damp}. Bottom two rows: same as above, but power spectra include Ly$\alpha$ damping for tracing of ionized regions through the simulation along the z-axis LOS.} \label{fig:cross-noise}
\end{figure*}

\subsection{21 cm -- Ly$\mathrm{\alpha}$ Cross-power Spectrum}\label{sec:cross-noise}
We now consider the detectability of the 21cm -- Ly$\mathrm{\alpha}$ cross-power spectrum, a signal enabling us to constrain the structure and evolution of ionized regions in the IGM during the EoR.

For a single mode $k$ and angle $\mu$, the variance estimate of the cross-power spectrum reads~\citep{2007ApJ...660.1030F,2009ApJ...690..252L} as
\begin{equation}
\sigma^2_\mathrm{21,Ly\mathrm{\alpha}} \left( k,\mu\right) = \frac{1}{2}\left[  P_\mathrm{21,Ly\mathrm{\alpha}}^2\left( k,\mu\right) + \sigma_{21}\left( k,\mu\right)\sigma_\mathrm{Ly\mathrm{\alpha}}\left( k,\mu\right)\right] .
\end{equation}
Here, $P_\mathrm{21,Ly\mathrm{\alpha}}\left(k,\mu \right)$ is the 21cm -- Ly$\mathrm{\alpha}$ cross-power spectrum. The variances of the 21cm and Ly$\mathrm{\alpha}$ auto spectra are $\sigma_{21}\left( k,\mu\right)$ and $\sigma_\mathrm{Ly\mathrm{\alpha}}\left( k,\mu\right)$, respectively, and both encompass cosmic variance and instrumental and thermal noise as defined in equations~(\ref{eq:sigma21}) and~(\ref{eq:sigmaLya}).
The variance $\sigma^2_\mathrm{21,Ly\mathrm{\alpha}} \left(k \right)$ for the full spherically averaged power spectrum here too is the sum over the upper-half plane of angles $\mu$, or equivalently $k$ modes with $k^2 = k^2_{\parallel}+k^2_{\perp}$, divided by the respective number of modes per bin, as in Equation~(\ref{eq:sigma-tot}).
Note that the 21cm brightness temperature $T_\mathrm{b}$ has been converted to brightness intensity $I_{21}$ for the cross-power spectra shown in this section, using Planck's law at observed frequency $\nu$ as
\begin{equation}
I_{21} \left( \nu , T_\mathrm{b}\right) = \frac{2h\nu^3}{c^2} \left( e^{\frac{h_\mathrm{P} \nu}{k_\mathrm{B} T_\mathrm{b}}} -1 \right)^{-1} ,  \label{eq:nuI21}
\end{equation}
with Boltzmann constant $k_\mathrm{B}$ and Planck's constant $h_\mathrm{P}$.

Figure~\ref{fig:cross-noise} shows the dimensionless 21cm -- Ly$\mathrm{\alpha}$ noise cross-power spectra at redshift $z=10$ and $z=7$ and the corresponding detectable S/N, including cosmic variance, thermal noise, and instrumental resolution effects. Instrument specifications of the 21cm experiment are taken as in Table~\ref{tab:exp} and for the Ly$\mathrm{\alpha}$ experiment we take CDIM specifications as in Table~\ref{tab:expL}. The two top rows show the result for the 21cm -- Ly$\mathrm{\alpha}$ noise cross-power spectra when including Ly$\alpha$ damping assuming the commonest filter scale as the typical size of an ionized region; see Section~\ref{sec:damp}. The two bottom rows depict the same, but the power spectra include Ly$\alpha$ damping in the tracing of ionized regions through the simulation along the z-axis LOS. Note the sensitivity of the turnover to positive cross-correlation at high $k$, as well as of the predicted S/N, to the modeling of Ly$\alpha$ damping (top versus bottom rows). Concerning the S/N various effects compete; for example, different error contributions and their redshift behavior seem to be won over by a stronger 21cm signal at large $k$ for $z=10$, leading to high S/N in the bottom left panel.

For both left and right panels in Figure~\ref{fig:cross-noise} a cut of $k_{\parallel}>0.5\,$Mpc$^{-1}$ for CDIM is applied to avoid the impact of limited spectral resolution in our Ly$\mathrm{\alpha}$ experiment, as described in the previous Section~\ref{sec:LyaNoise} and Appendix~\ref{app:SN}. The right panels in addition show the impact of foreground avoidance for the 21cm signal, where we cut the so-called foreground wedge as described in Section~\ref{sec:21noise} for a characteristic scale of $\theta_0\approx15^{\circ}$. Cutting away the foreground wedge means cutting away higher perpendicular modes $k_{\perp}$, which together with the cut of $k_{\parallel}>0.5\,$Mpc$^{-1}$ degrades the signal at $k$ above that scale, but leaves the shape of the cross-correlation signal mostly unaltered. 

Measuring 21cm fluctuations in the foreground window might be possible, though, by dedicated foreground modeling~\citep{Liu:2014yxa,Wolz:2015sqa}, which improves the prospect of detecting of the 21cm -- Ly$\mathrm{\alpha}$ cross-correlation signal at higher $k$. Alternatively, a higher instrumental resolution and an adjustment of instrument specifications might even render the turnover in the cross-correlation signal around a couple of Mpc$^{-1}$ from negative to positive to be detectable.
For the optimistic case of improved foreground avoidance, a detection of the 21cm -- Ly$\mathrm{\alpha}$ cross-correlation signal is feasible over one to two decades in scale, depending on assumptions, and reaches a detectability above 5$\sigma$ confidence over about one to two decades in scale. Detecting the cross-power spectrum at high redshift for use in a joint analysis with power spectra themselves is therefore feasible.
It is possible to measure the varying morphology of the cross-correlation signal at different redshifts, which in turn depends on the morphology and ionization fraction of the IGM during reionization, and therefore on reionization model parameters.

\section{Discussion}
 We demonstrate the feasibility of detecting cross-power spectra with future  intensity mapping probes, by simulating fluctuations in 21cm, Ly$\mathrm{\alpha}$ and H$\mathrm{\alpha}$ emission. Fast and seminumerical modeling of different tracers will be crucial when constraining the EoR, probing the ionized and neutral medium back to when the first galaxies started to ionize the medium around them. Making use of information other than power spectra themselves will help to break degeneracies and constrain reionization model parameters. 

We started by presenting modeling and power spectra for 21cm emission tracing the neutral IGM, for Ly$\mathrm{\alpha}$ galactic, diffuse IGM, and scattered IGM components, as well as H$\mathrm{\alpha}$ emission. Proceeding to the cross-power spectra between 21cm emission  and different Ly$\mathrm{\alpha}$ components, we showed the variation of the cross-power signal with some of the model parameters, laying the groundwork for future parameter studies. On top of that, the cross-power spectrum between 21cm emission and lines other than Ly$\mathrm{\alpha}$ can be used to extract further information on the state of the IGM, as shown for the cross-correlation with H$\mathrm{\alpha}$ emission. Here the relative strengths of different Ly$\mathrm{\alpha}$ emission components can be extracted from the cross-correlation signal. We show the detectability of the 21cm and Ly$\mathrm{\alpha}$ cross-correlation signals with future probes like SKA and CDIM, and also for the case when the Ly$\mathrm{\alpha}$ damping tail and foreground avoidance are included in the error calculations.

To extend this study, further parameter explorations and a refinement of foreground treatment, as well as the derivation of possible future parameter constraints involving accurate seminumerical modeling, are needed. Together with further adjustment of the modeling in light of high-redshift data, as well as hydronumerical simulations, this will bring us closer to extracting as much information as possible about the high-redshift universe from upcoming intensity mapping experiments.

\acknowledgments
A.C. acknowledges support from NSF CAREER AST-0645427 and AST-1313319, and the NASA grants NNX16AF39G and NNX16AF38G.
C.F. acknowledges support from NASA grants NASA NNX16AJ69G and NASA NNX16AF39G.
The computational analysis was performed using the High Performance Computing HPC@UCPH, HPC facility at the University of Copenhagen. This work was partially supported by the SFB-Transregio TR33 ''The Dark Universe" and the DNRF. 

%
%We thank all the people that have made this AASTeX what it is today.  This
%includes but not limited to Bob Hanisch, Chris Biemesderfer, Lee Brotzman,
%Pierre Landau, Arthur Ogawa, Maxim Markevitch, Alexey Vikhlinin and Amy
%Hendrickson.


%% for the paper.  Each keyword is check against the master list during
%% copy editing.  Individual instruments can be provided in parentheses,
%% after the keyword, but they are not verified.

%\vspace{5mm}
%\facilities{HST(STIS), Swift(XRT and UVOT), AAVSO, CTIO:1.3m,
%\software{IRAF, cloudy, IDL}



\clearpage
\appendix
\section{Comparison of Ly$\mathrm{\alpha}$ spectra: other work}\label{app:Lya}
\begin{figure}[!htb]
%\figurenum{2}
\includegraphics[width=0.45\columnwidth]{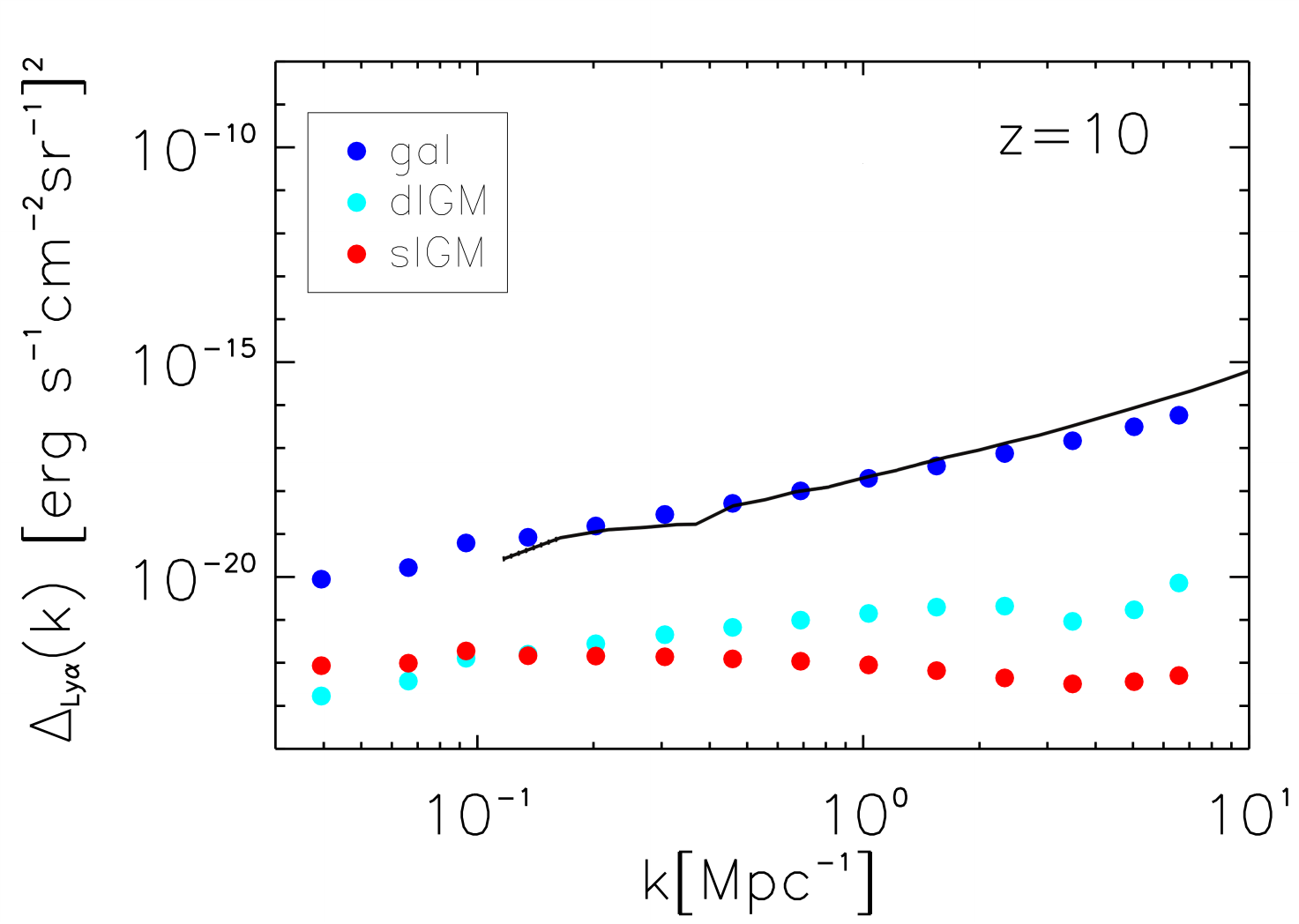}
\hfill
\includegraphics[width=0.45\columnwidth]{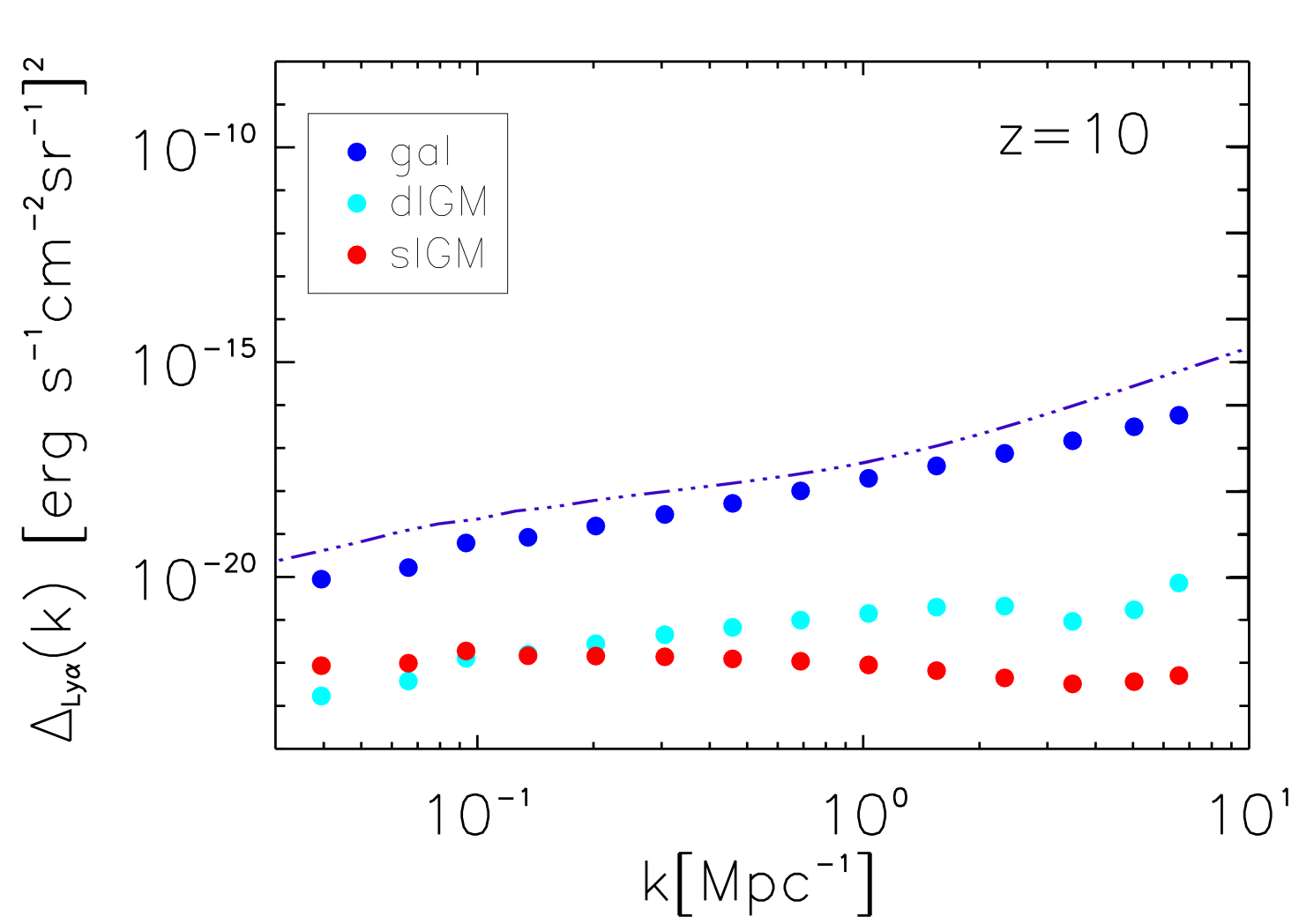}
\hfill
\includegraphics[width=0.45\columnwidth]{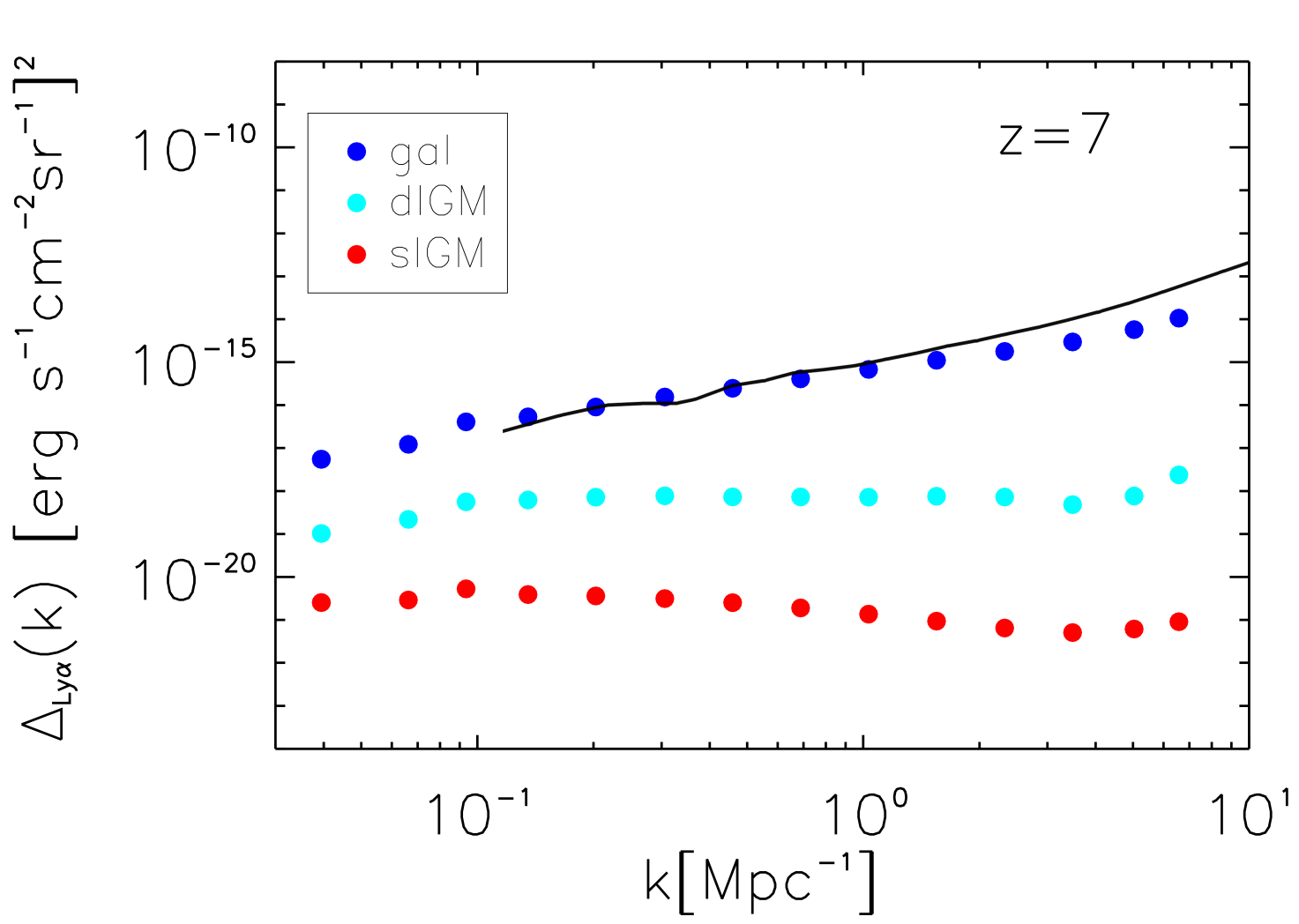}
\hfill
\includegraphics[width=0.45\columnwidth]{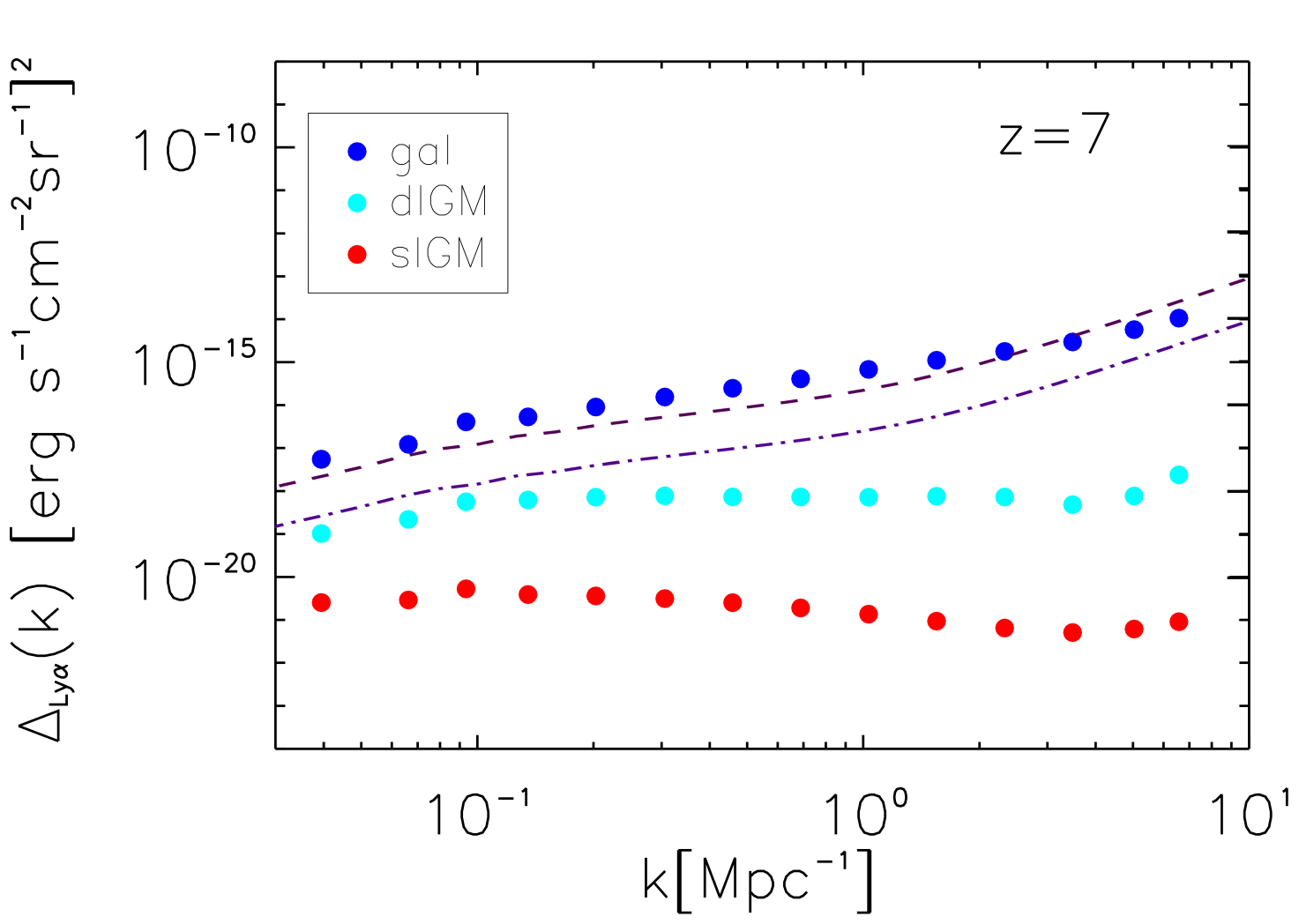}
\caption{Comparison of Ly$\mathrm{\alpha}$ power spectra in surface brightness $\left(\nu I_{\nu} \right)$ for galactic contribution, as well as diffuse and scattered IGM contributions (see Figure~\ref{fig:IGMgalz} in Section~\ref{sec:lya}), with spectra taken from~\citet{Silva12} (left, black lines) and~\citet{Pullen:2013dir} (right, top panel dash-dotted for $z=10$; bottom panel dashed for $z=6$ and dash-dotted for $z=8$).} \label{fig:compMP}
\end{figure}

Here we compare, for consistency, the Ly$\mathrm{\alpha}$ power spectra in surface brightness $\left(\nu I_{\nu} \right)$ obtained in this work for the galactic contribution, as well as diffuse and scattered IGM contributions (see Figure~\ref{fig:IGMgalz} in Section~\ref{sec:lya}), with Ly$\mathrm{\alpha}$ power spectra from other work. 
 Figure~\ref{fig:compMP} compares against the total galactic power spectrum from~\citet{Silva12} (black lines, left panels), and against the theoretical power spectrum for halo emission from~\citet{Pullen:2013dir} (dashed and dash-dotted lines, right panels), both at redshift $z=10$ (top) and $z=7$ (bottom). 
Encouragingly, the power spectra roughly agree with each other, especially given the differing approaches in modeling. 

In comparison to~\citet{Silva12}, who required an ionizing equilibrium by checking if the region's ionizing rate was equal to or higher than its recombination rate, we defined ionized regions via a fixed collapse fraction. In addition, diffuse IGM Ly$\alpha$ emission was taken into account in our study. Both our and the latter study made use of a seminumerical setup in the emission calculations, while~\citet{Pullen:2013dir} modeled the Ly$\alpha$ emission both by an approach based on the halo model (which we compare with here), assuming a Tinker fitting formula for the halo mass function~\citep{Tinker08}, as well as via an empirical model based on luminosity function measurements of Ly$\alpha$ emitters out to redshift $z\approx 8$. Also, the scattered IGM Ly$\alpha$ emission was neglected, which is included in our study. 

\clearpage
\section{S/N and mode cuts}\label{app:SN}
\begin{figure*}[!htb]
\plottwo{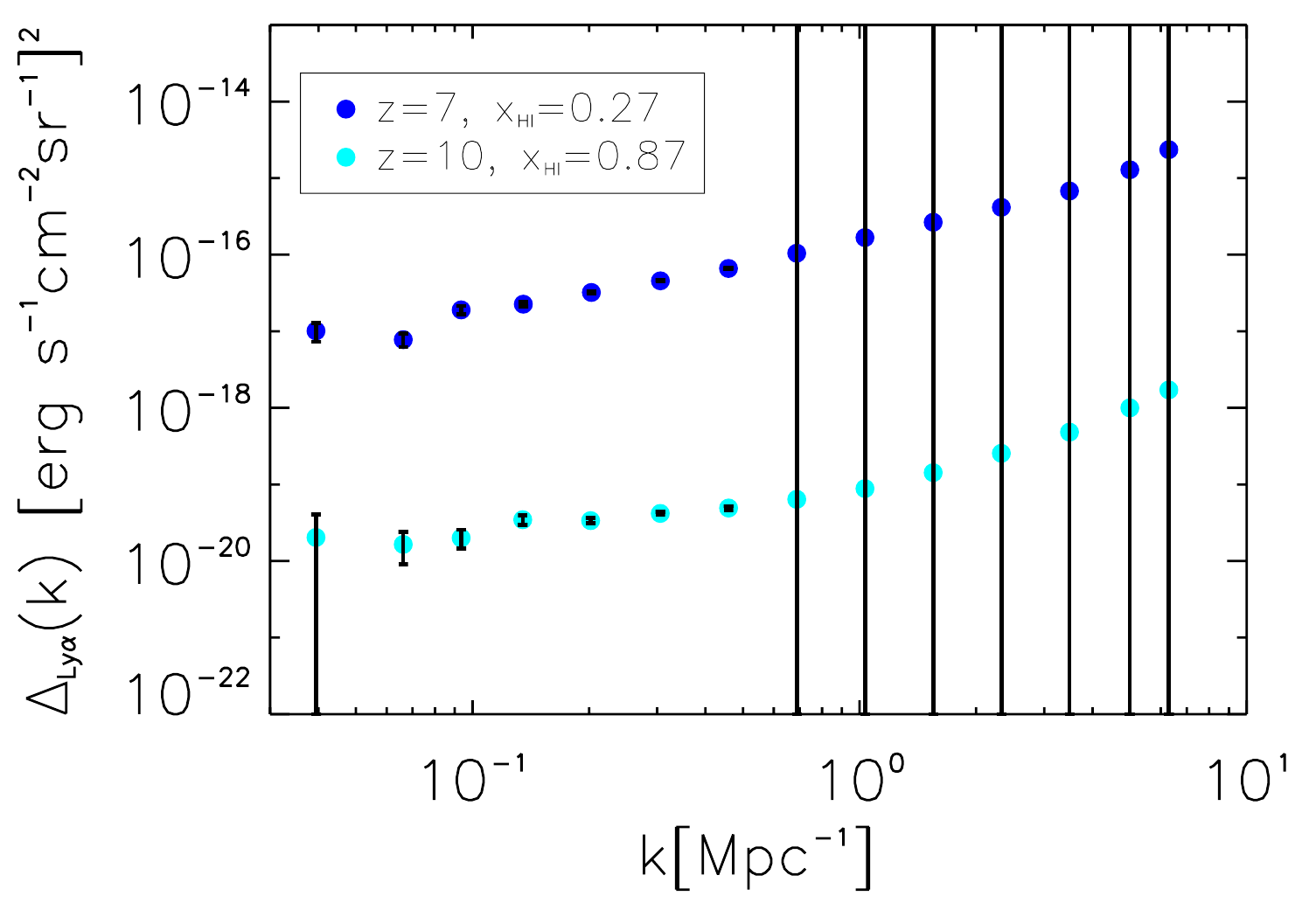}{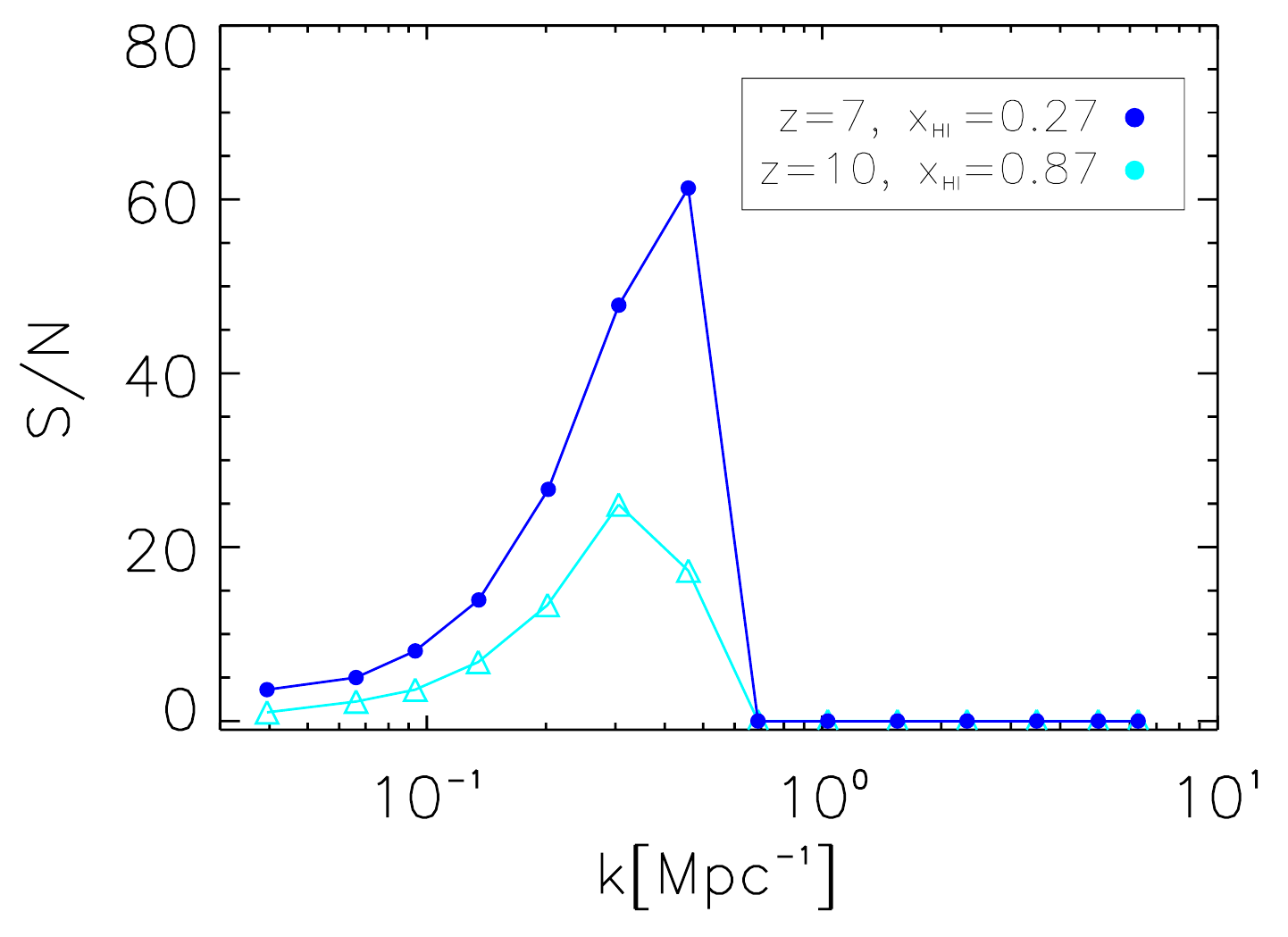}
\caption{Left: Ly$\mathrm{\alpha}$ noise power spectrum in surface brightness $\left(\nu I_{\nu} \right)$, including cosmic variance and thermal and instrumental noise for a CDIM-type survey. Right: corresponding detectability of the Ly$\mathrm{\alpha}$ power spectrum, showing the total S/N, with for example an S/N of 10 indicating a detection at 10$\sigma$ confidence; redshift $z=7$ and neutral fraction $\bar{x}_\mathrm{HI}=0.27$ in blue, $z=10$ and $\bar{x}_\mathrm{HI}=0.87$ in cyan.} \label{fig:lyasim-SN}
\end{figure*}

For completeness, we show here the Ly$\mathrm{\alpha}$ power spectra in surface brightness $\left(\nu I_{\nu} \right)$ for redshift $z=10$ and $z=7$ in Figure~\ref{fig:lyasim-SN} (left panel), including cosmic variance and thermal and instrumental noise, but before mode cuts have been applied. The sharp drop-off in S/N around $k=0.5\,$Mpc$^{-1}$ for CDIM (right panel) is due to the spectral resolution limit in parallel modes for CDIM. For SPHEREx, the corresponding sharp drop-off is at $k=0.06\,$Mpc$^{-1}$. We therefore chose for all plots shown in Sections~\ref{sec:LyaNoise} and~\ref{sec:cross-noise} to cut  all modes $k_{\parallel}<0.5\,$ Mpc$^{-1}$ for CDIM and  $k_{\parallel}<0.06\,$ Mpc$^{-1}$ for SPHEREx, around the mode where the S/N drops below one, in order to avoid instrumental noise dominating the signal. 

%% The reference list follows the main body and any appendices.s



%\vspace{1cm}
\clearpage
\bibliography{references}{} % if your bibtex file is called example.bib
\bibliographystyle{apj}

%\begin{thebibliography}{}
%
%%\bibitem[Corrales(2015)]{2015ApJ...805...23C} Corrales, L.\ 2015, \apj, 805, 23
%%\bibitem[Hanisch \& Biemesderfer(1989)]{1989BAAS...21..780H} Hanisch, R.~J., \& Biemesderfer, C.~D.\ 1989, \baas, 21, 780 
%%\bibitem[Lamport(1994)]{lamport94} Lamport, L. 1994, LaTeX: A Document Preparation System, 2nd Edition (Boston, Addison-Wesley Professional)
%%\bibitem[Schwarz et al.(2011)]{2011ApJS..197...31S} Schwarz, G.~J., Ness, J.-U., Osborne, J.~P., et al.\ 2011, \apjs, 197, 31  
%%\bibitem[Vogt et al.(2014)]{2014ApJ...793..127V} Vogt, F.~P.~A., Dopita, M.~A., Kewley, L.~J., et al.\ 2014, \apj, 793, 127  
%
%\end{thebibliography}



\end{document}